\documentclass[12pt]{article}

\usepackage{amssymb,amsmath,amsfonts,eurosym,geometry,ulem,graphicx,caption,color,setspace,sectsty,comment,footmisc,caption,pdflscape,array,hyperref, bbm}
\usepackage[authoryear, round]{natbib}
\usepackage{placeins}
\usepackage[utf8]{inputenc}
\usepackage{titling}
\usepackage{subcaption}
\usepackage{float}
\usepackage[table,xcdraw]{xcolor}
\usepackage[flushleft]{threeparttable}
\usepackage[english]{babel}
\usepackage{ragged2e}
\usepackage{amsthm}
\usepackage{bm}
\usepackage{booktabs}
\usepackage{adjustbox}
\usepackage{amsbsy}
\usepackage{fancyhdr}
\usepackage{econometrics}
\usepackage{parskip}
\usepackage{titletoc}

\usepackage{tikz}
\usepackage{pgfplots}
\usepgfplotslibrary{patchplots}
\pgfplotsset{compat=newest}

\newlength\figureheight
\newlength\figurewidth

\usepackage[T1]{fontenc}
\usepackage{sourcesanspro}
\usepackage{newpxmath}
\usepackage{microtype}
\usepackage[scaled]{helvet}
\usepackage{courier}

\usepackage[affil-it]{authblk}
\usepackage{enumerate}
\usepackage{relsize}
\usepackage{dsfont}
\usepackage{mathtools}
\usepackage{tabularx}
\usepackage{epstopdf}
\usepackage{rotating}
\usepackage{chngcntr}
\usepackage{apptools}
\usepackage{graphics}
\extrafloats{100}

\newtheorem{theorem}{Theorem}
\newtheorem{corollary}{Corollary}
\newtheorem{proposition}{Proposition}
\newtheorem{assumption}{Assumption}
\newtheorem{remark}{Remark}
\newtheorem{definition}{Definition}
\renewenvironment{proof}[1][Proof]{\noindent\textbf{#1.} }{\ \rule{0.5em}{0.5em}}

\DeclareFontFamily{U}{mathx}{\hyphenchar\font45}
\DeclareFontShape{U}{mathx}{m}{n}{<-> mathx10}{}
\DeclareSymbolFont{mathx}{U}{mathx}{m}{n}
\DeclareMathAccent{\widebar}{0}{mathx}{"73}

\newtheorem{mylem}{Lemma}

\theoremstyle{definition}{}
\theoremstyle{definition}{\newtheorem{exmp}{Example}}
\theoremstyle{definition}{}

\AtAppendix{\counterwithin{mythm}{section}}
\AtAppendix{\counterwithin{myprop}{section}}
\AtAppendix{\counterwithin{mylem}{section}}
\AtAppendix{\counterwithin{mydef}{section}}
\AtAppendix{\counterwithin{mycor}{section}}

\newcommand{\beginsupplement}{\setcounter{section}{0}
    \renewcommand{\thesection}{S\arabic{section}}\setcounter{table}{0}
    \renewcommand{\thetable}{S\arabic{table}}\setcounter{figure}{0}
    \renewcommand{\thefigure}{S\arabic{figure}}\setcounter{equation}{0}
    \renewcommand{\theequation}{S.\arabic{equation}}\setcounter{page}{1}
    \renewcommand{\thepage}{S.\arabic{page}}}

\newcolumntype{L}[1]{>{\raggedright\let\newline\\arraybackslash\hspace{0pt}}m{#1}}
\newcolumntype{C}[1]{>{\centering\let\newline\\arraybackslash\hspace{0pt}}m{#1}}
\newcolumntype{R}[1]{>{\raggedleft\let\newline\\arraybackslash\hspace{0pt}}m{#1}}

\DeclareCaptionLabelSeparator*{spaced}{\\[2ex]}
\def\E{{\mathbb E}}

\def\V{{\mathbb V}}

\newcommand{\Var}{{\mathbb{V}\text{ar}}}

\renewcommand{\va}{\mathbf a}

\renewcommand{\mJ}{\mathbf J}

\renewcommand{\mR}{\mathbf R}
\renewcommand{\vr}{\mathbf r}

\renewcommand{\mW}{\mathbf W}
\renewcommand{\vw}{\mathbf w}

\renewcommand{\vx}{\mathbf x}

\renewcommand{\mZ}{\mathbf Z}
\renewcommand{\vz}{\mathbf z}

\renewcommand{\vbeta}{\bm \beta}

\renewcommand{\vepsi}{\bm \epsi}

\renewcommand{\vtheta}{\bm \theta}
\renewcommand{\viota}{\bm \iota}

\renewcommand{\vmu}{\bm \mu}
\renewcommand{\vnu}{\bm \nu}

\renewcommand{\vomega}{\bm \omega}

\def\one{\mathds{1}}

\newcommand{\norm}[1]{\left\lVert#1\right\rVert}

\newcommand{\papertitle}{Sharpe Ratio Inference after Machine Learning}

\makeatletter
\IfPackageAtLeastTF{titling}{2026/10/12}{
  \posttitle{\par\end{center}\vskip 0.5em}
  \renewcommand{\@maketitle}{
    \newpage
    \null
    \vskip 2em
    \vspace*{\droptitle}
    \maketitlehooka
    {\@bspretitle \@title \@bsposttitle}
    \maketitlehookb
    {\@bspreauthor \@author \@bspostauthor}
    \maketitlehookc
    {\@bspredate \@date \@bspostdate}
    \maketitlehookd
    \par
    \vskip 1.5em}
}{}
\makeatother

\begin{document}

\begin{titlepage}
\singlespacing
\setlength{\affilsep}{0pt}
\title{\papertitle\thanks{We are very grateful to Peter Boswijk, Yi He and seminar and conference participants at the 2026 Netherlands Econometric Study Group, the 2026 Symposium on Econometric Theory and Applications at the University of Tokyo, the 2026 QFFE event at Aix-Marseille Universit\'e, the 2026 International Symposium on Nonparametric Statistics in Thessaloniki, the 2026 Vienna-Copenhagen Conference on Financial Econometrics at the University of Vienna, the 2026 Financial Econometrics Meets Machine Learning Conference at the University of Geneva, and the University of Amsterdam for useful comments and suggestions. All remaining errors are our own. An earlier version of this paper was circulated and presented under the title ``Debiased Tests for Sharpe Ratios of Machine Learning Portfolios". \emph{Email Addresses:} {\tt \href{mailto:s.c.barendse@uva.nl}{s.c.barendse@uva.nl}}, {\tt \href{mailto:m.cheng@uva.nl}{m.cheng@uva.nl}}} 
}
\author[1]{Sander Barendse}
\author[1,2]{Ming Cheng}
\affil[1]{\footnotesize Amsterdam School of Economics, University of Amsterdam}
\affil[2]{\footnotesize Tinbergen Institute}
\date{\small 
This version: November 20, 2026}

\setlength{\droptitle}{-2.3cm}
\maketitle
\thispagestyle{empty}
\vspace{-1em}
\begin{abstract}
\fontsize{12pt}{13.8pt}\selectfont
\noindent 
The economic value of return signals is often assessed using the Sharpe ratios of portfolios based on those signals. In modern asset pricing, return signals often depend on many features, including state variables and firm-specific characteristics, and machine-learning methods are increasingly used to estimate them. But how much confidence can we place in the estimated Sharpe ratios? We find that signal-estimation error can remain consequential for inference even in large asset panels. Moreover, pooling observations across assets does not necessarily make this uncertainty negligible when predictors contain common features or exhibit substantial cross-sectional dependence. As a remedy, we develop a debiased Sharpe-ratio estimator that accounts for signal-estimation error and provides valid confidence intervals and tests against benchmarks. Our target is the population Sharpe ratio of a specified portfolio rule evaluated at the true conditional expected returns given the features. The framework covers market-timing, long-short, and other signal-based portfolios, allowing the number of assets to grow with and greatly exceed the time-series sample size. The method combines orthogonal portfolio moments with time-series cross-fitting and remains valid under suitable estimation-error bounds for machine-learning forecasts. Empirically, we revisit the setting of \citet{gu2020empirical} using machine-learning estimators of the return signals. The strongest long-short Sharpe-ratio gains relative to market benchmarks remain significant after debiasing, providing statistical evidence of the economic value of their feature set. Conventional and debiased Sharpe-ratio estimates are generally similar, while the associated confidence intervals and tests can differ substantially. These differences vary across signal estimators and portfolio rules and do not uniformly weaken the evidence of portfolio performance.


\bigskip
\end{abstract}
\end{titlepage}
\pagebreak \newpage

\doublespacing

\section{Introduction} \label{sec:introduction}
Return predictability matters to investors when it can be translated into better portfolio decisions. Modern asset-pricing research seeks this information in large sets of state variables and firm-specific characteristics, increasingly using machine-learning (ML) methods to estimate the return signals. Portfolios based on these signals provide a way to assess their economic value, for example through their Sharpe ratios. But a high estimated Sharpe ratio need not provide strong statistical evidence of the signal's economic value. How much confidence can we place in the estimated Sharpe ratio when the return signal is itself estimated? This paper develops confidence intervals and tests that account for uncertainty in the estimated return signals.

Predictive return models translate information into investment decisions through portfolio weights that depend on predicted next-period returns. Market-timing portfolios scale exposure to an index with its predicted return \citep[see, e.g.,][]{CampbellThompson2008,welch2008comprehensive,gu2020empirical,kelly2023virtue}. In particular, \citet{CampbellThompson2008} show that predictive regressions with an out-of-sample R-squared as low as 1\% can generate economically meaningful gains relative to static portfolios. \citet{gu2020empirical}, henceforth GKX, use machine-learning methods to predict excess stock returns in a large asset panel. Using pooled estimation, which is becoming increasingly popular (see \citet{hellum2025power} and references therein), they obtain stock-level return predictions from a large feature set and aggregate these predictions to guide exposure to index portfolios. They find substantial Sharpe-ratio improvements for some forecasting methods relative to buy-and-hold strategies. GKX also construct long-short portfolios from the stock-level predictions, buying stocks with high predicted returns and shorting those with low predictions. These applications are part of a broader literature on the use of large feature sets and flexible models in asset pricing \citep[see, e.g.,][]{rapach2010out,kelly2019characteristics,feng2020taming,freyberger2020dissecting,kozak2020shrinking,bianchi2021bond,leippold2022machine,chen2024deep}; \citet{kelly2023financial} provide a recent survey.

The economic value of a feature set can be assessed through the population Sharpe ratio of a specified portfolio rule using the true return signals, by which we mean the conditional expected returns given those features. Holding the feature set and portfolio rule fixed, different forecasting methods provide alternative estimates of the same signals and, consequently, of the same Sharpe ratio. When signal-estimation error is consequential, conventional confidence intervals need not provide valid inference on this population Sharpe ratio. We develop a method that accounts for signal-estimation error and provides valid confidence intervals when the machine-learning forecasts meet suitable accuracy requirements.

We show that uncertainty from estimating the return signal can invalidate conventional inference on this target based on the standard Sharpe-ratio estimator. A large asset cross-section does not necessarily resolve the problem. For example, GKX employ a pooled-regression approach, which increases the number of observations available for training. Pooling is no panacea, however: common features or unrestricted cross-sectional dependence among asset-specific features can leave signal-estimation error too large for conventional inference on the population Sharpe ratio. The relevant issue is not whether an ML method predicts better than a low-dimensional benchmark, but whether its estimation error is sufficiently small to be ignored when assessing the economic value of the signal.

In this paper, we develop a debiased Sharpe-ratio estimator that is less sensitive to signal-estimation error for market-timing, long-short, and other signal-based portfolios. It provides valid confidence intervals for the population Sharpe ratio and for differences relative to benchmarks, under suitable conditions on signal estimation. Benchmark comparisons can be used to test whether a signal-based rule has a different Sharpe ratio from a buy-and-hold portfolio or another investment strategy. The procedure allows valid inference even when signal-estimation error is too large for conventional methods to ignore. Our framework accommodates large asset panels in which the number of assets grows with, and may greatly exceed, the time-series sample size.

Empirically, we revisit the GKX setting using machine-learning estimators of the return signals. Our results reinforce the economic case of the GKX feature set when used through the portfolio rules considered, as revealed through machine-learning estimation of the underlying return signals. These gains are concentrated in long--short portfolios, whereas no market--timing strategy significantly outperforms any benchmark under either procedure. Accounting for forecast estimation materially changes the benchmark evidence, particularly for some of the more flexible specifications. Although debiasing can substantially widen confidence intervals, especially for highly flexible specifications, it can also narrow them in other settings.These findings underscore that assessing the economic value of high-dimensional feature sets in modern asset-pricing applications requires careful consideration of signal-estimation uncertainty.

Our new methodology combines Neyman-orthogonal portfolio moments with a cross-fitting design adapted to time-series data. We construct orthogonal moment conditions for the first and second moments of portfolio returns, removing the first-order effect of signal-estimation error under suitable conditions. These moments yield a debiased estimator of the Sharpe-ratio, which is a function of the two portfolio moments. Our sample-splitting design is adapted to standard asset-pricing settings, accommodating fixed, moving, and expanding estimation windows. We establish asymptotic validity and uniform inference for independent observations and temporally dependent triangular arrays. Moreover, we show that the debiased estimator is semiparametrically efficient under independent and identically distributed observations. This demonstrates that accounting for signal-estimation error is compatible with efficient estimation, consistent with our finding that debiased confidence intervals are not uniformly wider than conventional ones. We also verify the required signal-estimation bounds for a pooled neural-network estimator. A simulation study shows that debiasing reduces false rejections and brings confidence-interval coverage closer to its intended level for market-timing and long-short portfolios.

Our analysis relates to the finance literature on estimation risk. \citet{kan2007optimal} characterize the consequences of using estimated means and covariances for optimal portfolio choice. \citet{kan2024sample} study how estimation risk separates the in-sample and out-of-sample Sharpe ratios of multifactor models. \citet{barillas2020model} develop model-comparison tests based on squared maximum Sharpe ratios, accounting for estimation error in mimicking-portfolio weights. We study signal-based rules whose weights depend on high-dimensional estimated return signals. This brings the estimation-risk question to portfolios formed from flexible conditional-mean models.

Our debiased estimator also generalizes classical Sharpe-ratio inference \citep{jobson1981performance,memmel2003performance,lo2002statistics,LedoitWolf2008}. These existing methods conduct inference on population Sharpe ratios from observed portfolio returns. We additionally account for uncertainty from estimating the return signals, targeting the Sharpe ratio of a portfolio rule evaluated at the true signals.
Although it is conceivable that resampling methods could lead to valid inference on our Sharpe ratio target, such methods have not yet been developed to the best of our knowledge and would be computationally expensive, requiring refitting of the machine-learning methods for each resampling iteration. \citet{liao2025uncertainty} develop a bootstrap for neural-network predictions, that reduces this cost by starting from the original fit and updating it on each resample, but do not relate it to Sharpe ratio inference. Our procedure requires no bootstrap replications and hence no repeated fitting or retuning for that purpose. Once forecasts satisfying the sample-splitting requirements and the required innovation-covariance input are available, implementation consists of evaluating the portfolio rule and its derivatives and computing the debiased sample moments and standard errors.

Finally, we build on the literature on double/debiased machine learning \citep{Chernozhukov2018Double}. Recent applications in finance include inference for conditional latent factors, risk premia, and stochastic discount factors \citep{gagliardini2026double}, and correction for selection in corporate emissions data \citep{chen2026green}. In a related and contemporaneous paper, \citet{quaini2026orthogonal} develops orthogonal inference for comparing univariate forecasts in a Diebold-Mariano framework, when the evaluation loss differs from the training objective, but does not treat the Sharpe ratio.

The remainder of the paper is organized as follows. Section~\ref{sec:theory} introduces the return model and signal-based portfolios, develops debiased inference for Sharpe ratios and their benchmark differences, and presents the IID asymptotic theory, neural-network rate conditions, and efficiency results. Sections~\ref{sec:simulation} and~\ref{sec:empirical} report the simulation and empirical evidence, respectively, and Section~\ref{sec:conclusion} concludes. The Supplement provides technical details, proofs, and additional results.

We use the following notation. Bold symbols denote vectors or matrices, and a prime denotes transpose. We write $\mathbf 0$, $\mathbf 1$ (also $\viota_N$ for the $N$-vector of ones), and $\mathbf I_d$ for zero vectors, vectors of ones, and the $d\times d$ identity matrix, with dimensions otherwise understood from context. The sets $\mathbb R$, $\mathbb N_0$, and $\mathbb N_+$ denote the real numbers, nonnegative integers, and positive integers; $\lfloor x\rfloor$ and $\lceil x\rceil$ denote floor and ceiling. For a vector $\mathbf a=(a_1,\ldots,a_d)'$, $\|\mathbf a\|_q=(\sum_{j=1}^d|a_j|^q)^{1/q}$ for $1\leq q<\infty$ and $\|\mathbf a\|_\infty=\max_j|a_j|$; an unspecified vector norm is Euclidean. For scalar- or vector-valued $f$ and $1\leq q<\infty$, $\|f(X)\|_{P,q}=(\mathbb E_P[\|f(X)\|_2^q])^{1/q}$ and $\|f(X)\|_{P,\infty}=\operatorname*{ess\,sup}_P\|f(X)\|_2$, with absolute value replacing the inner norm for scalar $f$. Here $\mathbb E_P$ and $\mathbb V_P$ denote expectation and variance under law $P$. For a matrix $\mathbf A$, $\|\mathbf A\|_{\mathrm{op}}=\sqrt{\lambda_{\max}(\mathbf A'\mathbf A)}$; $\lambda_{\min}$ and $\lambda_{\max}$ denote the smallest and largest eigenvalues of a symmetric matrix. We write $\operatorname{diag}(\mathbf a)$ for the diagonal matrix with diagonal $\mathbf a$, $\odot$ for the Hadamard product. For nonnegative sequences, $a_T\lesssim b_T$ means $a_T\leq Cb_T$ for all sufficiently large $T$, for an absolute constant $C>0$ independent of $T$, and $a_T\asymp b_T$ means that both $a_T\lesssim b_T$ and $b_T\lesssim a_T$ hold. The symbols $O$ and $o$ have their usual deterministic meanings; $O_P(a_T)$ and $o_P(a_T)$, for positive deterministic $a_T$, denote boundedness and convergence to zero, respectively, after division by $a_T$ in $P$-probability. Lowercase subscripts $p$ suppress the law, and uniformity over $P\in\mathcal P$ is stated explicitly. The symbol $\rightsquigarrow$ denotes weak convergence.

\FloatBarrier
\section{Theory} \label{sec:theory}
    \subsection{The return model}
    
    For each of the $T$ periods in our sample, we observe the excess returns on $N = N_T$ assets, together with state variables and asset-specific characteristics. We allow $N$ to grow with $T$ and potentially be much larger than it ($N \gg T$) to reflect the common asset pricing setup in which the number of assets available per month is proportional or larger than the total sample size. To emphasize the dependence of $N$ on $T$, we index all variables and parameters that are affected by it by $T$.
    
    For each time period $t = 1,\ldots,T$, we gather the excess returns in the vector
    \[\mathbf{R}_{T,t} =  (R_{1,t},\ldots,R_{N,t})',\]
    with the excess return of asset $i$ denoted by $R_{i,t}$, for $i=1,\ldots,N$.
    For each asset, we gather both the common and asset-specific features into the $p$-dimensional vector $\mathbf{Z}_{t}^{(i)}$,
    where $p = p_T$ is allowed to grow with $T$, and stack each of the asset-specific vectors in  $\mathbf{Z}_{T,t}:= \left({\mathbf{Z}_{t}^{(1)}}'\, \cdots\, {\mathbf{Z}_{t}^{(N)}}'\right)'$.  
    In forecasting scenarios, we predict the returns given past information, such that we let $\mathbf{Z}_{T,t}$ contain strictly lagged observations (but keep the time subscipt $t$ for ease of exposition). Finally, we collect the returns and vectors in  $\boldsymbol{W}_{T,t} = (\mathbf{R}_{T,t}',\mathbf{Z}_{T,t}')'$. 
    
    Let $\mathcal{P}$ be a class of distributions governing the
    joint distribution of asset returns and features. For each
    distribution $P\in\mathcal{P}$, the conditional mean and covariance functions
    below may depend on $P$, but we suppress this dependence to reduce notational
    clutter. We introduce the class $\mathcal{P}$ because our goal is uniform
    inference. In the double/debiased machine-learning literature, the
    data-generating process is unknown, and the aim is typically to obtain
    inference that is valid
    uniformly over a class of plausible data-generating processes, rather than only under one fixed distribution $P$.
    
    The following assumption specifies the return model and imposes two
    restrictions. First, it assumes a common signal function
    $\mu_0(\cdot)$ across assets, which motivates pooled regression. Second,
    it assumes a conditional covariance matrix that does not depend on the
    features and whose eigenvalues are uniformly bounded above and away from
    zero. Both restrictions can be relaxed as covered in the remark below. Let $\mathcal{M}_T$ denote a class of measurable signal functions,
    defined at all realizations of $\mathbf{Z}_{T,t}$ and returning
    $N_T$-dimensional signal vectors. Let $\mathcal{S}_T$ denote a class of
    $N_T\times N_T$ positive-semidefinite and symmetric matrices.
    
    \begin{assumption}[Conditional return model]
    \label{ass:conditional-return-model}
    For every $P\in\mathcal{P}$, sample size $T\geq 1$, and time index $t$,
    the returns satisfy
    \begin{align*}
    \mathbf R_{T,t}
    &=\boldsymbol\mu_{T,0}(\mathbf Z_{T,t})
      +\boldsymbol\varepsilon_{T,t},\\
    \boldsymbol\mu_{T,0}(\mathbf Z_{T,t})
    &=\bigl(\mu_0(\mathbf Z_t^{(1)}),\ldots,
             \mu_0(\mathbf Z_t^{(N_T)})\bigr)',\\
    \E_P[\boldsymbol\varepsilon_{T,t}\mid\mathbf Z_{T,t}]
    &=\mathbf 0,\\
    \E_P[\boldsymbol\varepsilon_{T,t}\boldsymbol\varepsilon_{T,t}'
      \mid\mathbf Z_{T,t}]
    &=\boldsymbol\Sigma_{T,0},
    \end{align*}
    with $\boldsymbol\mu_{T,0}\in\mathcal{M}_T$ and
    $\boldsymbol\Sigma_{T,0}\in\mathcal{S}_T$.
    The signal function $\mu_0(\cdot)$ and covariance matrix
    $\boldsymbol\Sigma_{T,0}$ may depend on $P$.
    Moreover, there are
    constants $0<\underline c\leq\overline c<\infty$ such that $\boldsymbol{\Sigma}_{T,0}$
    satisfies
    \[
    \underline c
    \leq \lambda_{\min}(\boldsymbol{\Sigma})
    \leq \lambda_{\max}(\boldsymbol{\Sigma})
    \leq \overline c,
    \]
    uniformly for all $P\in\mathcal{P}$ and $T$.
    \end{assumption}
    
    \begin{remark}[Approximate common signal function and feature-dependent covariance matrices]
    The common-signal restriction can be relaxed by allowing for
    asset-specific conditional mean functions $\mu_{0}^{(i)}(\cdot)$
    that are well approximated by the common signal function
    $\mu_{0}(\cdot).$ The $L_2$ distance used below to control the signal-estimation error also provides a natural measure of this approximation error resulting from the common-signal restriction. Combined, the approximation and estimation errors must satisfy the conditions imposed on the estimation error below.
    The constant conditional covariance restriction can be relaxed as well to allow feature-dependent conditional covariance matrices that are almost surely positive definite, at the cost of imposing an $L_2$ bound on the operator norm of its estimation error, which would significantly increase notational clutter.
    \end{remark}
    
    \subsection{Signal-based portfolios and their Sharpe ratios}
    We consider portfolio strategies whose weights are known functions of the signal $\boldsymbol{\mu}_{T,0}(\mathbf{Z}_{T,t})$ and other information in $\mathbf{Z}_{T,t}$, such as asset-specific index weights. Market-timing and long-short portfolios, as well as restricted versions, are included in this class of portfolio strategies. We provide specifications for these portfolio in the second part of this section.
    
    We define the portfolio weight vector as
    \[
    \boldsymbol{\omega}_{T}(\mathbf{Z}_{T,t},\boldsymbol{\mu}_{T,0}(\mathbf{Z}_{T,t})) =
    \left[\omega_{1}(\mathbf{Z}_{T,t},\boldsymbol{\mu}_{T,0}(\mathbf{Z}_{T,t})),\ldots,
    \omega_{N}(\mathbf{Z}_{T,t},\boldsymbol{\mu}_{T,0}(\mathbf{Z}_{T,t}))
    \right]',
    \]
    such that the strategy's Sharpe ratio is given by
    \begin{align*}
    \operatorname{SR}_{T,0}
    = \frac{\mathbb{E}_P[ \boldsymbol{\omega}_{T}(\mathbf{Z}_{T,t},
    \boldsymbol{\mu}_{T,0}(\mathbf{Z}_{T,t}))'\mathbf{R}_{T,t}]}
    {\sqrt{\mathbb{V}_P[\boldsymbol{\omega}_{T}(\mathbf{Z}_{T,t},
    \boldsymbol{\mu}_{T,0}(\mathbf{Z}_{T,t}))'\mathbf{R}_{T,t}]}}.
    \end{align*}
    The index $T$ reflects that the Sharpe ratio may depend on the sample size through the number of available assets $N$ (which grows with $T$).
    
    The following assumption restricts the class of portfolio strategies to those with continuously differentiable portfolio weight functions. The resulting Jacobian is a key ingredient in the debiased Sharpe-ratio estimator introduced in this paper.
    
    \begin{assumption}[Smooth portfolio weights]
    \label{ass:portfolio-rule}
    For each $T\geq 1$, the portfolio weight function
    $\boldsymbol\omega_{T}(\cdot,\cdot)$ is jointly measurable in its
    arguments. It is continuously differentiable in its signal argument on an
    open set, and its Jacobian,
    \[
     \mathbf J_{T}(\mathbf z,\mathbf m)
     :=\left.
     \frac{\partial\boldsymbol\omega_{T}
     (\mathbf z,\widetilde{\mathbf m})}
     {\partial\widetilde{\mathbf m}'}
     \right|_{\widetilde{\mathbf m}=\mathbf m}
     \in\mathbb R^{N\times N}, 
    \]
    is jointly measurable as a function of $(\mathbf z,\mathbf m)$.
    \end{assumption}
    
    \subsubsection{Market-timing portfolios}
    Market-timing strategies base the future position in an index portfolio on its signal
    \citep{CampbellThompson2008}. Let $\va_{T,t} = (a_{T,1,t},\ldots,a_{T,N,t})'$ denote the \textit{ex ante} index weight vector, observed at the start of  period $t$, where $0 \leq a_{i,t} \leq 1$ and $\sum_{i=1}^N a_{i,t} = 1$. Moreover, we impose a diversification condition $\max_i a_{T,i,t} \leq C/N$.
    The index weights are contained in $\mZ_{T,t}$, or simply deterministic, and therefore do not need to be estimated. 
    For example, $a_{i,t} = 1/N$ for equally-weighted market portfolios, $a_{i,t} = V_{i,t} / \sum_{i=1}^N V_{i,t}$ for value-weighted market portfolios, where $V_{i,t}$ is the market-value of asset $i$ at the beginning of the period $t$. For other indices, such as the S\&P500, the weights will depend on the indexation rules.
    
    The \textit{market-timing} strategy is given by
    \begin{align*}
    &\omega_{i}(\mathbf{Z}_{T,t},\boldsymbol{\mu}_{T,0}(\mathbf{Z}_{T,t})) = a_{i,t}\left(\sum_{i=1}^N a_{i,t}\, \mu_{0,i}(\mathbf{Z}_{T,t})\right)
    ,
    \end{align*}
    with Jacobian
    \begin{align*}
        \mJ_{T}(\mZ_{T,t},\vmu_{T,0}(\mZ_{T,t})) = \va_{T,t}\va_{T,t}^{\prime}.
    \end{align*}
    As we are interested in the Sharpe ratio of the market-timing portfolio, there is no need to divide the position by the conditional variance of the index return as in \citet{CampbellThompson2008}. This conditional variance is constant over time periods and will therefore drop out of the Sharpe ratio.

    \subsubsection{Long-short portfolios}
    Long-short strategies go long in the best performers in terms of the signal and short in the worst performers. Typically, the top and bottom deciles are used to categorize the best and worst performers and the portfolio is equally-weighted. The long-short portfolio weights are formally given by
    \begin{align*}
    &\omega_{i}(\mZ_t,\vmu_0(\mZ_t)) = h_i(\vmu_0(\mZ_t)),
    \end{align*}
    where, for any $\vmu\in\mathbb{R}^N$,
    \[h_i(\vmu) = \frac{10}{ N} \left(\one\left[\sum_{j=1}^N \one[\mu_i - \mu_j\leq 0 ] \leq 0.1N \right]  - \left(1- 
    \one\left[\sum_{j=1}^N \one[\mu_i - \mu_j\leq 0 ] \leq 0.9 N \right]\right)\right),\]
    with $\one[A]$ denoting the indicator function which takes value 1 if the event $A$ is true and zero otherwise. The specifications for long-short strategies that invest in narrower sets of top/bottom performers follow straightforwardly.
    
    This strategy's weights are nonsmooth due to the specification of the portfolio weight function relying on the indicator function. We introduce a smooth approximation of the long-short portfolio based on the calibrated logistic function, which is a common smooth approximator of the indicator function. Consider the logistic function $s(z;\tau)=(1+e^{-z/\tau})^{-1}$, with bandwidth $\tau > 0$. For each $\vmu \in \mathbb{R}^N$ and bandwidth $\tau>0$, we can find the unique threshold functions $c_+\left(\vmu\right)$ and $c_-\left(\vmu\right)$ such that
    \begin{align*}\label{eq:sim_ls_rule}
    p_i^+\left(\vmu;\tau \right)&=s\!\left(\vmu_i-c_+\left(\vmu\right) ;\tau\right),\qquad
    \sum_i p_i^+\left(\vmu;\tau \right)= \lfloor0.1N\rfloor\\
    p_i^-\left(\vmu;\tau \right)&=s\!\left(c_-\left(\vmu \right)-\vmu_i;\tau \right),\qquad
    \sum_i p_i^-\left(\vmu;\tau \right)=N-\lfloor0.9N\rfloor.
    \end{align*}
    The portfolio weight function of the smoothed long-short portfolio is subsequently given by
    \[\tilde{h}_i(\vmu;\tau ) = \frac{10}{N}\left(p_i^+(\vmu;\tau )-p_i^-(\vmu;\tau )\right).\]

    The $(i,j)$th entry of the Jacobian matrix is given by
    \[
    \frac{\partial\omega_i(\mZ_t,\vmu)}{\partial\mu_j}
    =
    \frac{10}{N\tau}
    \sum_{\sigma\in\{+,-\}}
    p_i^\sigma(1-p_i^\sigma)
    \left(
    \one[i=j]
    -
    \frac{p_j^\sigma(1-p_j^\sigma)}
    {\sum_{\ell=1}^N p_\ell^\sigma(1-p_\ell^\sigma)}
    \right),
    \qquad i,j=1,\ldots,N.
    \]
    
\begin{figure}[!htbp]
    \centering
    \includegraphics[width=0.78\linewidth]{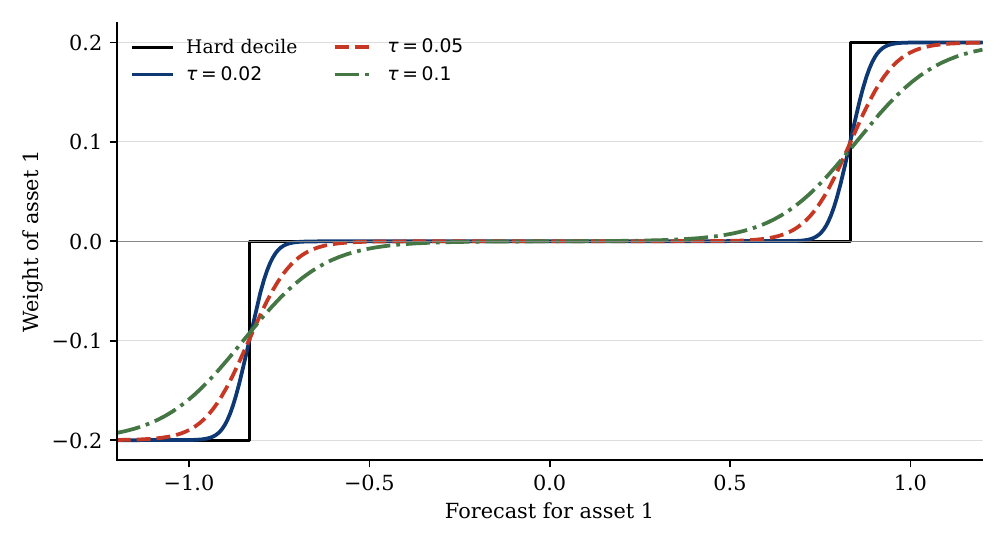}
    \caption{Nonsmooth and smoothed LS weights for one asset. There are $N=50$ assets. The other 49 forecasts are fixed and equally spaced on $[-1,1]$, while the forecast for asset 1 varies.}\label{fig:nonlinear:lsweights}
    \end{figure}

    The unique thresholds $c_-(\vmu_{T,0}(\mZ_{T,t}))$ and $c_+(\vmu_{T,0}(\mZ_{T,t}))$ are recalibrated numerically for each realization of $\vmu_{T,0}(\mZ_{T,t})$. Smaller $\tau$ produces a sharper approximation.
    Figure \ref{fig:nonlinear:lsweights} provides a visual comparison of the weight functions for nonsmooth and smoothed long-short portfolios. In practice, we choose a grid of candidate $\tau$ values, and choose a small value for which the estimated portfolio variance remains positive.
    
    \subsubsection{Other portfolios} 
    Other portfolios that are known functions of the signal $\vmu_{T,0}(\mZ_{T,t})$ and $\mZ_{T,t}$ can be treated using similar techniques as in the sequel. For example, we can impose gross-exposure contraints on the market-timing portfolio or put (smoothed) caps on the individual positions.
    
    \subsection{Motivating example:  standard Sharpe-ratio inference is biased under estimation error}

    The standard Sharpe-ratio estimator $\widehat{\mathrm{SR}}_T^\mathrm{std}$, defined in later Section \ref{sec:standard-sharpe-def}, can be biased in inference for $\mathrm{SR}_{T,0}$, resulting in invalid confidence intervals.
    To illustrate this, we consider an equally weighted market-timing portfolio of assets that satisfy a sparse return model with linear and nonlinear components that depend on asset-specific and common features, reflecting a modern asset-pricing setting that uses asset-specific features such as the dividend-price ratio and common features such as market volatility. In this example, the standard estimator provides biased inference, while the new debiased version  $\widehat{\operatorname{SR}}_T^{\mathrm{DB}}$, defined in later Section \ref{sec:DB-sharpe-def}, provides valid inference and confidence intervals.
    
    In this example, we estimate the model using pooled OLS with $p$ features, where $p < T$ and the design matrix has full column rank, so that OLS is well defined. The nonlinear component vanishes as the sample size grows, so the approximation error of the OLS estimator also vanishes. This mirrors neural network regression, in which approximation error can vanish even for nonlinear models with nonvanishing components. The advantage of this setting is that the first-order bias of the Sharpe-ratio estimator can be derived analytically. In later sections, we use simulations to show that bias also arises in a neural-network example.
    
    We use a fixed-window estimation strategy for the Sharpe ratio, to reduce notational clutter in this section, but similar results apply for moving-window and expanding-window approaches, like those discussed in Section~\ref{supp:oos-estimation-schemes}. Specifically, we first estimate the unobserved signal function $\vmu_{T,0}(\mZ_{T,t})$ on a training set $t=1,\ldots,T$. Subsequently, we compute each of the estimated counterparts to the portfolio weights $\vomega_{T,0}(\mZ_{T,t})$ and denote those $\hat{\vomega}_{T}(\mZ_{T,t})$. Finally, we estimate the Sharpe ratio in the typical fashion, using the plug-in sample moments on the test set, as
    \[
    \widehat{\operatorname{SR}}_T^{\operatorname{std}}
    =\frac{\frac{1}{T}\sum_{t=T+1}^{2T}
    \hat{\vomega}_{T}(\mZ_{T,t})'\mathbf{R}_{T,t}}
    {\sqrt{\frac{1}{T-1}\sum_{t=T+1}^{2T}
    \left(\hat{\vomega}_{T}(\mZ_{T,t})'\mathbf{R}_{T,t}-
    \frac{1}{T}\sum_{t=T+1}^{2T}
    \hat{\vomega}_{T}(\mZ_{T,t})'\mathbf{R}_{T,t}
    \right)^2}}.
    \]
    
    \begin{exmp}\label{exmp:linearpooled}
    \textbf{(Nonlinear model with pooled OLS).} Let $p<T$ and, for each date $t \geq 1$, let $C_t$ be a common standard normal factor, independent across dates. Let $Z_{i,t}$ and $W_{i,t,k}$, for $k=1,\ldots,p-2$, be mutually independent standard normal asset-specific regressors, independent of $C_t$ and independent across assets and months. Finally, define the third-order polynomial $H(c)=\frac{c^3-3c}{\sqrt{6}}$, such that $H(Z)$ has zero mean and unit variance for standard normal $Z$. The return model is given by
    \[
    R_{i,t}=\mu_{i,t}+\varepsilon_{i,t}, \qquad
    \varepsilon_{i,t}\sim\operatorname{NID}\bigl(0,\sigma_{\varepsilon,T}^2\bigr),
    \]
    where $\mu_{i,t}=\sqrt{\lambda}\,C_t+\sqrt{1-\lambda}\,Z_{i,t}+a_TH(C_t)$, and $\sigma_{\varepsilon,T}^2=19(1+a_T^2)$. The return errors are independent of all regressors. Moreover, $0<\lambda < 1$ and $a_T = c_a T^{-\alpha}$ denotes the vanishing coefficient on the nonlinear term, with $1/4 < \alpha < 1/2$ and $0 < c_a$. The error-variance level ensures that the population  $R^2$ is $5\%$.
    The pooled design is $\mZ_{t}^{(i)}=(C_t,Z_{i,t},W_{i,t,1},\ldots,W_{i,t,p-2})'$, and the coefficient vector is $\vbeta_{0}=(\sqrt\lambda,\sqrt{1-\lambda},0,\ldots,0)'$. Thus the conditional mean depends only on the first two regressors; $a_TH(C_t)$ is the omitted nonlinear component and the regressors $W_{i,t,k}$ are superfluous.
    \end{exmp}
    
    The pooled OLS estimator is
    \begin{align*}
    \hat\vbeta_T
    =\arg\min_{\vbeta\in\mathbb R^p}
    \frac{1}{NT}\sum_{i=1}^N\sum_{t=1}^T
    \left(R_{i,t}-{\mZ_{t}^{(i)}}'\vbeta\right)^2.
    \end{align*}

    For the debiased estimator $\widehat{\operatorname{SR}}_T^{\mathrm{DB}}$ 
     the conditional covariance matrix must be known or estimated using only the training sample. Here, we require it to satisfy $N^{-2}\mathbf1'\widehat{\boldsymbol\Sigma}_T\mathbf1=O_p(1)$, which is sufficient for covariance estimation to have a negligible effect at the root-$T$ scale, and which can be  satisfied for the sample covariance matrix of the training residuals even if $N$ is larger than $T$ (see Section \ref{sec:neural-net-cov-bounds}).
    
    The proposition below distinguishes the standard Sharpe-ratio estimator's first-order distortion in terms of  $a_T$ from the debiased estimator's second-order distortion. This result shows that inference based on the debiased estimator remains asymptotically centered despite approximation error when $a_T=T^{-\alpha}$ for any $\alpha > 1/4$, whereas the standard estimator has a diverging root-$T$ bias for any $\alpha<1/2$. The latter is problematic, as ML estimators may converge more slowly than the $\sqrt{T}$-rate $(\alpha<1/2)$, even when rates with $\alpha>1/4$ can be guaranteed. Consider the neural-network estimator: based on the neural-network rates in \cite{FarrellLiangMisra2021},  we can guarantee that the corresponding pooled-regression error is of order at most $T^{-\alpha}$ for some $1/4<\alpha<1/2$ in the setting with unrestricted cross-sectional correlation of the features (see Section~\ref{sec:neural-net-cov-bounds}).
    
    \begin{proposition}\label{prop:bias-var}
    Suppose the DGP in Example~\ref{exmp:linearpooled} holds and let $p/T\to\pi\in(0,1)$, $N/T\to\nu\in(0,\infty)$. Then, for the standard and debiased Sharpe-ratio estimators of the equally weighted market-timing portfolio, respectively,
    \begin{align*}
    \sqrt T\bigl(\widehat{\operatorname{SR}}_T^{\mathrm{std}}
    -\mathrm{SR}_{T,0}\bigr)
    &=Z_T
    +\frac{\sqrt3}{2\sqrt\lambda}\sqrt T\,a_T+o_p(1),\\
    \sqrt T\bigl(\widehat{\operatorname{SR}}_T^{\mathrm{DB}}
    -\mathrm{SR}_{T,0}\bigr)
    &=Z_T
    +\frac{29}{4\sqrt2\lambda}\sqrt T\,a_T^2+o_p(1),
    \end{align*}
    where $Z_T\rightsquigarrow N(0,3/4)$, and the debiased Sharpe-ratio estimator is defined in the sequel.
    \end{proposition}

\subsubsection{Sharpe-ratio expansions under cross-sectional dependence}

In Section~\ref{supp:factor-sharpe-expansions} we show that 
the same qualitative result as in Proposition~\ref{prop:bias-var} appears for features with unrestricted cross-sectional correlation. Indeed, we can replace the common factor $C_t$ by a factor-exposed and asset-specific regressor $C_{i,t}$, 
\[
C_{i,t}=\sqrt{\rho}\,F_t+\sqrt{1-\rho}\,\varsigma_{i,t},
\qquad \rho\in[0,1],
\]
where $F_t$ and $\varsigma_{i,t}$ are independent standard normal random variables, independent across dates, and the $\varsigma_{i,t}$ are also independent across assets. 
For any fixed $0 <\rho \leq 1$ the first-order and second-order biases in terms of $a_t$ remain.

    \subsection{The standard Sharpe-ratio estimator}\label{sec:standard-sharpe-def}
    
    The standard Sharpe-ratio estimator is a smooth map of the first and second
    plug-in sample moments. Let $\vtheta_{T,0}$ denote the vector containing the
    first and second oracle portfolio moments:
    \[
    \vtheta_{T,0}
    =
    \begin{bmatrix}
    \theta_{T,0,1}\\
    \theta_{T,0,2}
    \end{bmatrix}
    =
    \mathbb{E}\!
    \begin{bmatrix}
    \boldsymbol{\omega}_{T}
    (\mathbf{Z}_{T,t},\boldsymbol{\mu}_{T,0}(\mathbf{Z}_{T,t}))'
    \mathbf{R}_{T,t}
    \\
    \left(
    \boldsymbol{\omega}_{T}
    (\mathbf{Z}_{T,t},\boldsymbol{\mu}_{T,0}(\mathbf{Z}_{T,t}))'
    \mathbf{R}_{T,t}
    \right)^2
    \end{bmatrix}.
    \]
    It follows that the oracle Sharpe ratio $\mathrm{SR}_{T,0}$ is given by
    \[
    \mathrm{SR}_{T,0}=g(\vtheta_{T,0}),\quad\text{where }
    g(\vtheta)=\frac{\theta_1}{\sqrt{\theta_2-\theta_1^2}},
    \]
    for any $\vtheta\in\mathbb{R}^2$
    such that $\theta_2-\theta_1^2>0$.
    
    The sample-moment estimator of $\vtheta_{T,0}$, denoted
    $\hat\vtheta_{T}^{\mathrm{std}}$, is the plug-in GMM estimator based on the
    score function
    \[
    \boldsymbol{\psi}_T^\mathrm{std}(\vw,\boldsymbol{\theta},\boldsymbol{\mu})
    =
    \begin{bmatrix}
    \boldsymbol{\omega}_{T}
    (\vz,\boldsymbol{\mu}(\vz))'\vr\\
    \left\{
    \boldsymbol{\omega}_{T}
    (\vz,\boldsymbol{\mu}(\vz))'\vr
    \right\}^{2}
    \end{bmatrix}
    -\boldsymbol{\theta},
    \]
    for a generic observation $\vw=(\vz',\vr')'$ in the domain of
    $\mW_{T,t}$ and $\vmu\in\mathcal{M}_T$. Typically, sample splitting
    is used to estimate $\vmu_{T,0}$, and the plug-in sample moments are then
    computed over the evaluation sample. The
    standard Sharpe-ratio estimator follows as
    \[
    \widehat{\mathrm{SR}}_T^{\mathrm{std}}
    =
    g\!\left(\hat\vtheta_{T}^{\mathrm{std}}\right).
    \]
    In Proposition~\ref{prop:bias-var} above, we show that inference based on this estimator can be asymptotically biased. In Section~\ref{supp:neyman-violation}, we show that this score typically fails to satisfy Neyman orthogonality, a key ingredient for debiased inference.
    
    \subsection{The debiased Sharpe-ratio estimator}\label{sec:DB-sharpe-def}
    To facilitate valid inference, we introduce a debiased GMM estimator
    of $\boldsymbol\theta_{T,0}$, based on the novel score
    \begin{align*}
    \boldsymbol\psi_T^\mathrm{DB}(\vw,\boldsymbol\theta,\boldsymbol\eta)
    &=
    \boldsymbol\psi_T^\mathrm{std}(\vw,\boldsymbol\theta,\boldsymbol\mu)
    -
    \begin{bmatrix}
    \boldsymbol\mu(\mathbf z)'
    \mathbf J_{T,0}(\mathbf z,\boldsymbol\mu(\mathbf z))
    \{\boldsymbol\mu(\mathbf z)-\mathbf r\}\\
    2\boldsymbol\omega_{T,0}(\mathbf z,\boldsymbol\mu(\mathbf z))'
    \left\{
    \boldsymbol\mu(\mathbf z)\boldsymbol\mu(\mathbf z)'
    +\boldsymbol\Sigma
    \right\}
    \mathbf J_{T,0}(\mathbf z,\boldsymbol\mu(\mathbf z))
    \{\boldsymbol\mu(\mathbf z)-\mathbf r\}
    \end{bmatrix},
    \end{align*}
    for 
    $\vmu\in\mathcal{M}_T$ and
    $\boldsymbol\Sigma\in\mathcal{S}_T$, where
    $\boldsymbol\eta=(\boldsymbol\mu,\boldsymbol\Sigma)$.
    The debiasing terms rely on the signal function and the conditional covariance matrix of returns. Both must be estimated and we show in Section \ref{sec:neural-net-cov-bounds} that neural-network regression for the signal and the high-dimensional sample covariance matrix of return residuals work well and satisfy the theoretical restrictions for valid inference, for example.
    
    Under Assumption \ref{ass:conditional-return-model}, the moment condition
    \[
    \mathbb{E}\!\left[
    \boldsymbol\psi_T^\mathrm{DB}(
    \mW_{T,t},\boldsymbol\theta,\boldsymbol\eta_{T,0})
    \right]
    =\mathbf 0
    \]
    is satisfied uniquely at
    $\boldsymbol\theta=\boldsymbol\theta_{T,0}$, since the expected score is
    affine in $\boldsymbol\theta$. Moreover, the score satisfies the
    Neyman-orthogonality property with respect to the full nuisance parameter
    $\boldsymbol\eta$, evaluated at
    \[
    \boldsymbol\eta_{T,0}
    =(\boldsymbol\mu_{T,0},\boldsymbol\Sigma_{T,0}).
    \]
    
    The debiased GMM estimator, denoted
    $\hat\vtheta_{T}^{\mathrm{DB}}$, is the plug-in GMM estimator based on the
    debiased moments, and the
    debiased Sharpe-ratio estimator is then given by
    \[
    \widehat{\mathrm{SR}}_T^{\mathrm{DB}}
    =
    g\!\left(\hat\vtheta_{T}^{\mathrm{DB}}\right).
    \]

    \begin{remark}
        All our results extend straightforwardly to any other statistic that can be written as a suitable function $g(\hat{\vtheta}_{T}^{\mathrm{DB}})$. Examples are the certainty equivalent return under mean-variance utility, parametric normal Value-at-Risk, Expected Shortfall and/or return on risk-adjusted capital. We provide an overview of their functions in Table \ref{tab:other-performance-measures} below.
    \end{remark}
\begin{table}[htbp]
\centering
\caption{Performance and risk measures as functions of portfolio moments.}
\label{tab:other-performance-measures}
\small
\setlength{\tabcolsep}{6pt}
\renewcommand{\arraystretch}{1.55}
\begin{tabular}{lc}
\toprule
Measure & Function $g(\vtheta)$ \\
\midrule
Mean--variance certainty equivalent return
  & $\displaystyle
     \theta_1-\frac{\gamma}{2}(\theta_{2} - \theta_{1}^{2})$ \\

Gaussian value-at-risk
  & $\displaystyle
     -\theta_1+z_q\sqrt{\theta_{2} - \theta_{1}^{2}}$ \\

Gaussian expected shortfall
  & $\displaystyle
     -\theta_1+\frac{\phi(z_q)}{1-q}\sqrt{\theta_{2} - \theta_{1}^{2}}$ \\

Return on risk-adjusted capital
  & $\displaystyle
     \frac{\theta_1}{z_q\sqrt{\theta_{2} - \theta_{1}^{2}}}$ \\
\bottomrule
\end{tabular}
\par\smallskip
\begin{minipage}{0.95\linewidth}
\footnotesize
\textit{Notes.}
$\gamma > 0$ is the mean--variance risk-aversion coefficient.
For confidence level $q\in(1/2,1)$, $z_q$ is the $q$-quantile
of the standard normal distribution and $\phi$ is its density. Value-at-Risk and Expected Shortfall are computed using loss returns. Risk-adjusted capital is Gaussian Value-at-Risk in excess of expected loss.
\end{minipage}
\end{table}

    \subsection{Sample-splitting and the final debiased estimator}
    
    Our final estimator $\widehat{\boldsymbol\theta}_{T}^\mathrm{DB}$ of $\boldsymbol\theta_{T,0}$ is an average of the debiased plug-in GMM estimators obtained in a sample-splitting design that allows for standard time series training-test splits, e.g. fixed-window, moving-window, and expanding-window estimation schemes.
    We provide full details in Section~\ref{supp:oos-estimation-schemes}.
    We maintain strict training--evaluation separation by estimating the signal and covariance matrix using only observations strictly preceding the evaluation sample.
    
    \subsection{Estimation of the Sharpe-ratio differential}
    
    Often, we are interested in estimating the difference in Sharpe ratios between
    strategies. We denote by $\boldsymbol\nu_{T,0}(\mathbf{Z}_{T,t})$ the
    known, non-estimated weight vector of the portfolio against which we
    want to compare the market-timing strategy, such as buy-and-hold and equally
    weighted strategies. As in the preceding subsection, we may use a GMM
    estimator of the first and second portfolio moments, whose population vector
    is denoted by
    $\boldsymbol\tau_{T,0}=(\tau_{T,0,1},\tau_{T,0,2})'$. Given that
    $\boldsymbol\nu_{T,0}(\mathbf{Z}_{T,t})$ is known for such portfolios, we use
    the standard score for the benchmark strategy:
    \[
    \boldsymbol\psi_T^{\mathrm{bench}}(\vw,\boldsymbol\tau)
    =
    \begin{bmatrix}
    \boldsymbol\nu_{T,0}(\vz)'\vr\\
    \left\{
    \boldsymbol\nu_{T,0}(\vz)'\vr
    \right\}^{2}
    \end{bmatrix}
    -\boldsymbol\tau.
    \]
    The estimator of the benchmark moments is then
    \[
    \hat{\boldsymbol\tau}_{T}
    =
    \frac{1}{T_{\mathrm{eval}}}
    \sum_{t\in\mathcal{T}_{\mathrm{eval}}}
    \begin{bmatrix}
    \boldsymbol\nu_{T,0}(\mathbf{Z}_{T,t})'\mathbf{R}_{T,t}\\
    \left\{
    \boldsymbol\nu_{T,0}(\mathbf{Z}_{T,t})'\mathbf{R}_{T,t}
    \right\}^{2}
    \end{bmatrix},
    \]
    where $\mathcal{T}_{\mathrm{eval}}$ denotes the evaluation-period indices.
    The Sharpe-ratio differential is given by
    \[
    \Delta\mathrm{SR}_{T,0}
    =
    g(\boldsymbol\theta_{T,0})-g(\boldsymbol\tau_{T,0}),
    \]
    and its debiased estimator by
    \[
    \widehat{\Delta\mathrm{SR}}_{T}^{\mathrm{DB}}
    =
    g\!\left(\hat{\boldsymbol\theta}_{T,0}^{\mathrm{DB}}\right)
    -
    g\!\left(\hat{\boldsymbol\tau}_{T}\right).
    \]

    \subsubsection{Valid confidence intervals }
In the following sections, we formally establish that the confidence intervals below are valid under sufficient conditions that can, for example, be satisfied using a pooled neural-network estimator of the return signal. These confidence intervals are based on the asymptotic normality of the new debiased estimators. We derive results for the IID data setting, as well as a time-series setting with $\beta$-mixing triangular-array rows. All oracle and estimated variances and covariance matrices used throughout the paper, including their HAC and non-HAC versions, are defined in Section~\ref{supp:cov-def}.

Let $z_{1-\alpha/2}$ denote the $(1-\alpha/2)$-quantile of the standard normal distribution, for example  $z_{1-\alpha/2} \approx 1.96$ for $\alpha = 5\%$. For a fixed nonzero vector $\mathbf c\in\mathbb R^2$, asymptotic two-sided $(1-\alpha)$-confidence intervals for $\mathbf c'\boldsymbol\theta_{T,0}$, $\mathrm{SR}_{T,0}$, and $\Delta\mathrm{SR}_{T,0}$ are, respectively,
\begin{align*}
\mathbf c'\widehat{\boldsymbol\theta}_T
&\;\pm\; z_{1-\alpha/2}
\sqrt{\frac{\mathbf c'\widehat{\mathbf V}_T\mathbf c}
{T_{\mathrm{eval}}}},\\
\widehat{\mathrm{SR}}_T^{\mathrm{DB}}
&\;\pm\; z_{1-\alpha/2}
\sqrt{\frac{\widehat v_{T,\mathrm{SR}}}
{T_{\mathrm{eval}}}},\\
\widehat{\Delta\mathrm{SR}}_T^{\mathrm{DB}}
&\;\pm\; z_{1-\alpha/2}
\sqrt{\frac{\widehat v_{T,\Delta}}
{T_{\mathrm{eval}}}},
\end{align*}
where $T_{\mathrm{eval}}$ denotes the total sample size of the evaluation folds (see Section~\ref{supp:oos-estimation-schemes}) and the variance and covariance estimators are defined in Section~\ref{supp:cov-def}. Under the conditions of Theorem~\ref{thm:cross-fitted-smooth-score-clt}, Lemma~\ref{lem:sharpe-hac-delta-method}, and Corollary~\ref{cor:sharpe-differential-clt}, respectively, these intervals have coverage $1-\alpha+o(1)$ uniformly over $P\in\mathcal P$.

For the time-series setting, described in Section~\ref{supp:time-series-results}, use the corresponding HAC covariance estimates. The confidence intervals become
\begin{align*}
\mathbf c'\widehat{\boldsymbol\theta}_T^\mathrm{DB}
&\;\pm\; z_{1-\alpha/2}
\sqrt{\frac{\mathbf c'\widehat{\mathbf V}_{T,0}^{\mathrm{HAC}}\mathbf c}
{T_{\mathrm{eval}}}},\\
\widehat{\mathrm{SR}}_T^{\mathrm{DB}}
&\;\pm\; z_{1-\alpha/2}
\sqrt{\frac{\widehat v_{T,\mathrm{SR}}^{\mathrm{HAC}}}
{T_{\mathrm{eval}}}},\\
\widehat{\Delta\mathrm{SR}}_T^{\mathrm{DB}}
&\;\pm\; z_{1-\alpha/2}
\sqrt{\frac{\widehat v_{T,\Delta}^{\mathrm{HAC}}}
{T_{\mathrm{eval}}}}.
\end{align*}
These intervals too have coverage $1-\alpha+o(1)$ uniformly over $P\in\mathcal P$, under the additional conditions noted in Section~\ref{supp:time-series-results}.

\subsection{Asymptotic normality of the debiased estimators} 
We now establish the uniform asymptotic normality of the debiased estimators of the portfolio moments and Sharpe ratio,  justifying the associated asymptotic confidence intervals. In this section, we prove the result for IID data (Assumption~\ref{ass:iid-rows}), while Section~\ref{supp:time-series-results} extends it to time-series data. Specifically, that section allows for $\beta$-mixing triangular-array rows and develops HAC-based inference. 

Below, we provide a set of sufficient conditions, among which Assumption~\ref{ass:signal-and-nuisance-rates} is particularly important because it provide sufficient error bounds on the estimators of the signal vector and the conditional covariance matrix of returns. In Section~\ref{sec:neural-net-cov-bounds}, we show that these rates can be satisfied by a pooled neural-network estimator of the signal and a sample covariance matrix of the training residuals, for $N$ polynomial in $T$, $N \asymp T^\xi$, for any fixed $\xi > 0$. The assumption also imposes uniform bounds on the data. Such boundedness conditions are common in theoretical analyses of machine-learning estimators (see \citet{FarrellLiangMisra2021}).

    \begin{assumption}[Signal and nuisance-estimation bounds]
    \label{ass:signal-and-nuisance-rates}
    For all $P\in\mathcal{P}$ and $T \geq 1$, 
    there is a constant $C<\infty$, independent of $P$ and $T$, such that:
    \begin{align*}
    P\!\left(\|\mathbf W_T\|_\infty\leq C\right)=
    P(\|\vmu_{T,0}(\mW_T)\|_\infty \leq C) &= 1. 
    \end{align*}
    
    Moreover, there exist deterministic sequences
    $r_{\mu,T}\geq0$, $r_{\Sigma,T}\geq0$, and $\Delta_T\geq 0$, independent of $P$ and with $\Delta_T\to0$, such that for all $P \in\mathcal{P}$ and folds $k=1,\ldots,K$, it holds with $\mathbb{P}_P$-probability at least $1-\Delta_T$, that
    \begin{align*}
    P(\|\widehat{\boldsymbol\mu}_T^{(k)}(\mW_T)\|_\infty \leq C) &= 1,\\
    \left\| \widehat{\boldsymbol\mu}_{T,k}(\mW_T)
    -\boldsymbol\mu_{T,0}(\mW_T)
    \right\|_{P,2}
    &\leq r_{\mu,T},\\
    \left\|
    \widehat{\boldsymbol\Sigma}_{T,k}
    -\boldsymbol\Sigma_{T,0}
    \right\|_{\mathrm{op}}
    &\leq r_{\Sigma,T}.
    \end{align*}
    Here $\mathbb P_P$ denotes probability with respect to the training
    sample, and $\|\cdot\|_{P,2}$ is the population $L_2(P)$ norm for a
    fixed fitted estimator.
    
    Finally, the rates satisfy
    \begin{align*}
    \frac{r_{\Sigma,T}}{N_T}\longrightarrow 0, \qquad 
    \sqrt{T}\left(\frac{r_{\mu,T}}{\sqrt{N_T}}\right)^2
    \longrightarrow 0, \qquad 
    \sqrt{T}\,\left(\frac{r_{\Sigma,T}}{N_T}\right)
    \left(\frac{r_{\mu,T}}{\sqrt{N_T}}\right)
    \longrightarrow 0.
    \end{align*}
    \end{assumption}

    \begin{assumption}[Independent and identically distributed rows and covariance matrix regularity]
    \label{ass:iid-rows}
    For every $T\geq 1$, the sequence
    $\{\mathbf W_{T,t}:t\geq1\}$
    is independent and identically distributed under every $P\in\mathcal P$.
    \end{assumption}
    
    \begin{assumption}
    \label{ass:cov-reg}
    The asymptotic covariance matrices $\mathbf V_{T,0}$ (defined in Section~\ref{supp:cov-def}) have eigenvalues uniformly bounded away from zero and from above. There is a constant $0< \underline c \leq \overline c < \infty$, independent of  $P$ and $T$, such that
    $
    \underline c \leq \lambda_{\min}(\mathbf V_{T,0}) \leq \lambda_{\max}(\mathbf V_{T,0}) \leq \overline c,
    $
    for all $P\in\mathcal{P}$ and sufficiently large $T$.
    \end{assumption}
    
    The following result shows that the debiased portfolio moments of the market-timing and smoothed long-short portfolios are asymptotically standard normal once studentized by the consistent covariance matrix estimator $\widehat{\mathbf V}_{T}$ (see Section~\ref{supp:cov-def} for a definition).
    
    \begin{theorem}[Uniform weak convergence of the cross-fitted smooth-score estimator]
    \label{thm:cross-fitted-smooth-score-clt}
    Suppose Assumptions~\ref{ass:conditional-return-model}-\ref{ass:cov-reg} hold. Then, for the market-timing portfolio satisfying
    $\max_i a_{T,i,t}\leq C/N_T$ and for the smoothed long-short portfolio with fixed $\tau\in(0,\infty)$, it holds
    uniformly for all $P\in\mathcal{P}$, 
    \begin{align}
     \widehat{\mathbf V}_{T}^{-1/2}\sqrt{T_{\mathrm{eval}}}
     \{\widehat{\boldsymbol\theta}_{T}-\boldsymbol\theta_{T,0}\}
     \rightsquigarrow N(\mathbf0,\mathbf I_2),
     \label{eq:cross-fitted-smooth-score-clt}
    \end{align}
    in the bounded-Lipschitz metric.
    \end{theorem}
    
    Finally, the debiased Sharpe-ratio estimator is asymptotically normal, as established in Lemma~\ref{lem:sharpe-hac-delta-method}. This additionally requires the portfolio variance to be bounded away from zero and the first and second moments of the portfolio return to be bounded. An analogous result holds for the estimator of the Sharpe-ratio differential, as established in Corollary~\ref{cor:sharpe-differential-clt}.
    
    \begin{assumption}[Uniform regularity of the Sharpe-ratio]
    \label{ass:sharpe-target-regularity}
    There exist constants 
    $\underline v_{\mathrm{SR}}>0$ and $C<\infty$, independent of $P$ and $T$, such that
    \begin{align*}
    \theta_{T,0,2}-\theta_{T,0,1}^2
     \geq\underline v_{\mathrm{SR}}\qquad\text{and}\qquad
    \norm{\boldsymbol\theta_{T,0}}_2\leq C,
    \end{align*}
    for all $P\in\mathcal{P}$ and sufficiently large $T$.
    \end{assumption}
        
    \begin{mylem}[Uniform delta method for the Sharpe ratio]
    \label{lem:sharpe-hac-delta-method}
    Suppose the Assumptions~\ref{ass:conditional-return-model}-\ref{ass:sharpe-target-regularity} hold. Then,
    \begin{align*}
\sqrt{T_\mathrm{eval}}\frac{\widehat{\mathrm{SR}}^\mathrm{DB}_{T}
     -\mathrm{SR}_{T,0}}{\sqrt{\widehat v_{T,\mathrm{SR}}}}
     &\rightsquigarrow N(0,1),
    \end{align*}
    uniformly in the bounded-Lipschitz metric.
    \end{mylem}

    \subsection{Example: Signals estimated by pooled neural-network regression}\label{sec:neural-net-cov-bounds}
    
    We show that a pooled neural-network estimator of the signals in combination with a sample covariance matrix of the return residuals satisfy the conditions for debiased inference in Assumption~\ref{ass:signal-and-nuisance-rates}. 
    For ease of exposition, consider the fixed-window scheme with $K=2$ folds, so that $T=2T_{\mathrm{block}}$. 
    The first fold, $t=1,\ldots,T_{\mathrm{block}}$, is used for training,
    and the second fold, $t=T_{\mathrm{block}}+1,\ldots,T$, is used for
    evaluation. We suppress the fold superscript because there is only one
    nuisance estimate.
    
    We use the popular ReLU multilayer perceptron class of neural networks, see \citet{FarrellLiangMisra2021}, with width and depth that may increase with $T$. We adapt the conditions on the signal function $\mu_0$ and its neural-network estimator in their paper to the pooled-regression setting. We estimate the common signal function
    $\mu_0$ in Assumption~\ref{ass:conditional-return-model} by the pooled
    least-squares neural-network estimator
    \begin{align}
    \widehat\mu_T
    &\in
    \underset{\mu\in\mathcal N_T}{\arg\min}\;
    \frac{1}{NT_{\mathrm{block}}}
    \sum_{t=1}^{T_{\mathrm{block}}}\sum_{i=1}^{N}
    \left\{R_{i,t}-\mu\!\left(\mathbf Z_t^{(i)}\right)\right\}^{2}.
    \label{eq:pooled-nn-estimator}
    \end{align}
    The corresponding estimator of the signal vector is
    \begin{align}
    \widehat{\boldsymbol\mu}_T(\mathbf Z_{T,t})
    &:=
    \left(
    \widehat\mu_T(\mathbf Z_t^{(1)}),\ldots,
    \widehat\mu_T(\mathbf Z_t^{(N)})
    \right)'.
    \label{eq:pooled-nn-vector-estimator}
    \end{align}
    The following assumption presents the pooled-regression counterparts to Assumptions 1 and 2 in \citet{FarrellLiangMisra2021}. Note that the assumption leaves the cross-sectional dependence of the features unrestricted. This is important in the modern asset-pricing setting as the feature set typically contains common state variables as well as asset-specific features with moderate to high correlation for many asset pairs.

    \begin{assumption}\label{ass:FLM-main-text}
    The following conditions hold for all $t=1,\ldots,T$, and uniformly for all $P \in \mathcal{P}$ and $T$ sufficiently large.
    \begin{enumerate}
        \item The true signal function $\mu_0$ satisfies (i) $\max_{i=1,\ldots,N}\|\mu_0(\mathbf{Z}_T^{(i)})\|_{P,\infty}\leq M$, for some finite $M$, and (ii) for some  $\beta\in\mathbb N_+$,
    \[
    \mu_0\in\mathcal W_C^{\beta,\infty}([-1,1]^p)
    :=\left\{
    \mu:
    \max_{|\boldsymbol\alpha|\leq\beta}{\mathrm{ess}\sup}_{\vz\in [-1,1]^p}
    \left|D^{\boldsymbol\alpha}\mu(\vz)\right|_{\infty}
    \leq 1
    \right\},
    \]
    where $D^{\boldsymbol\alpha}\mu$ denotes the weak partial derivative
    associated with the multi-index
    $\boldsymbol\alpha\in\mathbb N_0^p$, with
    $|\boldsymbol\alpha|=\sum_{j=1}^p\alpha_j$;
    
    \item The estimator $\widehat \mu_T$ is in $\mathcal N_T$, where $\mathcal N_T$ is a class of scalar-output ReLU multilayer perceptrons with the architecture having width $H\asymp T^{p/[2(\beta+p)]}\log^2T$ and depth $L\asymp\log T$ and restricted to functions satisfying
    $\sup_\mathbf{z}\|\mu(\mathbf{z})\|_\infty\leq 2M$;
    \item The return errors $\varepsilon_{T,i}$ are independent. The feature vectors $\mathbf{Z}_t^{(i)}$ are continuously distributed and the feature dimension $p$ is fixed. The data is bounded as follows 
    $
    P\!\left(\|\mathbf Z_T\|_\infty\leq 1\right)=1
    $
    and
        $
    P\!\left(\|\mathbf R_T\|_\infty\leq M\right)=1.
    $
    \end{enumerate}
    \end{assumption}

    The debiased estimator requires an estimator of $\boldsymbol \Sigma_{T,0}$ and we use the sample covariance matrix of the return residuals.
    Define the fitted return-residual
    vectors and their training-sample mean by
    \begin{align*}
    \widehat{\boldsymbol\varepsilon}_{T,t}
    &:=\mathbf R_{T,t}
    -\widehat{\boldsymbol\mu}_T(\mathbf Z_{T,t})
    \quad \text{and} \quad 
    \overline{\boldsymbol\varepsilon}_T
    :=\frac{1}{T_{\mathrm{block}}}
    \sum_{t=1}^{T_{\mathrm{block}}}
    \widehat{\boldsymbol\varepsilon}_{T,t}.
    \end{align*}
    The sample covariance matrix of the return residuals is defined as 
    \begin{align}
    \widehat{\boldsymbol\Sigma}_T
    &:=\frac{1}{T_{\mathrm{block}}}
    \sum_{t=1}^{T_{\mathrm{block}}}
    \left(
    \widehat{\boldsymbol\varepsilon}_{T,t}
    -\overline{\boldsymbol\varepsilon}_T
    \right)
    \left(
    \widehat{\boldsymbol\varepsilon}_{T,t}
    -\overline{\boldsymbol\varepsilon}_T
    \right)'.
    \label{eq:nn-residual-sample-covariance}
    \end{align}
    
    The following lemma provides guarantees about the estimation rates for this pair of estimators. 
    The proof employs a new pooled-regression version of the neural-network rate results in \citet{FarrellLiangMisra2021}, presented in Section \ref{sec:supp-FLM-pooled}, and the sample covariance matrix bounds for sub-Gaussian random vectors in Theorem 4.7.1 and Exercise 4.7.3 in \citet{vershynin2018high}.
    The lemma implies that, for $N \asymp T^\xi$ with any fixed $\xi>0$, the rate restrictions for debiased inference in Assumption~\ref{ass:signal-and-nuisance-rates} are satisfied if $\beta>p$.
    This restriction requires the oracle signal function $\mu_0$ to be smooth enough relative to the amount of features. The results can be extended to features with time-series dependence using the results in \citet{brown_statistical_2024}.
    
    \begin{mylem}[Pooled neural-network and residual-covariance rates]
    \label{lem:pooled-nn-residual-covariance-rates}
    Let Assumptions~\ref{ass:conditional-return-model},  \ref{ass:iid-rows}, and \ref{ass:FLM-main-text} hold.
    Then, with probability at least $1-\Delta_T$,
    \begin{align*}
    P\!\left(
    \|\widehat{\boldsymbol\mu}_T(\mathbf Z_T)\|_\infty
    \leq2M\right)&=1,\quad 
    \left\|
    \widehat{\boldsymbol\mu}_T-\boldsymbol\mu_{T,0}
    \right\|_{P,2}
    \leq r_{\mu,T},\quad 
    \left\|
    \widehat{\boldsymbol\Sigma}_T-\boldsymbol\Sigma_{T,0}
    \right\|_{\mathrm{op}}
    \leq r_{\Sigma,T},
    \end{align*}
    where 
    \begin{align*}
    r_{\mu,T}
    &:=\sqrt{Nq_T},\\
    r_{\Sigma,T}
    &:=b_{\Sigma,T}
    +2\sqrt{(C+b_{\Sigma,T})Nq_T}
    +Nq_T,\\
    \Delta_T
    &:=\exp\!\left\{
    -T_{\mathrm{block}}^{p/(\beta+p)}
    \log^8T_{\mathrm{block}})
    \right\}+4\exp(-N),\\
    q_T
    &:=C\left\{
    T_{\mathrm{block}}^{-\beta/(\beta+p)}
    \log^7T_{\mathrm{block}}\,\log(NT_{\mathrm{block}})
    +\frac{\log\log T_{\mathrm{block}}}{T_{\mathrm{block}}}
    \right\},\\
    b_{\Sigma,T}
    &:=C\left\{
    \sqrt{\frac{N}{T_{\mathrm{block}}}}
    +\frac{N}{T_{\mathrm{block}}}
    \right\}.
    \end{align*}
    \end{mylem}
    
    \subsection{Semiparametric efficiency of the debiased estimator}
    In Section~\ref{sec:supp:iid-efficiency}, we show that $\widehat{\boldsymbol\theta}_T^\mathrm{DB}$ is semiparametrically efficient for IID data, and that this efficiency carries over to the Sharpe-ratio estimator $\widehat{\mathrm{SR}}_T^\mathrm{DB}$. This result implies that no other regular estimator of $\mathrm{SR}_{T,0}$ in the relevant model has a smaller asymptotic variance than $\widehat{\mathrm{SR}}_T^\mathrm{DB}$. 
    It provides a strong argument for the use of the debiased estimator: not only does it yield valid asymptotic confidence intervals, it also yields asymptotically shortest Wald confidence intervals. Importantly, even if the standard estimator $\widehat{\mathrm{SR}}_T^\mathrm{std}$ is regular in the relevant model, it will not yield a smaller asymptotic confidence interval than the debiased estimator.

\FloatBarrier
\section{Monte Carlo simulations} \label{sec:simulation}
To assess the finite-sample performance of our testing procedure, we perform a Monte Carlo simulation in a pooled neural network regression setting. We adapt the DGP in \hyperref[exmp:linearpooled]{Example 1} to accommodate the required theoretical conditions in the previous section. In particular, we ensure that the returns, regressors and conditional mean functions are bounded. We furthermore change the conditional mean function to be nonlinear in two active regressors. For $(\mR_{t})_{t=1}^{T}$ we simulate
\begin{equation*}\label{eq:sim_nonlinear_dgp}
R_{it}=\mu_{0}(\vx_{it})+\varepsilon_{it},\qquad
\varepsilon_{it}\stackrel{\mathrm{iid}}{\sim} \sigma\mathrm{clip}(N(0,1), -10,10),
\end{equation*}
where the disturbances $(\vepsi_{t})_{t=1}^{T}$ are independent across every index.  We fix $p=25$ and $T = 780$ and impose the following structure on the predictor vector
\[
\vx_{it}=(U_{t1},U_{it2},\ldots,U_{it,25})'.
\]
Every element of $\vx_{it}$ is marginally uniformly distributed on $[0,1]$. The first predictor $U_{t1}$ is common to all assets $i$ at every $t$. The remaining $p-1$ predictors are independent across assets and predictors, and all dates are independent.

We use
\begin{equation*}\label{eq:sim_basis}
L(u)=\sqrt{3}(2u-1),\qquad
H(u)=\frac{\sqrt{7}}{2}\{5(2u-1)^3-3(2u-1)\},
\end{equation*}
to denote standardised Legendre polynomials of order one and three. The functional form of the conditional mean function is then given by
\begin{align*}\label{eq:sim_nonlinear_mean}
\mu_{0}(\vx_{it})
=&\left\{\sqrt{\lambda}L(U_{t1})+aH(U_{t1})\right\}\\
&+\frac{\sqrt{1-\lambda}}{\sqrt{1+a^{2}}}\left\{\sqrt{0.99}L(U_{it2})+0.1 L(U_{it3}) - aH(U_{it2})\right\},
\end{align*}
where the first bracketed term represents a common factor component, while the second bracketed term is intended to capture idiosyncratic variation. Note that the DGP is sparse; only the first three of the $p$ features actually enter into the DGP.

The common component has variance $\lambda+a^2$, while the idiosyncratic signal component has variance $1-\lambda$. The parameter $a$ is calibrated such that we have a predictive $R^{2}$ of 5\%, i.e.
\begin{equation*}\label{eq:sim_noise_calibration}
\operatorname{Var}(\mu_{0,it})=1+a^2,\qquad
\sigma^2=19(1+a^2) \implies 
R^2_{\mathrm{pred}}=\frac{\operatorname{Var}(\mu_{0,it})}{\operatorname{Var}(R_{it})}=0.05.
\end{equation*}
Thus across settings we have retained the same signal-to-noise ratio. For $i \neq j$ we further have,
\begin{equation*}\label{eq:sim_correlations}
\operatorname{Corr}(\mu_{0,it},\mu_{0,jt})
=\frac{\lambda+a^2}{1+a^2},\qquad
\operatorname{Corr}(R_{it},R_{jt})
=\frac{\lambda+a^2}{20(1+a^2)}.
\end{equation*}
We vary the number of assets $N\in\{50,150,250,500\}$ and proportion of variance explained by the common component $\lambda\in\{0.25,0.40,0.55\}$. These loadings correspond to signal correlations of approximately $0.252$, $0.402$, and $0.551$, and return correlations of $0.0126$, $0.0201$, and $0.0276$.

\subsection{Neural network training}\label{sec:sim_estimation}
We use 360 dates for fitting, 60 for validation, and the final 360 for out-of-sample evaluation. We fit a pooled feedforward neural network (FNN) with one hidden layer of 32 ReLU units. To speed up computation, we initialise its weights and biases such that its initial predictions equal those of pooled OLS. Appendix~\ref{sec:app_sim_training} contains details on the initialisation, as well as specifics on the training procedure.

We use the portfolio rules and inference procedure in Section~\ref{sec:theory}, with the passive equally weighted (EW) index as the benchmark for both strategies whenever we estimate a difference of Sharpe ratios. The MT portfolio rule tracks the EW index. The return covariance nuisance is estimated by the uncentered residual second moment over the 420 fitting and validation dates. We report two-sided $5\%$ rejection frequencies against the matching population Sharpe ratios, computed as described in Appendix~\ref{sec:app_sim_nulls}.

\subsection{Rejection frequencies}\label{sec:sim_results}
Table~\ref{tab:nonlinear:primary} reports every combination of $N$ and $\lambda$, using the same DGP and training rule throughout. We choose $\tau=0.05$ for the exposition in the main text. Each result is computed using 1,000 Monte Carlo replications, with the conventional and debiased tests evaluated on the same simulated samples. Appendix~\ref{sec:app_sim_size} reports rejection frequencies for several other values of $\tau$. The long--short results are sensitive to the bandwidth, consistent with the trade-off illustrated in Figure~\ref{fig:nonlinear:lsweights}. Small values of $\tau$ make the weights change sharply near the cutoffs, which can produce large Jacobian entries and make the debiasing correction unstable in finite samples. Large values of $\tau$ smooth the weights to the point that they can depart substantially from the original decile rule.

With the FNN predictions, conventional standalone MT rejection increases from $6.9$--$10.5\%$ at $N=50$ to $34.0$--$54.4\%$ at $N=500$. Debiased MT rejection is $4.5$--$7.4\%$ across the full grid. At $\lambda=0.40$, for example, conventional rejection is $10.5$, $24.9$, $35.3$, and $42.0\%$ as $N$ increases, compared with $5.3$, $4.5$, $5.1$, and $6.2\%$ for the debiased procedure. The fixed number of fitting dates and the common nonlinear signal mean that more assets do not mechanically remove common prediction error. At the same time, the index's idiosyncratic return-noise variance falls as $\sigma^2/N$.

\begin{table}[!htbp]
\centering
\small
\begin{threeparttable}
\caption{Simulation rejection frequencies}
\label{tab:nonlinear:primary}
\setlength{\tabcolsep}{4pt}
\begin{tabular}{rrrrrrrrrr}
\toprule
& & \multicolumn{2}{c}{MT} & \multicolumn{2}{c}{MT--EW} & \multicolumn{2}{c}{LS} & \multicolumn{2}{c}{LS--EW} \\
\cmidrule(lr){3-4}\cmidrule(lr){5-6}\cmidrule(lr){7-8}\cmidrule(lr){9-10}
$\lambda$ & $N$ & C & D & C & D & C & D & C & D \\
\midrule
0.25 & 50 & 6.9 & 5.3 & 5.5 & 4.5 & 18.6 & 9.9 & 14.3 & 8.2 \\
0.25 & 150 & 25.3 & 5.8 & 11.7 & 6.6 & 15.8 & 6.4 & 12.5 & 6.7 \\
0.25 & 250 & 37.9 & 6.1 & 15.9 & 4.9 & 12.0 & 6.9 & 10.6 & 6.1 \\
0.25 & 500 & 54.4 & 7.4 & 22.3 & 5.5 & 13.2 & 6.6 & 11.4 & 5.5 \\
\addlinespace
0.40 & 50 & 10.5 & 5.3 & 7.2 & 5.3 & 19.5 & 10.6 & 12.8 & 8.0 \\
0.40 & 150 & 24.9 & 4.5 & 10.1 & 4.7 & 14.6 & 7.4 & 13.4 & 7.1 \\
0.40 & 250 & 35.3 & 5.1 & 14.9 & 5.9 & 13.4 & 6.0 & 10.9 & 5.7 \\
0.40 & 500 & 42.0 & 6.2 & 17.2 & 5.5 & 14.4 & 4.8 & 12.6 & 5.0 \\
\addlinespace
0.55 & 50 & 10.2 & 4.5 & 6.3 & 4.6 & 19.8 & 10.9 & 15.2 & 7.9 \\
0.55 & 150 & 22.4 & 5.6 & 9.2 & 4.6 & 18.4 & 6.2 & 13.2 & 5.5 \\
0.55 & 250 & 30.2 & 6.2 & 13.6 & 6.8 & 16.5 & 7.5 & 12.2 & 5.9 \\
0.55 & 500 & 34.0 & 4.9 & 14.4 & 6.7 & 15.2 & 4.9 & 13.5 & 5.3 \\
\bottomrule
\end{tabular}
\begin{tablenotes}[flushleft]\footnotesize
\item Entries are rejection percentages for two-sided 5\% tests based on 1,000 Monte Carlo replications. C and D denote conventional and debiased tests.
\end{tablenotes}
\end{threeparttable}
\vspace{\baselineskip}
\end{table}

For standalone LS at $\tau=0.05$, conventional rejection is $12.0$--$19.8\%$. Debiased rejection is $4.8$--$7.5\%$ for $N\geq150$ and $9.9$--$10.9\%$ for $N=50$. At $\lambda=0.40$ and $N=500$, the conventional and debiased values are $14.4\%$ and $4.8\%$, respectively. Thus the grid illustrates a clear reduction in the effect of estimation on LS inference as well as on MT inference, with additional finite-sample variation in the smallest cross-section.

The same table reports the results for the Sharpe ratio difference tests against an EW index. Conventional MT--EW rejection ranges from $5.5$ to $22.3\%$, compared with $4.5$--$6.8\%$ after debiasing. Conventional LS--EW rejection ranges from $10.6$ to $15.2\%$, compared with $5.0$--$8.2\%$ after debiasing.

Overall, the simulation exercise indicate that indeed our test remains valid for both portfolios under consideration, especially so when the effective sample size in terms of the number of assets $N$ increases. In this setting, the size of the standard test need not move closer to its nominal value, whereas for the debiased test we generally find that as $N$ increases, we get closer to the correct coverage rate.

\subsection{Distributions of standard and debiased Sharpe-ratio estimators }\label{sec:sim_distributions}
\begin{figure}[!htbp]
\centering
\includegraphics[width=\linewidth]{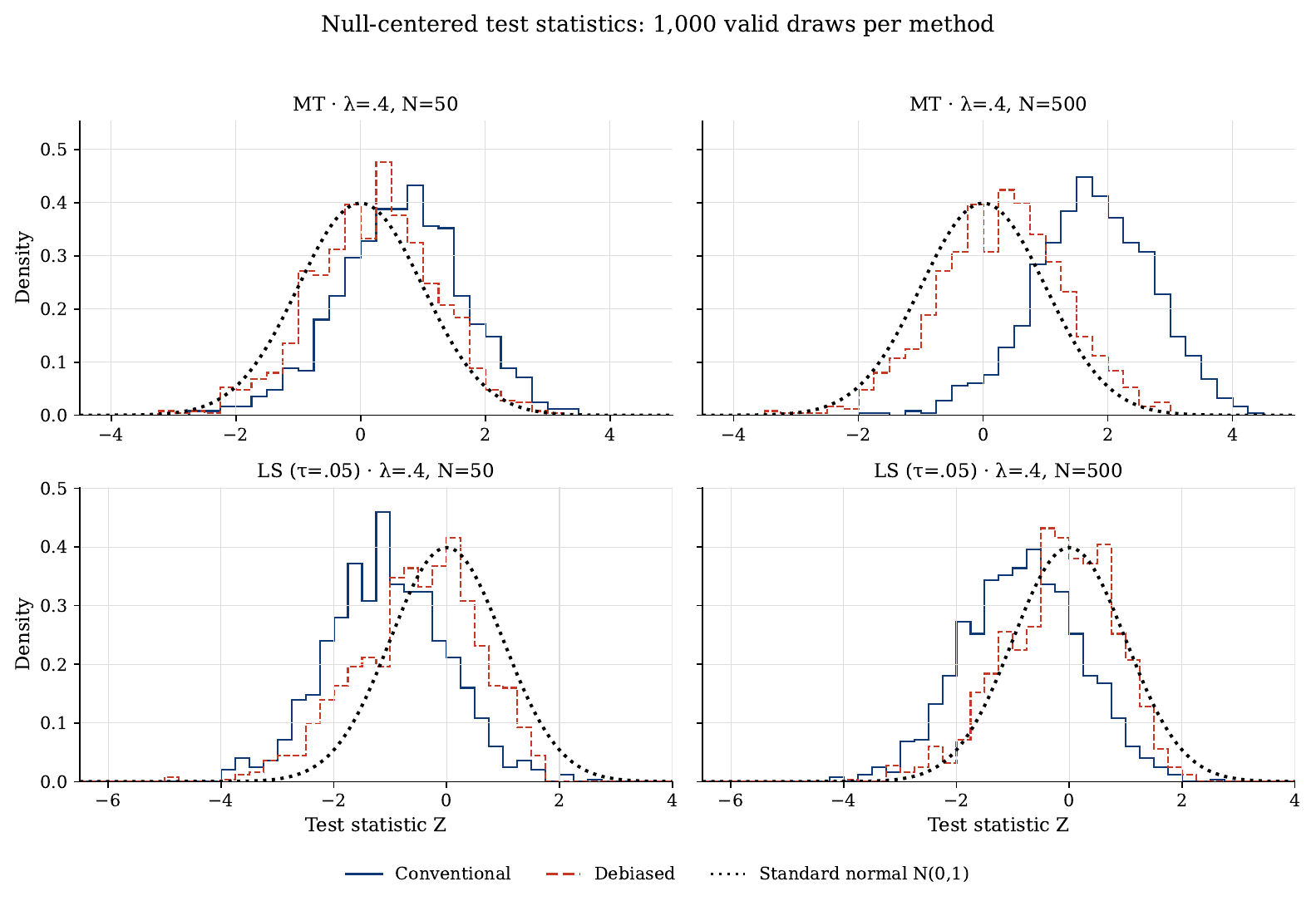}
\caption{Test-statistic histograms for MT and smoothed LS at $\tau=0.05$, with $\lambda=0.40$ and $N\in\{50,500\}$. The dotted curve is the standard normal density.}\label{fig:nonlinear:histograms}
\end{figure}

To further illustrate the potential bias of the standard test and its corresponding confidence intervals, we show the distribution of the test statistics in Figure~\ref{fig:nonlinear:histograms} for some highlighted settings. 
The conventional MT distribution moves to the right of the standard normal reference as $N$ increases, whereas the debiased distribution stays much closer to the posited asymptotic distribution. The conventional LS statistics move in the opposite direction in these examples. The correction reduces the shift in both cases. This difference in signs emphasizes that the sign of the bias term is generally unknown, due to the fact that it is a non-linear transformation of both portfolio moments. The results suggest that our proposed procedure removes this bias. Figures~\ref{fig:nonlinear:hist_mt_full} and~\ref{fig:nonlinear:hist_ls_full} in the Appendix show results for other DGP values.

\FloatBarrier
\section{Empirical Application} \label{sec:empirical}
\subsection{Data and ML methods}
We follow GKX in combining monthly CRSP excess returns with 94 firm characteristics, their interactions with eight lagged macroeconomic predictors \citep{welch2008comprehensive}, and two-digit SIC indicators\footnote{The firm characteristics and macroeconomic data are available on \href{https://dachxiu.chicagobooth.edu/}{Dacheng Xiu's} and \href{https://sites.google.com/view/agoyal145}{Amit Goyal's} websites.}. We use their monthly median imputation and rank transformation of the characteristics. Our sample extends through December 2021. The out-of-sample period is January 1987--December 2021, covering 420 months and about 2.86 million stock--month observations. For missing SIC codes, we carry the most recent code forward within each firm, fill leading gaps from the first available code, and assign the remaining gaps to an unknown-industry category. Following GKX's annual validation scheme, we initially fit on 1957--1974, validate on 1975--1986, and evaluate in 1987. Each year the fitting sample expands and the 12-year validation window rolls forward. We select hyperparameters by pooled validation mean squared error and carry the selected fit into the next evaluation year.

To estimate the conditional mean, we use the OLS-3, elastic-net, principal component regression, partial least squares, and NN1--NN5 model classes described by GKX. We use squared loss and follow GKX's elastic-net mixing parameter and penalty range. Motivated by \citet{shen2025can}, we replace GKX's neural-network $\ell_1$ penalty with an $\ell_2$ penalty. We retain their network widths and depths, ensemble three fitted networks, and use the tuning grids in Section~\ref{supp:empirical-tuning}. For each forecasting method and evaluation year, we estimate the contemporaneous innovation covariance as the uncentered second moment of monthly residual vectors from the selected model over the combined training and validation sample. We set missing stock--month residuals to zero and use the same covariance estimate for all 12 months of the following evaluation year. 

\begin{figure}[!htbp]
\centering
\includegraphics[width=\linewidth]{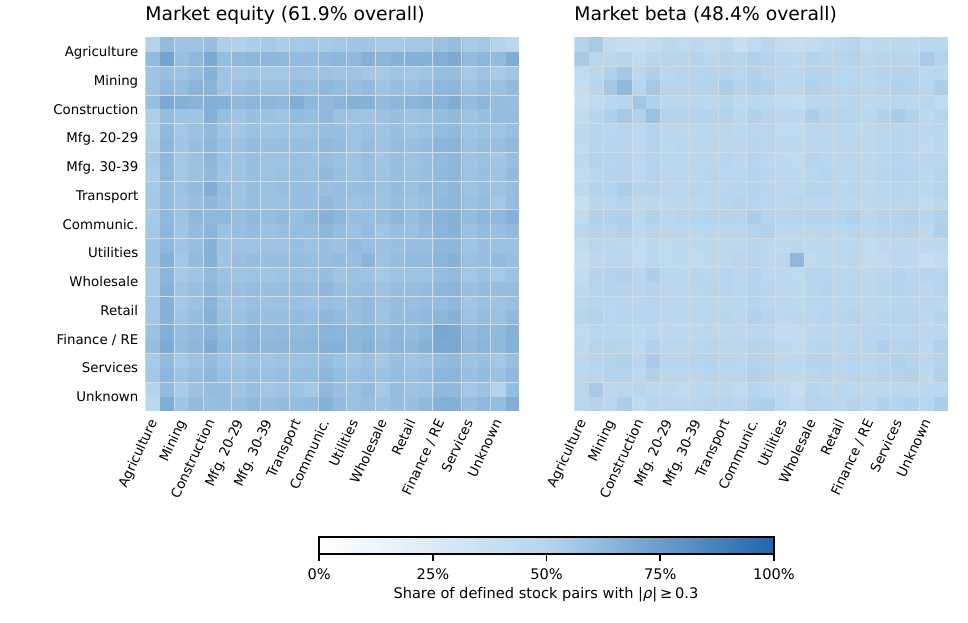}
\caption{Cross-stock co-movement in firm characteristics, 1987--2021. The left and right panels show market equity and market beta, respectively. Each cell gives the share of stock pairs within or between industry--size groups whose absolute time-series correlation is at least 0.3. We correlate the same monthly rank-transformed characteristic across stocks using at least 120 common months. Each industry is split into smaller- and larger-stock groups in that order on both axes. Grey cells have no defined correlations.}
\label{fig:empirical_characteristic_commonality}
\end{figure}

Figure~\ref{fig:empirical_characteristic_commonality} shows that firm characteristics exhibit sizable correlations for a substantial fraction of stock pairs. For example, among stock pairs with at least 120 common months and nonzero variation in both histories, $61.9\%$ of market-equity correlations and $48.4\%$ of market-beta correlations have absolute value at least $0.3$. Similar patterns appear for dividend-to-price ($49.0\%$), book-to-market ($40.0\%$), and idiosyncratic volatility ($51.6\%$). These results extend across industries and suggest widespread cross-sectional dependence in the features. Hence, under the assumptions of Section~\ref{sec:neural-net-cov-bounds} and the stated inference conditions, our neural-network error bounds guarantee asymptotically valid debiased confidence intervals, but cannot guarantee the validity of the conventional intervals.

\subsection{Market-timing and long-short portfolios vs.~benchmarks}
We generate portfolios using the market-timing and smoothed long--short rules of Section~\ref{sec:theory}. We time the S\&P 500 using forecasts for the matched constituents, weighted by lagged market equity. We use the smoothing constant $\tau=0.001$ for our long-short portfolio, which we select on the basis of the results reported in Table~\ref{tab:review_ls_tau_t}. Our passive benchmarks are the CRSP equal-weighted and value-weighted indices and the matched, value-weighted S\&P 500 portfolio on the observed panel. We report annualized Sharpe ratios and use five-lag Newey--West standard errors \citep{newey1987simple} for two-sided tests at the nominal 5\% level.
We report both $t$-statistics and their change, $\Delta t$, computed as the debiased minus the conventional statistics.

\subsubsection{Benchmarks}
The annualized Sharpe ratios of the CRSP equal-weighted, CRSP value-weighted, and matched S\&P 500 benchmarks are $0.517$, $0.564$, and $0.608$, respectively (Table~\ref{tab:review_passive_benchmarks}). The level tests compare the Sharpe ratios of these benchmarks to zero and show that all are significantly different from zero.

\subsubsection{S\&P 500 timing}
Table~\ref{tab:gkx_sp_mt} shows that conventional inference rejects a zero Sharpe ratio for all nine S\&P 500 timing strategies, while debiased inference rejects for six. The decision changes for NN2, NN4, and NN5 even though their annualized point estimates move by at most $0.017$. For example, the NN2 estimate moves from $0.492$ to $0.487$, and its standard error increases from $0.196$ to $0.273$. The resulting $t$-statistic falls from $2.50$ to $1.78$. This pattern is consistent with the trade-off between forecast flexibility and estimation error. A flexible specification can capture richer return signals, and the precision of the resulting portfolio performance estimate depends on how well the economically relevant part of the signal is learned. In these timing results, several more flexible neural networks show the largest increases in uncertainty once forecast estimation is accounted for.

\begin{table}[!htbp]
\centering
\caption{Sharpe ratios and zero-Sharpe tests for S\&P 500 timing}
\label{tab:gkx_sp_mt}
\small
\begin{adjustbox}{max width=\textwidth}
\begin{tabular}{lrrrrr}
\toprule
& $R^2_{\mathrm{oos}}$ & \multicolumn{2}{c}{Conventional} & \multicolumn{2}{c}{Debiased} \\
\cmidrule(lr){3-4}\cmidrule(lr){5-6}
Method & (\%) & $\widehat{SR}$ & $t$ & $\widehat{SR}$ & $t$ \\
\midrule
OLS-3 & 0.20 & 0.455 & 2.55 & 0.442 & 3.59 \\
ENet & 0.30 & 0.525 & 2.77 & 0.549 & 2.92 \\
PCR & 0.24 & 0.509 & 3.18 & 0.547 & 3.70 \\
PLS & 0.24 & 0.470 & 2.74 & 0.475 & 3.22 \\
\midrule
NN1 & 0.33 & 0.613 & 2.78 & 0.665 & 3.24 \\
NN2 & 0.12 & 0.492 & 2.50 & 0.487 & 1.78 \\
NN3 & 0.16 & 0.589 & 3.06 & 0.588 & 2.36 \\
NN4 & 0.07 & 0.512 & 2.57 & 0.496 & 1.75 \\
NN5 & 0.08 & 0.521 & 2.64 & 0.505 & 1.80 \\
\bottomrule
\end{tabular}
\end{adjustbox}
\begin{minipage}{0.98\linewidth}
\vspace{6pt}\footnotesize
\textit{Notes:} The out-of-sample period is January 1987--December 2021 ($T=420$). Sharpe ratios are annualized by $\sqrt{12}$. All $t$-statistics use Newey--West HAC standard errors with five lags. Conventional inference uses realized returns and their squares; debiased inference uses the orthogonal first- and second-moment scores. $R^2_{\mathrm{oos}}$ is the pooled stock-level forecasting $R^2$, in percent, relative to a zero-return forecast; it uses the full forecasting universe rather than only matched index constituents.
\end{minipage}
\end{table}

Across the timing strategies, the Sharpe-ratio point estimates remain close. The debiased standard errors are $29$--$43\%$ larger for NN2--NN5 and smaller for NN1 and the linear specifications. Some $t$-statistics rise after debiasing, including that of OLS-3. 

\begin{figure}[!htbp]
\centering
\includegraphics[width=\linewidth]{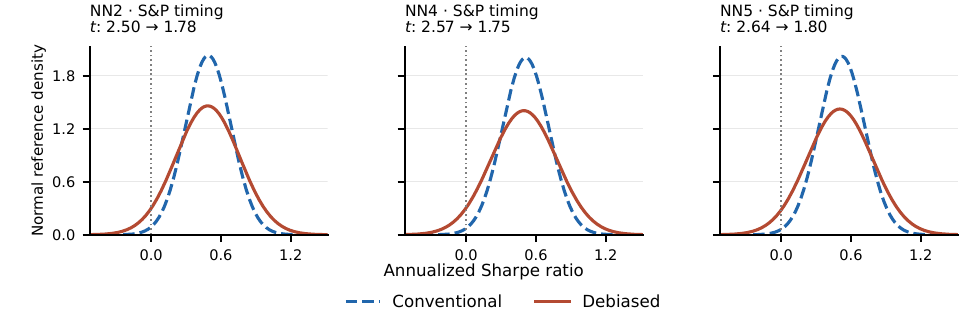}
\caption{Inference on the Sharpe ratio of S\&P 500 timing. The panels show the three strategies whose conventional and debiased tests of zero Sharpe disagree at the nominal two-sided 5\% level. Each normal reference density has mean equal to the annualized Sharpe estimate and standard deviation equal to its Newey--West standard error. The vertical line marks zero.}
\label{fig:empirical_sharpe_reference_curves}
\end{figure}

Figure~\ref{fig:empirical_sharpe_reference_curves} presents normal reference densities based on the conventional and debiased Sharpe-ratio estimates and their standard errors. Although the two densities are centered at roughly the same value, the increased dispersion after debiasing is pronounced for neural networks with richer architectures. This illustrates how inference can change with little movement in the point estimates.

The benchmark comparisons in Table~\ref{tab:review_sp_differences} give a more demanding test of the economic value of timing. Under either procedure, none of these timing strategies significantly outperforms any of the three passive benchmarks. For NN4 and NN5, the conventional tests indicate significant underperformance relative to the matched S\&P 500, while the debiased tests fail to reject equality. For NN4, the estimated gap widens from $-0.096$ to $-0.111$ and its standard error increases from $0.045$ to $0.103$. The $t$-statistic accordingly moves from $-2.15$ to $-1.08$. Table~\ref{tab:review_sp_diff_estimates} in the Supplement reports the full set of estimated gaps.

\begin{table}[!htbp]
\centering
\caption{$t$-statistics for Sharpe differences between S\&P 500 timing and passive benchmarks}
\label{tab:review_sp_differences}
\small
\setlength{\tabcolsep}{5pt}
\begin{adjustbox}{max width=\linewidth}
\begin{tabular}{lrrrrrrrrr}
\toprule
 & \multicolumn{3}{c}{CRSP EW} & \multicolumn{3}{c}{CRSP VW} & \multicolumn{3}{c}{Observed-panel S\&P} \\
\cmidrule(lr){2-4} \cmidrule(lr){5-7} \cmidrule(lr){8-10}
Method & Conv. & Deb. & $\Delta t$ & Conv. & Deb. & $\Delta t$ & Conv. & Deb. & $\Delta t$ \\
\midrule
OLS-3 & $-0.31$ & $-0.42$ & $-0.11$ & $-0.61$ & $-0.76$ & $-0.15$ & $-0.86$ & $-1.04$ & $-0.18$ \\
ENet & $0.03$ & $0.12$ & $+0.09$ & $-0.15$ & $-0.06$ & $+0.09$ & $-0.33$ & $-0.23$ & $+0.09$ \\
PCR & $-0.03$ & $0.13$ & $+0.16$ & $-0.24$ & $-0.08$ & $+0.16$ & $-0.44$ & $-0.29$ & $+0.15$ \\
PLS & $-0.18$ & $-0.17$ & $+0.01$ & $-0.37$ & $-0.37$ & $0.00$ & $-0.55$ & $-0.57$ & $-0.01$ \\
\midrule
NN1 & $0.58$ & $0.99$ & $+0.41$ & $0.37$ & $0.88$ & $+0.51$ & $0.04$ & $0.53$ & $+0.49$ \\
NN2 & $-0.19$ & $-0.17$ & $+0.01$ & $-0.92$ & $-0.66$ & $+0.27$ & $-1.75$ & $-1.11$ & $+0.64$ \\
NN3 & $0.59$ & $0.48$ & $-0.11$ & $0.36$ & $0.25$ & $-0.12$ & $-0.27$ & $-0.21$ & $+0.06$ \\
NN4 & $-0.05$ & $-0.13$ & $-0.08$ & $-0.94$ & $-0.62$ & $+0.31$ & $-2.15$ & $-1.08$ & \textbf{+1.06} \\
NN5 & $0.03$ & $-0.08$ & $-0.11$ & $-0.81$ & $-0.57$ & $+0.24$ & $-2.09$ & $-1.05$ & \textbf{+1.05} \\
\bottomrule
\end{tabular}
\end{adjustbox}
\begin{minipage}{0.98\linewidth}
\vspace{6pt}\footnotesize
\textit{Notes:} Conv. and Deb. report conventional and debiased $t$-statistics for a zero strategy-minus-benchmark Sharpe difference. Standard errors use the joint strategy--benchmark Newey--West covariance with five lags. $\Delta t$ is the descriptive change, debiased minus conventional, computed before rounding. Bold changes indicate a different decision in the two-sided 5\% test.
\end{minipage}
\end{table}

\FloatBarrier

\subsubsection{Long--short portfolios}
Long--short portfolios provide a more optimistic picture of the economic value of return signals. Table~\ref{tab:gkx_ls_tau001} shows that the conventional and debiased procedures both reject a zero Sharpe ratio for the same seven smoothed long--short strategies. All seven strategies have estimated Sharpe ratios above those of the benchmarks, although only five significantly outperform all three benchmarks under both procedures. Conventional and debiased point estimates are generally similar, although NN2's annualized Sharpe ratio increases from 1.467 to 1.938 after debiasing. The deeper neural-network architectures, NN4 and NN5, again yield low estimated Sharpe ratios, potentially reflecting the difficulty of estimating highly parameterized models with the available data.

\begin{table}[!htbp]
\centering
\caption{Sharpe ratios and zero-Sharpe tests for smoothed long--short ($\tau=0.001$)}
\label{tab:gkx_ls_tau001}
\small
\begin{adjustbox}{max width=\textwidth}
\begin{tabular}{lrrrrr}
\toprule
& $R^2_{\mathrm{oos}}$ & \multicolumn{2}{c}{Conventional} & \multicolumn{2}{c}{Debiased} \\
\cmidrule(lr){3-4}\cmidrule(lr){5-6}
Method & (\%) & $\widehat{SR}$ & $t$ & $\widehat{SR}$ & $t$ \\
\midrule
OLS-3 & 0.20 & 0.791 & 4.01 & 0.764 & 4.04 \\
ENet & 0.30 & 1.692 & 7.80 & 1.697 & 8.82 \\
PCR & 0.24 & 1.624 & 7.02 & 1.601 & 7.64 \\
PLS & 0.24 & 1.879 & 7.43 & 1.815 & 8.39 \\
\midrule
NN1 & 0.33 & 2.202 & 9.90 & 2.172 & 10.95 \\
NN2 & 0.12 & 1.467 & 5.60 & 1.938 & 6.63 \\
NN3 & 0.16 & 0.952 & 4.51 & 0.937 & 5.20 \\
NN4 & 0.07 & 0.111 & 0.55 & 0.343 & 0.37 \\
NN5 & 0.08 & -0.007 & -0.04 & -0.084 & -0.22 \\
\bottomrule
\end{tabular}
\end{adjustbox}
\begin{minipage}{0.98\linewidth}
\vspace{6pt}\footnotesize
\textit{Notes:} The out-of-sample period is January 1987--December 2021 ($T=420$). Sharpe ratios are annualized by $\sqrt{12}$. All $t$-statistics use Newey--West HAC standard errors with five lags. Conventional inference uses realized returns and their squares; debiased inference uses the orthogonal first- and second-moment scores. $R^2_{\mathrm{oos}}$ is the pooled stock-level forecasting $R^2$, in percent, relative to a zero-return forecast; it uses the full forecasting universe rather than only matched index constituents.
\end{minipage}
\end{table}

The benchmark tests in Table~\ref{tab:review_ls_diff_main} show that elastic net, PCR, PLS, NN1, and NN2 significantly outperform each of the three passive benchmarks under both procedures. For NN1 against the matched S\&P 500, the annualized Sharpe gap is 1.594 conventionally and 1.564 after debiasing, with corresponding $t$-statistics of 5.52 and 5.72. Thus, debiasing leaves NN1's large estimated advantage and strong evidence of outperformance essentially unchanged. For elastic net, PCR, PLS, and NN1, significant outperformance persists across the reported bandwidth grid for $\tau$. Table~\ref{tab:review_ls_diff_estimates} reports the full set of point estimates at the main bandwidth, and Section~\ref{supp:empirical-bandwidth} reports the sensitivity results. Together, these results show that return signals based on the GKX feature set can deliver substantial and statistically significant improvements in risk-adjusted performance relative to passive benchmarks.

\begin{table}[!htbp]
\centering
\caption{$t$-statistics for long--short Sharpe differences from passive benchmarks ($\tau=0.001$)}
\label{tab:review_ls_diff_main}
\small
\setlength{\tabcolsep}{5pt}
\begin{adjustbox}{max width=\linewidth}
\begin{tabular}{lrrrrrrrrr}
\toprule
 & \multicolumn{3}{c}{CRSP EW} & \multicolumn{3}{c}{CRSP VW} & \multicolumn{3}{c}{Observed-panel S\&P} \\
\cmidrule(lr){2-4} \cmidrule(lr){5-7} \cmidrule(lr){8-10}
Method & Conv. & Deb. & $\Delta t$ & Conv. & Deb. & $\Delta t$ & Conv. & Deb. & $\Delta t$ \\
\midrule
OLS-3 & $0.88$ & $0.79$ & $-0.08$ & $0.76$ & $0.67$ & $-0.08$ & $0.61$ & $0.53$ & $-0.08$ \\
ENet & $3.58$ & $3.83$ & $+0.25$ & $3.67$ & $3.94$ & $+0.27$ & $3.59$ & $3.86$ & $+0.27$ \\
PCR & $3.31$ & $3.46$ & $+0.16$ & $3.26$ & $3.42$ & $+0.16$ & $3.17$ & $3.32$ & $+0.15$ \\
PLS & $3.88$ & $4.05$ & $+0.17$ & $3.93$ & $4.13$ & $+0.20$ & $3.87$ & $4.07$ & $+0.20$ \\
\midrule
NN1 & $5.43$ & $5.65$ & $+0.22$ & $5.59$ & $5.80$ & $+0.21$ & $5.52$ & $5.72$ & $+0.20$ \\
NN2 & $2.56$ & $3.67$ & $+1.11$ & $2.56$ & $3.73$ & $+1.18$ & $2.49$ & $3.68$ & $+1.19$ \\
NN3 & $1.41$ & $1.55$ & $+0.14$ & $1.29$ & $1.48$ & $+0.19$ & $1.17$ & $1.34$ & $+0.17$ \\
NN4 & $-1.66$ & $-0.18$ & $+1.48$ & $-1.87$ & $-0.23$ & $+1.64$ & $-2.01$ & $-0.28$ & \textbf{+1.73} \\
NN5 & $-2.08$ & $-1.46$ & \textbf{+0.61} & $-2.19$ & $-1.53$ & \textbf{+0.66} & $-2.34$ & $-1.61$ & \textbf{+0.72} \\
\bottomrule
\end{tabular}
\end{adjustbox}
\begin{minipage}{0.98\linewidth}
\vspace{6pt}\footnotesize
\textit{Notes:} Conv. and Deb. report conventional and debiased $t$-statistics for a zero strategy-minus-benchmark Sharpe difference. Standard errors use the joint strategy--benchmark Newey--West covariance with five lags. $\Delta t$ is the descriptive change, debiased minus conventional, computed before rounding. Bold changes indicate a different decision in the two-sided 5\% test.
\end{minipage}
\end{table}

The conventional tests find significant underperformance for NN4 against the matched S\&P 500 and for NN5 against all three benchmarks. The debiased tests fail to reject equality in these four comparisons, implying that the estimation error of these methods is too great to meaningfully rank the two strategies.  

\begin{figure}[!htbp]
\centering
\includegraphics[width=\linewidth]{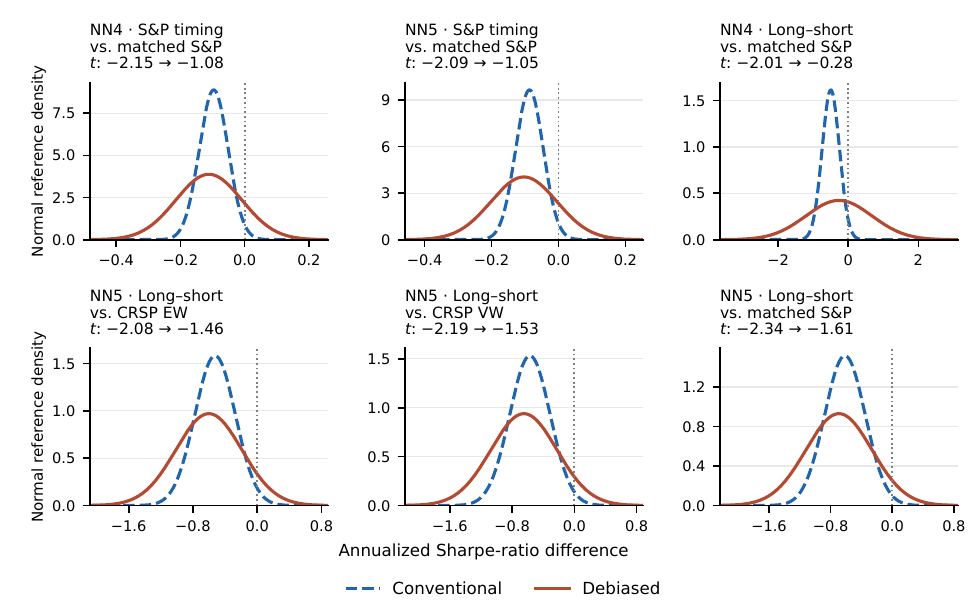}
\caption{Inference on Sharpe differences from passive benchmarks. The panels show the six comparisons whose conventional and debiased tests disagree at the nominal two-sided 5\% level. The long--short rule uses $\tau=0.001$. Each Gaussian reference density is centered at the annualized strategy-minus-benchmark Sharpe estimate, with standard deviation equal to the corresponding Newey--West standard error computed from the joint scores. The vertical line marks a zero difference.}
\label{fig:empirical_difference_reference_flips}
\end{figure}

Figure~\ref{fig:empirical_difference_reference_flips} displays normal reference densities for the Sharpe-ratio differences in all six comparisons where the conventional and debiased tests disagree at the nominal two-sided 5\% level, including four long--short and two market--timing comparisons. For the deeper neural-network architectures, NN4 and NN5, the debiased reference densities are substantially wider than their conventional counterparts.

\FloatBarrier

\FloatBarrier
\section{Conclusion}
We develop Sharpe-ratio inference for market-timing and long-short portfolios constructed from estimated
return signals. Neyman-orthogonal corrections to portfolio moments account estimation error, while chronological cross-fitting accommodates
temporal dependence. The resulting confidence intervals apply to the Sharpe
ratio and its difference from an observed benchmark under suitable learning
rates in large asset panels.

The simulations illustrate why a large cross-section need not eliminate
estimation effects and show how debiasing improves the calibration of
Sharpe-ratio inference. In the empirical application, conventional and
debiased point estimates are generally similar, while their standard errors
and significance conclusions can differ substantially. The estimated normal
reference distributions make these differences visible without reducing
the comparison to whether a test crosses a significance threshold. The
results support accounting for estimated signals when assessing portfolio
performance, alongside the return variation measured by conventional
procedures.
 \label{sec:conclusion}

\FloatBarrier
\clearpage
\onehalfspacing
\setlength\bibsep{5pt}
\bibliographystyle{ecta}
\bibliography{references}

\pagebreak

\bigskip

\appendix
\beginsupplement
\setcounter{figure}{0}
\setcounter{table}{0}
\counterwithin{equation}{section}
\counterwithin{theorem}{section}
\counterwithin{corollary}{section}
\counterwithin{proposition}{section}
\counterwithin{assumption}{section}
\counterwithin{remark}{section}
\counterwithin{definition}{section}
\counterwithin{mythm}{section}
\counterwithin{myprop}{section}
\counterwithin{mylem}{section}
\counterwithin{mydef}{section}
\counterwithin{mycor}{section}

\begin{center}
{\large\bf SUPPLEMENTARY MATERIAL}
\end{center}

\begin{description}

\item[Supplementary Appendix:] The Supplementary Appendix contains technical details and proofs,
additional theoretical results, and supplementary simulation
and empirical analyses.

\startcontents[appendix]
\printcontents[appendix]{}{1}{}
\end{description}

\pagebreak

\pagebreak

\begin{center}
    \textbf{\Large Supplementary Appendix to `\papertitle'}\\

    Authors: Sander Barendse \& Ming Cheng\\
    \thedate

\end{center}

\section{Technical details}\label{sec:tech-detail}
\subsection{Notation}
We retain the notation of the main text and use the following additional
conventions. For a matrix $\mathbf A=(a_{ij})$, write
$\|\mathbf A\|_{\max}=\max_{i,j}|a_{ij}|$; for symmetric matrices,
$\mathbf A\succeq\mathbf B$ means that $\mathbf A-\mathbf B$ is
positive semidefinite. We write $Q\ll P$ for absolute continuity,
$dQ/dP$ for the Radon--Nikodym derivative,
$P^{\otimes n}$ for the $n$-fold product law, and $P_{\mathbf Z}$ for
the marginal law of $\mathbf Z$ under $P$.
For a square matrix, $\operatorname{tr}(\mathbf A)$ denotes its trace.
For a specified class $\mathcal P$, $X_T=o_P(1)$ uniformly over
$P\in\mathcal P$ means that
$\sup_{P\in\mathcal P}P(\|X_T\|>\varepsilon)\to0$ for every
$\varepsilon>0$; uniform $O_P(1)$ means that
$\lim_{M\to\infty}\limsup_{T\to\infty}
\sup_{P\in\mathcal P}P(\|X_T\|>M)=0$.
Uniform orders at a positive deterministic scale are defined by division
by that scale.

\subsection{Sharpe-ratio expansions under cross-sectional factor dependence}
\label{supp:factor-sharpe-expansions}

Consider Example~\ref{exmp:linearpooled}, replacing $C_t$ by
\[
C_{i,t}=\sqrt{\rho}\,F_t+\sqrt{1-\rho}\,\varsigma_{i,t},
\qquad \rho\in[0,1],
\]
where $F_t$ and $\varsigma_{i,t}$ are independent standard normal random variables, independent across dates and also independent across assets in case of the idiosyncratic errors. We retain the remaining assumptions of the example, including $a_T=c_aT^{-\alpha}$ with $1/4<\alpha<1/2$, $p/T\to\pi\in(0,1)$, and $N/T\to\nu\in(0,\infty)$, and impose the same rate on the sample covariance matrix.
For every fixed $\rho\in(0,1]$, it can be shown that
\begin{align*}
\sqrt T\bigl(\widehat{\mathrm{SR}}_T^{\mathrm{std}}-\mathrm{SR}_{T,0}\bigr)
&=Z_T+\frac{\sqrt3\,\rho}{2\sqrt\lambda}\sqrt T\,a_T+o_p(1),\\
\sqrt T\bigl(\widehat{\mathrm{SR}}_T^{\mathrm{DB}}-\mathrm{SR}_{T,0}\bigr)
&=Z_T+\frac{29\rho^2}{4\sqrt2\lambda}\sqrt T\,a_T^2+o_p(1),
\end{align*}
where $Z_T\rightsquigarrow N(0,3/4)$. 
Hence, any fixed factor exposure in a feature retains the first-order bias in the standard estimator, while the debiased estimator remains centered asymptotically.

\subsection{The standard score is not Neyman-orthogonal}\label{supp:neyman-violation}
The standard score $\boldsymbol{\psi}_T^\mathrm{std}$ does not satisfy
Neyman orthogonality, which is key ingredient for debiased inference (see \citet{chernozhukov2017double} for example). Indeed, the Gateaux derivative at
$\boldsymbol{\mu}_{T,0}$ takes value
\begin{align*}
&\left.
\frac{d}{d\kappa}
\mathbb{E}\!\left[
\boldsymbol{\psi}_T^\mathrm{std}
(\mW_{T,t},\boldsymbol{\theta},
 \boldsymbol{\mu}_{T,0}+\kappa\mathbf h)
\right]
\right|_{\kappa=0}\\
&\quad =
\mathbb{E}\!
\begin{bmatrix}
\mathbf h(\mathbf Z_{T,t})'
\mathbf J_{T,0}\!\left(
\mathbf Z_{T,t},
\boldsymbol{\mu}_{T,0}(\mathbf Z_{T,t})
\right)'
\mathbf R_{T,t}
\\
2\left\{
\boldsymbol{\omega}_{T}\!\left(
\mathbf Z_{T,t},
\boldsymbol{\mu}_{T,0}(\mathbf Z_{T,t})
\right)'\mathbf R_{T,t}
\right\}
\mathbf h(\mathbf Z_{T,t})'
\mathbf J_{T,0}\!\left(
\mathbf Z_{T,t},
\boldsymbol{\mu}_{T,0}(\mathbf Z_{T,t})
\right)'
\mathbf R_{T,t}
\end{bmatrix}
.
\end{align*}
This derivative is typically nonzero for at least some admissible
$\mathbf h$ under realistic modeling assumptions. 

\subsection{Out-of-sample estimation schemes}\label{supp:oos-estimation-schemes}
We partition the sample into a fixed number $K$ of consecutive folds. For ease of exposition, suppose $T=K\,T_\mathrm{block}$, for some fold size  $T_\mathrm{block} \geq 1$, and let
\begin{align*}
 I_{T,k}=\{(k-1)T_\mathrm{block}+1,\ldots,kT_\mathrm{block}\},\qquad k=1,\ldots,K.
\end{align*}
This partitioning strategy may be relaxed to allow varying fold sizes, as long each fold size is proportional to $T$, but we refrain from doing so to reduce notational clutter.

Let the fixed training-window length satisfy
$1\leq K_{\mathrm{train}}\leq K-1$, and define the evaluation window length and the total number of evaluation folds as 
\begin{align*}
T_{\mathrm{eval}}=K_{\mathrm{eval}}T_\mathrm{block}, \qquad  K_{\mathrm{eval}}&=K-K_{\mathrm{train}},
\end{align*}
For each evaluation fold, the nuisance estimator
\begin{align*}
 \widehat{\boldsymbol\eta}_T^{(k)}
 =\bigl(\widehat{\boldsymbol\mu}_T^{(k)},
 \widehat{\boldsymbol\Sigma}_T^{(k)}\bigr)
 \in\mathcal M_T\times\mathcal S_T
\end{align*}
is measurable with respect to the observations indexed by its training
set $I_{T,k}^-$, where the three illustrative training schemes are
\begin{align*}
 I_{T,k}^-&=\bigcup_{\ell=1}^{K_{\mathrm{train}}}I_{T,\ell}
 &&\text{(fixed window)},\\
 I_{T,k}^-&=\bigcup_{\ell=k-K_{\mathrm{train}}}^{k-1}I_{T,\ell}
 &&\text{(moving window)},\\
 I_{T,k}^-&=\bigcup_{\ell=1}^{k-1}I_{T,\ell}
 &&\text{(expanding window)}.
\end{align*}
The intermediate estimator solves for each fold $k=1,\ldots,K$ 
\begin{align*}
 \frac1{T_\mathrm{block}}\sum_{t\in I_{T,k}}\boldsymbol\psi_{T}^\mathrm{DB}
 (\mathbf W_{T,t},\breve{\boldsymbol\theta}_{T}^\mathrm{DB}(I_{T,k}),
 \widehat{\boldsymbol\eta}_T(I_{T,k}^-))=\mathbf0.
\end{align*}
Since the score is affine in its target argument,
\begin{align*}
 \breve{\boldsymbol\theta}_{T}^\mathrm{DB}(I_{T,k})
 &=\boldsymbol\theta_{T,0}(P)+\frac1{T_\mathrm{block}}\sum_{t\in I_{T,k}}
 \boldsymbol\psi_{T}^\mathrm{DB}(\mathbf W_{T,t},\boldsymbol\theta_{T,0}(P),
 \widehat{\boldsymbol\eta}_T(I_{T,k}^-)).
\end{align*}
The final estimator of $\boldsymbol\theta_{T,0}$ is 
\begin{align*}
 \widehat{\boldsymbol\theta}_{T}^\mathrm{DB}
 =\frac1{K_{\mathrm{eval}}}\sum_{k\geq K_\mathrm{train}+1}
 \breve{\boldsymbol\theta}_{T}^\mathrm{DB}(I_{T,k}).
\end{align*}

\subsection{Covariance and variance specifications}\label{supp:cov-def}

\subsubsection{Oracle specifications}
Let 
\begin{align*}
\mathbf V_{T,0}
&:=\E_P\!\left[
\boldsymbol\psi_T^{\mathrm{DB}}
\bigl(\mathbf W_{T,t},\boldsymbol\theta_{T,0},\boldsymbol\eta_{T,0}\bigr)
\boldsymbol\psi_T^{\mathrm{DB}}
\bigl(\mathbf W_{T,t},\boldsymbol\theta_{T,0},\boldsymbol\eta_{T,0}\bigr)'\right],\\
\mathbf B_{T,0}
&:=\E_P\!\left[
\boldsymbol\psi_T^{\mathrm{bench}}
\bigl(\mathbf W_{T,t},\boldsymbol\tau_{T,0}\bigr)
\boldsymbol\psi_T^{\mathrm{bench}}
\bigl(\mathbf W_{T,t},\boldsymbol\tau_{T,0}\bigr)'\right],\\
\mathbf C_{T,0}
&:=\E_P\!\left[
\boldsymbol\psi_T^{\mathrm{DB}}
\bigl(\mathbf W_{T,t},\boldsymbol\theta_{T,0},\boldsymbol\eta_{T,0}\bigr)
\boldsymbol\psi_T^{\mathrm{bench}}
\bigl(\mathbf W_{T,t},\boldsymbol\tau_{T,0}\bigr)'\right]
\end{align*}
denote the strategy's covariance matrix, the benchmark's covariance matrix, and the cross-covariance matrix, respectively. The specification of each of these matrices and other population variances and covariances below may depend on $P$.

For the Sharpe-ratio variances, define
\begin{align*}
\mathbf d_{T,\theta}
&:=\nabla g(\boldsymbol\theta_{T,0}),
&
\mathbf d_{T,\tau}
&:=\nabla g(\boldsymbol\tau_{T,0}),
&
\mathbf d_{T,\Delta}
&:=\begin{bmatrix}
\mathbf d_{T,\theta}\\
-\mathbf d_{T,\tau}
\end{bmatrix},
\end{align*}
where
\begin{align*}
\nabla g(\mathbf x)
&=\frac{1}{(x_2-x_1^2)^{3/2}}
\begin{bmatrix}
x_2\\
-x_1/2
\end{bmatrix},
&&\qquad x_2-x_1^2>0.
\end{align*}

Table~\ref{tab:oracle-covariance-variance} provides the oracle specifications for the covariance matrices corresponding to the portfolio moments and the variances corresponding to the Sharpe ratio and Sharpe-ratio differences.

\begin{table}[htbp]
\centering
\caption{Oracle asymptotic covariance matrices and variances for
moment and Sharpe-ratio inference under IID observations.}
\label{tab:oracle-covariance-variance}
\begingroup
\small
\setlength{\tabcolsep}{4pt}
\renewcommand{\arraystretch}{1.55}
\begin{tabular}{lc}
\toprule
Target & Oracle specification\\
\midrule

Strategy moments \par
\ensuremath{\boldsymbol\theta_{T,0}}
&
\ensuremath{
\begin{aligned}
\mathbf V_{T,0}
\end{aligned}}
\\[0.5ex]

Extended moments \par
\ensuremath{
(\boldsymbol\theta_{T,0}',\boldsymbol\tau_{T,0}')'}
&
\ensuremath{
\begin{aligned}
\boldsymbol\Xi_{T,0}
&=\begin{bmatrix}
\mathbf V_{T,0}&\mathbf C_{T,0}\\
\mathbf C_{T,0}'&\mathbf B_{T,0}
\end{bmatrix}
\end{aligned}}
\\[0.5ex]

Sharpe ratio \par
\ensuremath{\mathrm{SR}_{T,0}}
&
\ensuremath{
\begin{aligned}
v_{T,0,\mathrm{SR}}
&:=\mathbf d_{T,\theta}'
\mathbf V_{T,0}\mathbf d_{T,\theta}
\end{aligned}}
\\[0.5ex]

Sharpe-ratio difference \par
\ensuremath{\Delta\mathrm{SR}_{T,0}}
&
\ensuremath{
\begin{aligned}
v_{T,0,\Delta}
&:=\mathbf d_{T,\Delta}'
\boldsymbol\Xi_{T,0}\mathbf d_{T,\Delta}\\
\end{aligned}}
\\
\bottomrule
\end{tabular}
\endgroup
\end{table}

\subsubsection{Time-series oracle specifications}

The time-series counterparts replace the IID covariance matrices by their long-run counterparts in Table~\ref{tab:oracle-covariance-variance}.

We first define the autocovariance matrices 
\begin{align*}
\mathbf V_{T,0}(h)
&:=\E_P\Bigl[
\boldsymbol\psi_T^{\mathrm{DB}}(\mathbf W_{T,t},\boldsymbol\theta_{T,0},\boldsymbol\eta_{T,0})
\boldsymbol\psi_T^{\mathrm{DB}}(\mathbf W_{T,t+h},\boldsymbol\theta_{T,0},\boldsymbol\eta_{T,0})'\Bigr],\\
\mathbf B_{T,0}(h)
&:=\E_P\Bigl[
\boldsymbol\psi_T^{\mathrm{bench}}(\mathbf W_{T,t},\boldsymbol\tau_{T,0})
\boldsymbol\psi_T^{\mathrm{bench}}(\mathbf W_{T,t+h},\boldsymbol\tau_{T,0})'\Bigr],\\
\mathbf C_{T,0}(h)
&:=\E_P\Bigl[
\boldsymbol\psi_T^{\mathrm{DB}}(\mathbf W_{T,t},\boldsymbol\theta_{T,0},\boldsymbol\eta_{T,0})
\boldsymbol\psi_T^{\mathrm{bench}}(\mathbf W_{T,t+h},\boldsymbol\tau_{T,0})'\Bigr],\\
\mathbf D_{T,0}(h)
&:=\E_P\Bigl[
\boldsymbol\psi_T^{\mathrm{bench}}(\mathbf W_{T,t},\boldsymbol\tau_{T,0})
\boldsymbol\psi_T^{\mathrm{DB}}(\mathbf W_{T,t+h},\boldsymbol\theta_{T,0},\boldsymbol\eta_{T,0})'\Bigr].
\end{align*}
At lag zero, these matrices satisfy
\begin{align*}
\mathbf V_{T,0}(0)=\mathbf V_{T,0} \quad 
\mathbf B_{T,0}(0) =\mathbf B_{T,0},\quad
\mathbf C_{T,0}(0) = \mathbf D_{T,0}'(0)=\mathbf C_{T,0}.
\end{align*}
The long-run counterparts are 
\begin{align*}
\mathbf V_{T,0}^\mathrm{LR}
&:= \mathbf V_{T,0}(0)
+\sum_{h=1}^{\infty}
\{\mathbf V_{T,0}(h)
+\mathbf V_{T,0}(h)'\},\\
\mathbf B_{T,0}^\mathrm{LR}
&:= \mathbf B_{T,0}(0)
+\sum_{h=1}^{\infty}
\{\mathbf B_{T,0}(h)
+\mathbf B_{T,0}(h)'\},\\
\mathbf C_{T,0}^\mathrm{LR}
&:= \mathbf C_{T,0}(0)
+\sum_{h=1}^{\infty}
\{\mathbf C_{T,0}(h)
+\mathbf D_{T,0}(h)'\}.
\end{align*}

\subsubsection{Covariance estimation}

We first discuss the HAC estimator of the long-run covariance matrices. The IID estimator follows as a special case. We treat the case for $\mathbf V_{T,0}^{\mathrm{LR}}$. The other cases follow from similar steps.

For the cross-fitting designs outlined in Section~\ref{supp:oos-estimation-schemes}, let the estimated counterparts of the autocovariances for each fold $k>K_\mathrm{train}$ be 
\begin{align*}
\widehat{\mathbf V}_{T,k}(h)
&:=\frac1{T_\mathrm{block}}
\sum_{t=(k-1)T_\mathrm{block}+h+1}^{kT_\mathrm{block}}
\boldsymbol\psi_T^{\mathrm{DB}}\bigl(\mathbf W_{T,t},
\widehat{\boldsymbol\theta}_T,
\widehat{\boldsymbol\eta}_T(I_{T,k}^-)\bigr)\\
&\hspace{27mm}\times
\boldsymbol\psi_T^{\mathrm{DB}}\bigl(\mathbf W_{T,t-h},
\widehat{\boldsymbol\theta}_T,
\widehat{\boldsymbol\eta}_T(I_{T,k}^-)\bigr)',
\qquad h=0,\ldots,m_T.
\end{align*}
For predefined lag length $m_T$ and weight sequence $\{w_{T_\mathrm{block},h}\}$, the block HAC estimator is
\begin{align*}
\widehat{\mathbf V}_{T}^{\mathrm{HAC}}
&:=\frac1{K_{\mathrm{eval}}}\sum_{k>K_\mathrm{train}}
\widehat{\mathbf V}_{T,k}^{\mathrm{HAC}},
\end{align*}
where
\begin{align*}
\widehat{\mathbf V}_{T,k}^{\mathrm{HAC}} = \widehat{\mathbf V}_{T,k}(0)
+\sum_{h=1}^{m_T}w_{T_\mathrm{block},h}
\{\widehat{\mathbf V}_{T,k}(h)
+\widehat{\mathbf V}_{T,k}(h)'\}.
\end{align*}
Under IID observations we omit all positive-lag terms, giving the special case
\begin{align*}
\widehat{\mathbf V}_{T}
&:=\frac1{K_{\mathrm{eval}}}\sum_{k>K_\mathrm{train}}
\widehat{\mathbf V}_{T,k}(0).
\end{align*}
The same estimation strategy gives $\widehat{\mathbf B}_{T}^{\mathrm{HAC}}$, $\widehat{\mathbf B}_{T}$, $\widehat{\mathbf C}_{T}^{\mathrm{HAC}}$, and $\widehat{\mathbf C}_{T}$. Making the relevant replacements in Table~\ref{tab:oracle-covariance-variance}, we moreover obtain HAC and IID variance estimators of $v_{T,0,\mathrm{SR}}$ and $v_{T,0,\Delta}$, which we denote $\widehat{v}_{T,\mathrm{SR}}^\mathrm{HAC}$, $\widehat{v}_{T,\mathrm{SR}}$, $\widehat{v}_{T,\Delta}^\mathrm{HAC}$, and $\widehat{v}_{T,\Delta}$, respectively.

\subsection{Asymptotic normality of the Sharpe-ratio differential estimator}

\begin{corollary}[Asymptotic normality of the debiased estimator of the Sharpe-ratio differential]
    \label{cor:sharpe-differential-clt}
    Let Assumptions~\ref{ass:conditional-return-model}-\ref{ass:sharpe-target-regularity} hold. Moreover, suppose the following conditions hold, uniformly over $P\in\mathcal P$ and sufficiently large $T$.
    \begin{enumerate}
        \item The known benchmark weights satisfy
    \[
    P\left(\norm{\vnu_{T,0}(\mathbf Z_{T,t})}_\infty 
    \leq \frac{C}{N} \right) = 1;
    \]
    \item The benchmark portfolio has uniformly positive variance: 
    \[
    \tau_{T,0,2}-\tau_{T,0,1}^{2}\geq \underline v_{\mathrm{bench}}>0,
    \qquad
    \|\boldsymbol\tau_{T,0}\|_2\leq C;
    \]
    \item For some constants
    $0<\underline v_\Delta\leq\overline v_\Delta<\infty$,
    \[
    \underline v_\Delta
    \leq v_{T,\Delta}
    \leq \overline v_\Delta.
    \]
    \end{enumerate}
  Then, 
    \[
    \sqrt{T_{\mathrm{eval}}}
    \frac{
    \widehat{\Delta\mathrm{SR}}_{T}^{\mathrm{DB}}
    -\Delta\mathrm{SR}_{T,0}}
    {\sqrt{\widehat{v}_{T,\Delta}}}
    \rightsquigarrow N(0,1).
    \]
    \end{corollary}

\subsection{Semi-parametric efficiency}
\label{sec:supp:iid-efficiency}
We discuss semiparametric efficiency of the debiased moment conditions in the IID case. We follow similar arguments as in \citet{newey1990semiparametric}, extended to triangular arrays.
Let $P_T$ denote the true law of one out-of-sample observation. The comparison class allows IID evaluation observations with joint law
$Q_T$, generated by
\[
\mathbf R_{T,t}
=\boldsymbol\mu_{T,Q}(\mathbf Z_{T,t})
 +\boldsymbol\varepsilon_{T,Q,t},
\qquad
\mathbb E_{Q_T}[\boldsymbol\varepsilon_{T,Q,t}
 \mid\mathbf Z_{T,t}]=\mathbf0,
\]
subject to conditions we detail below. Assumption~\ref{ass:conditional-return-model} is
imposed at the true law $P_T$.
We compare estimators that are regular under local alternatives
$Q_T\ll P_T$ for which $Q_T^{\otimes T_{\mathrm{eval}}}$ is contiguous
to $P_T^{\otimes T_{\mathrm{eval}}}$.
Only the evaluation-observation distribution is perturbed and the
independent initial-training distribution is held fixed. All limits are as $T\to\infty$, with the sample-splitting scheme fixed. 

We impose the following regularity conditions, which we verify below
for MT and smoothed LS with fixed $\tau>0$.

\begin{assumption}[Regularity for IID efficiency]
\label{ass:iid-efficiency-regularity}
\begin{enumerate}
\item[(E1)] \emph{IID observations and bounded returns.}
Each observed triangular array row $\{\mW_{T,t} = (\mR_{T,t}^{\prime}, \mZ_{T,t}^{\prime})^{\prime}: t = 1, \dots , T\}$ is IID with common law $P_{T}$, 
$T_{\mathrm{eval}}\to\infty$, and
$\|\mathbf R_{T,t}\|_\infty\le M$ on the common return support, for a
constant $M$ independent of $T$.

\item[(E2)] \emph{Uniform bounds on the portfolio transformation.}
The known rule is continuously differentiable on a neighbourhood of
$[-M,M]^{N_T}$. There is a constant $C$, independent of $T$, such that
for every relevant $\mathbf z$ and
$\mathbf m,\mathbf m+\mathbf v\in[-M,M]^{N_T}$,
\begin{align*}
\|\boldsymbol\omega_T(\mathbf z,\mathbf m)\|_1&\le C,\\
\|\mathbf J_T(\mathbf z,\mathbf m)\mathbf v\|_1
&\le C\|\mathbf v\|_\infty,\\
\|\boldsymbol\omega_T(\mathbf z,\mathbf m+\mathbf v)
 -\boldsymbol\omega_T(\mathbf z,\mathbf m)
 -\mathbf J_T(\mathbf z,\mathbf m)\mathbf v\|_1
&\le C\|\mathbf v\|_\infty^2.
\end{align*}
All portfolio scaling is included in $\boldsymbol\omega_T$: the return is $\boldsymbol\omega_T'\mathbf R$, with no
additional factor outside the weights.

\item[(E3)] \emph{Asymptotic linearity and score covariance.}
Write
\[
\boldsymbol\psi_{T,0}(\mathbf w)
=\boldsymbol\psi_T^{\mathrm{DB}}
 (\mathbf w,\boldsymbol\theta_{T,0},\boldsymbol\eta_{T,0}),
\qquad
\mathbf V_{T,0}
=\mathbb E_{P_T}[\boldsymbol\psi_{T,0}\boldsymbol\psi_{T,0}'].
\]
The estimator satisfies
\begin{equation}
\sqrt{T_{\mathrm{eval}}}
\{\widehat{\boldsymbol\theta}_T^{\mathrm{DB}}-\boldsymbol\theta_{T,0}\}
=\frac1{\sqrt{T_{\mathrm{eval}}}}
 \sum_{t\in\mathcal T_{\mathrm{eval}}}
 \boldsymbol\psi_{T,0}(\mathbf W_{T,t})+o_P(1),
\label{eq:eff-triangular-al}
\end{equation}
and $\lambda_{\min}(\mathbf V_{T,0})\ge c>0$ for all sufficiently large
$T$, with $c$ independent of $T$.
The covariance upper bound follows from (E1)--(E2).
\end{enumerate}
\end{assumption}

Under (E1)--(E2), the oracle portfolio return and both components of
$\boldsymbol\psi_{T,0}$ are uniformly bounded. The eigenvalue bound in
(E3) also makes
$\mathbf V_{T,0}^{-1/2}\boldsymbol\psi_{T,0}$ uniformly bounded.

\begin{proposition}[Efficiency of transformed portfolios]
\label{prop:iid-efficiency-moments}
Let $\boldsymbol\omega_T$ be any known portfolio-weight function.
Grant Assumptions~\ref{ass:conditional-return-model} and
\ref{ass:portfolio-rule} at $P_T$, and
Assumption~\ref{ass:iid-efficiency-regularity}.
Then $\widehat{\boldsymbol\theta}_T^{\mathrm{DB}}$ is regular and
semiparametrically efficient for the first two oracle portfolio moments
in the stated comparison model. In particular, under every admissible
local alternative,
\[
\mathbf V_{T,0}^{-1/2}\sqrt{T_{\mathrm{eval}}}
\{\widehat{\boldsymbol\theta}_T^{\mathrm{DB}}
  -\boldsymbol\theta_{T,0}(Q_T)\}
\rightsquigarrow N(\mathbf0,\mathbf I_2).
\]
\end{proposition}

\begin{proof}
Set $\mathbf U_T=\mathbf V_{T,0}^{-1/2}\boldsymbol\psi_{T,0}$.
Thus $\mathbb E_{P_T}\mathbf U_T=\mathbf0$ and
$\mathbb E_{P_T}[\mathbf U_T\mathbf U_T']=\mathbf I_2$.
For a law $Q$, let $\boldsymbol\mu_{T,Q}$ and
$\boldsymbol\Omega_{T,Q}$ denote its conditional mean and conditional
second-moment matrix. Its oracle moments use
$\boldsymbol\mu_{T,Q}$ in the same known portfolio rule and take
expectations under $Q$.

\emph{Step 1: derivatives along a regular parametric submodel.}
For this calculation fix a row $T$ and take a regular parametric
submodel $Q_{T,\rho}$ through $P_T$, with density $p_{T,\rho}$ and
score $s_T=\left.\partial\log p_{T,\rho}/\partial\rho\right|_0$.
It suffices here to start with bounded scores.
Differentiating
\[
\mathbb E_{Q_{T,\rho}}\!\left[
\mathbf b(\mathbf Z)'
\{\mathbf R-\boldsymbol\mu_{T,Q_{T,\rho}}(\mathbf Z)\}\right]=0
\]
for arbitrary bounded test functions $\mathbf b$, Leibniz' integral
rule gives
\[
\mathbb E_{P_T}\!\left[
\mathbf b(\mathbf Z)'
\{\boldsymbol\varepsilon_Ts_T-\dot{\boldsymbol\mu}_{T,s}\}\right]=0.
\]
Hence
\[
\dot{\boldsymbol\mu}_{T,s}(\mathbf Z)
=\mathbb E_{P_T}[\boldsymbol\varepsilon_Ts_T\mid\mathbf Z].
\]

Let $q_{T,\rho}(\mathbf w)
=\boldsymbol\omega_T(\mathbf z,
 \boldsymbol\mu_{T,Q_{T,\rho}}(\mathbf z))'\mathbf r$.
For $k=1,2$, applying Leibniz' rule once more yields
\begin{align*}
\left.\frac{d}{d\rho}\theta_{T,0,k}(Q_{T,\rho})\right|_0
&=\left.\frac{d}{d\rho}
 \int q_{T,\rho}(\mathbf w)^k p_{T,\rho}(\mathbf w)
 \,d\lambda(\mathbf w)\right|_0\\
&=\mathbb E_{P_T}[q_{T,0}^k s_T]
 +\mathbb E_{P_T}[kq_{T,0}^{k-1}\dot q_{T,0}],
\end{align*}
where the chain rule gives
$\dot q_{T,0}=\mathbf R'\mathbf J_{T,0}
\dot{\boldsymbol\mu}_{T,s}$.
Conditions (E1)--(E2) justify these interchanges for the bounded
submodels used here. Conditioning on $\mathbf Z$ now gives
\begin{align*}
\dot\theta_{T,1}
&=\mathbb E_{P_T}[q_{T,0}s_T]
 +\mathbb E_{P_T}[\boldsymbol\mu_{T,0}'\mathbf J_{T,0}
                 \dot{\boldsymbol\mu}_{T,s}]\\
&=\mathbb E_{P_T}[\psi_{T,0,1}s_T],\\
\dot\theta_{T,2}
&=\mathbb E_{P_T}[q_{T,0}^2s_T]
 +2\mathbb E_{P_T}[\boldsymbol\omega_{T,0}'
   \boldsymbol\Omega_{T,0}\mathbf J_{T,0}
   \dot{\boldsymbol\mu}_{T,s}]\\
&=\mathbb E_{P_T}[\psi_{T,0,2}s_T].
\end{align*}
At the truth,
$\boldsymbol\Omega_{T,0}
=\boldsymbol\mu_{T,0}\boldsymbol\mu_{T,0}'+\boldsymbol\Sigma_{T,0}$,
so these are precisely the debiased scores we introduce. Their expectation is zero.

Every bounded mean-zero function is the score of a submodel
$dQ_{T,\rho}/dP_T=1+\rho s_T$ for sufficiently small $|\rho|$.
Their mean-square closure is $L_2^0(P_T)$.
Consequently, the scores belong to the model's tangent space and their
projection onto that space equals the scores themselves.
This verifies the influence-function and projection calculations in
Theorems~2.2 and~3.1 of \citet{newey1990semiparametric}.
The following steps establish the required limiting statements when
$N_T$ can grow with $T$.

\emph{Step 2: a bound uniform in the number of assets.}
Fix $T$, take $Q\ll P_T$, and suppress the row index for this step. Expanding the portfolio rule using (E2) and subtracting the linear
terms identified in Step~1 gives
\[
\|\boldsymbol\theta_0(Q)-\boldsymbol\theta_0(P)
 -\mathbb E_Q\boldsymbol\psi_0\|
\le C\mathbb E_{Q_{\mathbf Z}}\!\left[
 \|\boldsymbol\mu_Q-\boldsymbol\mu_0\|_\infty^2
 +\|\boldsymbol\mu_Q-\boldsymbol\mu_0\|_\infty
  \|\boldsymbol\Omega_Q-\boldsymbol\Omega_0\|_{\max}\right].
\]
Here and below $C$ is independent of $T$ and $N_T$.
Write
$H(P,Q)^2=\int(\sqrt{dP}-\sqrt{dQ})^2$, and let
$h_{\mathbf z}=H(P(\mathbf R\mid\mathbf z),Q(\mathbf R\mid\mathbf z))$.
Bounded returns and the Cauchy-Schwarz inequality imply
\[
\|\boldsymbol\mu_Q(\vz)-\boldsymbol\mu_0(\vz)\|_\infty
 +\|\boldsymbol\Omega_Q(\vz)-\boldsymbol\Omega_0(\vz)\|_{\max}
\le C h_{\mathbf z}.
\]
The triangle inequality for square-root densities gives
\[
\mathbb E_{Q_{\mathbf Z}}h_{\mathbf Z}^2
\le\{H(P,Q)+H(P_{\mathbf Z},Q_{\mathbf Z})\}^2
\le4H(P,Q)^2.
\]
Consequently,
\begin{equation}
\|\boldsymbol\theta_{T,0}(Q)-\boldsymbol\theta_{T,0}
 -\mathbb E_Q\boldsymbol\psi_{T,0}\|
\le C H(P_T,Q)^2,
\label{eq:eff-uniform-moment-remainder}
\end{equation}
with $C$ independent of $T$ and $N_T$.

\emph{Step 3: the limit under local alternatives.}
Contiguity implies $T_{\mathrm{eval}}H(P_T,Q_T)^2=O(1)$, and also ensures that the negligible remainder in
\eqref{eq:eff-triangular-al} remains negligible under $Q_{T}$
(see Section~14.5 and Lemma~6.4 of \citet{van2000asymptotic}).
Combining this with Step~2 and (E3) yields
\[
\sqrt{T_{\mathrm{eval}}}\,\mathbf V_{T,0}^{-1/2}
\{\widehat{\boldsymbol\theta}_T^{\mathrm{DB}}
 -\boldsymbol\theta_{T,0}(Q_T)\}
=\frac1{\sqrt{T_{\mathrm{eval}}}}\sum_t
 \{\mathbf U_T(\mathbf W_{T,t})-\mathbb E_{Q_T}\mathbf U_T\}
 +o_{Q_T}(1).
\]
Since $H(P_T,Q_T)\to0$ and $\mathbf U_T$ is uniformly bounded,
$\operatorname{Var}_{Q_T}(\mathbf U_T)=\mathbf I_2+o(1)$.
The bounded triangular-array CLT therefore gives the stated normal
limit and establishes regularity.

\emph{Step 4: the information bound and its attainment.}
We construct a local submodel and verify local asymptotic normality in the sense of Definition~7.14 of \citet{van2000asymptotic}. For each fixed $\mathbf h\in\mathbb R^2$, consider the submodel
\[
\frac{dQ_{T,\mathbf h}}{dP_T}
=1+\frac{\mathbf h'\mathbf U_T}{\sqrt{T_{\mathrm{eval}}}}.
\]
These are positive densities for all sufficiently large $T$, since
$\mathbf U_T$ is bounded and centred.
A log-likelihood expansion gives, uniformly for bounded $\mathbf h$,
\[
\log\frac{dQ_{T,\mathbf h}^{\otimes T_{\mathrm{eval}}}}{dP_T^{\otimes T_{\mathrm{eval}}}}
=\mathbf h'\frac1{\sqrt{T_{\mathrm{eval}}}}\sum_t\mathbf U_T(\mathbf W_{T,t})
 -\tfrac12\mathbf h'\mathbf h+o_P(1).
\]
The sum converges to $N(\mathbf0,\mathbf I_2)$, so these are contiguous
local alternatives with limiting information $\mathbf I_2$.
Moreover, $\mathbb E_{Q_{T,\mathbf h}}\mathbf U_T
=\mathbf h/\sqrt{T_{\mathrm{eval}}}$ exactly, and Step~2 gives
\[
\sqrt{T_{\mathrm{eval}}}\,\mathbf V_{T,0}^{-1/2}
\{\boldsymbol\theta_{T,0}(Q_{T,\mathbf h})
 -\boldsymbol\theta_{T,0}\}
=\mathbf h+o(1).
\]
By Theorems~9.3--9.4 and Proposition~8.4 of
\citet{van2000asymptotic}, the standardized limiting error of any
regular competitor has the same distribution as $\mathbf G+\mathbf B$,
where $\mathbf G\sim N(\mathbf0,\mathbf I_2)$ and $\mathbf B$ is a
random vector independent of $\mathbf G$.
Hence, whenever the limiting covariance is finite,
\[
\operatorname{Cov}(\mathbf G+\mathbf B)
=\mathbf I_2+\operatorname{Cov}(\mathbf B)
\succeq\mathbf I_2.
\]
The normal limit in Step~3 shows that the debiased estimator
attains this bound.
\end{proof}

\begin{samepage}
\begin{corollary}[Sharpe ratios and benchmark differences]
\label{cor:iid-efficiency-sharpe}
Let $\boldsymbol\omega_T$ be any known portfolio rule satisfying
the conditions of Proposition~\ref{prop:iid-efficiency-moments}.
Grant Assumption~\ref{ass:sharpe-target-regularity} for its moment vector $\boldsymbol\theta_{T,0}$.
Then the corresponding debiased Sharpe-ratio estimator is regular
and semiparametrically efficient in the same comparison model.
Under the additional conditions of
Corollary~\ref{cor:sharpe-differential-clt}, the same
holds for its differential against a known benchmark.
The variance-bound sequences are the corresponding influence-function
variances.
\end{corollary}

\begin{proof}
Apply the uniform Taylor argument of
Lemma~\ref{lem:sharpe-hac-delta-method} to the moment expansions
above, jointly with the benchmark moments for the differential.
This gives the corresponding asymptotic-linear representations
and preserves Step~2's uniform $O(H(P_T,Q)^2)$ population remainder.
The standardized scalar influence functions are uniformly bounded,
so Steps~3--4 establish regularity and attainment of the respective
variance bounds.
\end{proof}
\par
\end{samepage}

\subsubsection{Verification of conditions (E1) - (E3) for the MT and smooth LS portfolios.}
Assumptions~\ref{ass:signal-and-nuisance-rates}--\ref{ass:cov-reg}
and the asymptotic-linear expansion in the proof of
Theorem~\ref{thm:cross-fitted-smooth-score-clt} provide (E1) and (E3).
The portfolio bounds in the proof of
Lemma~\ref{lem:expected-score-difference} imply
$\|\boldsymbol\omega_T\|_1\le C$ and
$\|\mathbf J_T\mathbf v\|_1\le C\|\mathbf v\|_\infty$.
The MT Taylor remainder is zero. For smoothed LS, symmetry,
$\ell_1$ duality and the contracted derivative bound in the proof of
Lemma~\ref{lem:expected-score-difference} give
\[
\|D\mathbf J_T[\mathbf v]\mathbf v\|_1
\le\frac{C_\tau}{N_T}\|\mathbf v\|_2^2
\le C_\tau\|\mathbf v\|_\infty^2.
\]
Taylor's formula gives (E2), uniformly in $T$ and $N_T$ for fixed
$\tau>0$.
Thus Proposition~\ref{prop:iid-efficiency-moments} applies to both rules.
Corollary~\ref{cor:iid-efficiency-sharpe} applies under
Assumption~\ref{ass:sharpe-target-regularity} and, for the differential,
the additional conditions of Corollary~\ref{cor:sharpe-differential-clt}.

\subsection{Time-series inference}\label{supp:time-series-results}

In this section, we show that the inference result in Theorem~\ref{thm:cross-fitted-smooth-score-clt} carries over to data that satisfies a triangular-array structure with absolutely regular rows. The delta-method result for the Sharpe-ratio in Lemma~\ref{lem:sharpe-hac-delta-method} goes through if we suppose the conditions of Theorem~\ref{thm:cross-fitted-time-series-clt} and replace the IID variance estimator by its HAC version.
\\
\begin{assumption}[Stationary and uniformly exponentially beta-mixing rows]
\label{ass:beta-mixing-rows}
For every sample size, the row process
\begin{align*}
 \{\mathbf W_{T,t}:t\geq1\}
\end{align*}
is strictly stationary for all $P \in \mathcal{P}$. Define the absolute-regularity coefficient 
\begin{align*}
 \beta_{P,T}(h)=\sup_{s\geq1}\beta_P\bigl(
 &\sigma(\mathbf W_{T,t}:1\leq t\leq s),
 \sigma( 
\mathbf W_{T,t}:t\geq s+h)\bigr),
 \qquad h\geq1.
\end{align*}
There are constants $C_\beta<\infty$ and
$0<\rho_\beta<1$ such that
\begin{align*}
 \sup_{P\in\mathcal P}\sup_{T\geq1}\beta_{P,T}(h)
 \leq C_\beta\rho_\beta^h,\qquad h\geq1.
\end{align*}
\end{assumption}

\begin{remark}[Factor-model justification]
Suppose, for measurable maps $g_T$,
\begin{align*}
\mathbf W_{T,t}
=g_T(\mathbf F_t,\mathbf U_{T,t}),
\end{align*}
where the fixed-dimensional factor process $\{\mathbf F_t\}$ is strictly stationary and
exponentially beta-mixing with constants uniform over the distribution
class. Suppose the innovation vectors $\mathbf U_{T,t}$ are iid over time and independent
of the entire factor process, but leave their contemporaneous cross-sectional
dependence arbitrary. Each row process $\{\mathbf W_{T,t}\}_t$ is therefore strictly stationary.
Given that adding independent iid innovations leaves the factor process's
beta-mixing coefficients unchanged at positive lags, and measurable
transformations cannot increase those coefficients, we obtain the same exponential
bound for $\mathbf W_{T,t}$ as that of $\mathbf F_t$, irrespective of the dimension of $\mathbf W_{T,t}$.
\end{remark}

\begin{assumption}[Time-series oracle covariance regularity]
\label{ass:ts-oracle-covariance}
For all sufficiently large $T$,
there exist constants $0<\underline\ell\leq\overline\ell<\infty$ such that
\begin{align*}
 \inf_{P\in\mathcal P}\lambda_{\min}(\mathbf V_T^\mathrm{LR})
 \geq\underline\ell,
 \qquad\sup_{P\in\mathcal P}\lambda_{\max}(\mathbf V_T^\mathrm{LR})
 \leq\overline\ell.
\end{align*}
\end{assumption}

\begin{assumption}[HAC bandwidth and lag weights]
\label{ass:ts-hac-tuning}
Let $\delta_T = 
\dfrac{r_{\mu,T}}{\sqrt{N_T}} +\dfrac{r_{\Sigma,T}}{N_T} $. The deterministic bandwidth
$m_T\in\{0,\ldots,T_\mathrm{block}-1\}$ and deterministic weights satisfy
\begin{align*}
 m_T&\to\infty,\qquad m_T=o(T^{1/3}),
 \qquad m_T\delta_T\to0,\\
 w_{T_\mathrm{block},0}&=1,\qquad
 \sup_T\max_{0\leq h\leq m_T}|w_{T_\mathrm{block},h}|\leq C,
 \qquad w_{T_\mathrm{block},h}\to1\quad\text{for every fixed }h\geq0.
\end{align*}
\end{assumption}

\begin{theorem}[Uniform time-series weak convergence with HAC studentization]
\label{thm:cross-fitted-time-series-clt}
Maintain the assumptions of
Theorem~\ref{thm:cross-fitted-smooth-score-clt}, replacing
Assumption~\ref{ass:iid-rows} by Assumption~\ref{ass:beta-mixing-rows}.
Suppose additionally that Assumptions~\ref{ass:ts-oracle-covariance}
and \ref{ass:ts-hac-tuning} hold. Then
\begin{align}
 (\widehat{\mathbf V}_T^\mathrm{HAC})^{-1/2}\sqrt{n_T}
 \{\widehat{\boldsymbol\theta}_{T,\tau}-\boldsymbol\theta_0(P)\}
 \rightsquigarrow\mathcal N(\mathbf0,\mathbf I_2)
 \label{eq:time-series-hac-studentization}
\end{align}
uniformly over $P\in\mathcal P$ in the bounded-Lipschitz metric.
\end{theorem}

\section{Proofs} \label{sec:app_proofs}
This section provides proofs to the results in the main text and Section \ref{sec:tech-detail}.

\subsection{Proposition~\ref{prop:bias-var}}

The pooled-OLS score identity gives $\hat\vbeta_T-\vbeta_{T,0}=(\sum_{i,t}\mZ_t^{(i)}{\mZ_t^{(i)}}')^{-1}\sum_{i,t}\mZ_t^{(i)}\{a_TH(C_t)+\varepsilon_{i,t}\}$. In combination with population orthogonality of the regressors and concentration of the sample Gram matrix, we obtain $\hat\beta_{C,T}-\sqrt\lambda=O_p(a_T/\sqrt T+1/\sqrt{NT})$, $\hat\beta_{Z,T}-\sqrt{1-\lambda}=O_p(1/\sqrt{NT})$, and $\|\hat\vbeta_{W,T}\|_2=O_p(\sqrt{p/(NT)})$, where the subscripts identify the corresponding regressors.

The plug-in sample moments at $\vbeta_{T,0}$ have exact linear score representations, so an iid triangular-array CLT and the first-order delta method apply at the limiting moments $(\lambda,3\lambda^2)'$ (see the proofs of Theorem~\ref{thm:cross-fitted-smooth-score-clt} and Lemma~\ref{lem:sharpe-hac-delta-method}). Hence, $Z_T=-(4\sqrt{2T})^{-1}\sum_{t=T+1}^{2T}(C_t^4-6C_t^2+3)$, whose limiting variance is $3/4$. The identities $\E[C_tH(C_t)]=0$, $\E[C_t^3H(C_t)]=\sqrt6$, and $\E[C_t^2H(C_t)^2]=7$ give moment distortions $(O_p(a_T^2),-2\sqrt6\lambda^{3/2}a_T+O_p(a_T^2))'$ and debiased moment distortions $(-a_T^2,-35\lambda a_T^2)'+o_p(a_T^2)$. Applying $\nabla g(\lambda,3\lambda^2)=(3/(2\sqrt2\lambda),-1/(4\sqrt2\lambda^2))'$ and using $\sqrt T\,a_T^2\to0$ gives the result.

\subsection{Theorem~\ref{thm:cross-fitted-smooth-score-clt}}
We introduce the following simplified notation.
\begin{align*}
 \boldsymbol\xi_{T,0,t}
 &: =\boldsymbol\psi_{T}^\mathrm{DB}(\mathbf W_{T,t},
\boldsymbol\theta_{T,0},\boldsymbol\eta_{T,0}),\\
 \mathbf D_{T,k}(\mathbf w)
 &:=
 \boldsymbol\psi_{T}^\mathrm{DB}(\mathbf w,
\boldsymbol\theta_{T,0},\widehat{\boldsymbol\eta}_T^{(k)})
 -\boldsymbol\psi_{T}^\mathrm{DB}(\mathbf w,
\boldsymbol\theta_{T,0},\boldsymbol\eta_{T,0}),\\
 \mathbf D_{T,k,t}&:=\mathbf D_{T,k}(\mathbf W_{T,t}),\\
 \overline{\mathbf D}_{T,k}
 &:=\E_{P}[\mathbf D_{T,k}(\mathbf W_T)],
\end{align*}
where the expectation on the final line is with respect to the marginal distribution of the fresh observation $\mathbf W_T$.

\textit{Step 1 (Oracle CLT).}
The conditional return model implies that both oracle correction terms
have conditional expectation zero, implying
$\E_P[\boldsymbol\xi_{T,0,t}]=\mathbf0$. Moreover, under Assumptions~\ref{ass:iid-rows}-\ref{ass:cov-reg}, the scores are iid and we have 
$\norm{\mathbf V_{T,0}^{-1/2}}_{\mathrm{op}}
 \leq\underline c^{-1/2},$ uniformly over $\mathcal{P}$. Moreover, Lemma~\ref{lem:score-moment-bounds} implies that the Lyapunov condition holds, such that the Lindeberg condition is satisfied. Hence, a triangular-array CLT (Theorem 2.27 in \citet{van2000asymptotic} paired with the Cram\'{e}r-Wold device) gives
\begin{align}
 \mathbf V_{T,0}^{-1/2}
 \frac1{\sqrt{T_{\mathrm{eval}}}}
 \sum_{k > K_\mathrm{train}}\sum_{t\in I_{T,k}}
 \boldsymbol\xi_{T,0,t}
 \rightsquigarrow N(\mathbf0,\mathbf I_2).
 \label{eq:oracle-smooth-score-clt}
\end{align}
for all $P\in\mathcal{P}$. The uniform assumptions imply a CLT along every possibly drifting sequence $\{P_T\}$, such that the sequential characterization then yields uniform convergence.

\textit{Step 2 (Decomposition and centered remainder).}
The estimator has the decomposition
\begin{align*}
 &\sqrt{T_{\mathrm{eval}}}\{
 \widehat{\boldsymbol\theta}_{T}-\boldsymbol\theta_{T,0}\}\\
 &\quad =\frac1{\sqrt{T_{\mathrm{eval}}}}
 \sum_{k > K_\mathrm{train}}\sum_{t\in I_{T,k}}
 \boldsymbol\xi_{T,0,t}
 +\frac1{\sqrt{T_{\mathrm{eval}}}}
 \sum_{k > K_\mathrm{train}}\sum_{t\in I_{T,k}}
 (\mathbf D_{T,k,t}-\overline{\mathbf D}_{T,k})
 +\frac{T_\mathrm{block}}{\sqrt{T_{\mathrm{eval}}}}
 \sum_{k > K_\mathrm{train}}\overline{\mathbf D}_{T,k}.
\end{align*}
To control the centered remainder, define $
 \mathbf S_{T,k}:=\frac1{\sqrt{T_{\mathrm{eval}}}}
 \sum_{t\in I_{T,k}}(\mathbf D_{T,k,t}-\overline{\mathbf D}_{T,k}),$ and the events $
 A_{T,k}:=\{\norm{\mathbf D_{T,k}(\mathbf W_T)}_{P,2}
 \leq\delta_T\}$. Each event is measurable with respect to its fold's own training
sigma-field. Conditioning on the training data, the evaluation observations are iid and independent of the estimated nuisance, and on $A_{T,k}$ we have 
\begin{align*}
 \E_P[\norm{\mathbf S_{T,k}}_2^2]
 \leq\frac{T_\mathrm{block}}{T_{\mathrm{eval}}}
 \norm{\mathbf D_{T,k}(\mathbf W_T)}_{P,2}^2.
\end{align*}
In light of the bounds provided in Lemma~\ref{lem:score-moment-bounds}, Conditional Chebyshev (by which we mean application of the Chebyshev inequality to the conditional probability and Lemma 2.1 in \citet{chernozhukov2017double}), the triangle inequality, and a union bound imply, for any $\varepsilon>0$, that 
\begin{align*}
 \sup_{P\in\mathcal P}P\left(
 \norm{\sum_{k > K_\mathrm{train}}\mathbf S_{T,k}}_2
 >\varepsilon\right)
 =o(1).
\end{align*}

\textit{Step 3 (Bias, approximation, and conclusion).}
Note that 
\begin{align*}
 \norm{\frac{T_\mathrm{block}}{\sqrt{T_{\mathrm{eval}}}}
 \sum_{k > K_\mathrm{train}}\overline{\mathbf D}_{T,k}}_2
 &\leq\sqrt{T_{\mathrm{eval}}}
 \max_{k > K_\mathrm{train}}\norm{\overline{\mathbf D}_{T,k}}_2 \leq\sqrt T\max_{k > K_\mathrm{train}}
 \norm{\overline{\mathbf D}_{T,k}}_2.
\end{align*}
In light of the bound in Lemma~\ref{lem:expected-score-difference}, Conditional Chebyshev and a union bound give, for any $\varepsilon>0$, that 
\begin{align*}
 \sup_{P\in\mathcal P}P\left(
 \norm{\frac{T_\mathrm{block}}{\sqrt{T_{\mathrm{eval}}}}
 \sum_{k > K_\mathrm{train}}\overline{\mathbf D}_{T,k}}_2
 >\varepsilon\right)
 =o(1).
\end{align*}

\textit{Step 4 (Consistent covariance estimation and studentization).}
This follows from steps similar to Steps 4 and 5 of the time-series proof (Theorem~\ref{thm:cross-fitted-time-series-clt}), after removing all autocovariances from the HAC estimator (setting $m_T = 0$).

\subsection{Lemma~\ref{lem:sharpe-hac-delta-method}}
All stochastic bounds and limits below are uniform over
$P\in\mathcal P$. The proof of Theorem \ref{thm:cross-fitted-smooth-score-clt}, gives
\begin{align*}
\sqrt{T_\mathrm{eval}}\norm{\widehat{\boldsymbol\theta}_{T}
 -\boldsymbol\theta_{T,0}}_2&=O_P(1),\\
 \norm{\widehat{\mathbf V}_T-\mathbf V_{T,0} }_{\mathrm{op}}&=o_P(1),\\
 \mathbf V_{T,0}^{-1/2}\sqrt{T_\mathrm{eval}}
 \{\widehat{\boldsymbol\theta}_{T}-\boldsymbol\theta_{T,0}\}
 &\rightsquigarrow N(\mathbf0,\mathbf I_2).
\end{align*}

\textit{Step 1 (Uniform Taylor expansion).}
Assumption~\ref{ass:sharpe-target-regularity} places all vectors $\vtheta_{T,0}$ 
in a compact subset of $\mathcal D = \{\vtheta : \theta_2 - \theta_1^2 > 0\}$. Consequently, there is an
$\epsilon_0>0$ such that their closed $\epsilon_0$-neighborhood is
contained in $\mathcal D$ and the gradient and Hessian of $g$ are
bounded there. With probability tending to one uniformly, the estimator
and the line segment joining it to the target lie in this neighborhood.
Taylor's theorem therefore bounds the remainder by
\begin{align*}
 &\sqrt{T_\mathrm{block}}\left|
 g(\widehat{\boldsymbol\theta}_{T})
 -g(\boldsymbol\theta_{T,0})
 -\nabla g(\boldsymbol\theta_{T,0})'
 (\widehat{\boldsymbol\theta}_{T}-\boldsymbol\theta_{T,0})
 \right|\\
 &\quad\leq C\sqrt{T_\mathrm{block}}
 \norm{\widehat{\boldsymbol\theta}_{T}
 -\boldsymbol\theta_{T,0}}_2^2
 =O_P(T_\mathrm{block}^{-1/2})=o_P(1).
\end{align*}
Moreover, the bounded Hessian ensures 
\begin{align*}
 \norm{\nabla g(\widehat{\boldsymbol\theta}_{T})
 -\nabla g(\boldsymbol\theta_{T,0})}_2=o_P(1).
\end{align*}

\textit{Step 2 (Scalar variance consistency and nondegeneracy).}
The oracle gradient $g(\boldsymbol\theta_{T,0})$ is uniformly bounded and bounded away from zero.
For the latter claim, its first component satisfies
\begin{align*}
 \frac{\theta_{T,0,2}}
 {\{\theta_{T,0,2}-\theta_{T,0,1}^2\}^{3/2}}
 \geq\frac{1}
 {\sqrt{\theta_{T,0,2}-\theta_{T,0,1}^2}},
\end{align*}
which is bounded away from zero by bounded oracle moments.
The upper and lower eigenvalue bounds for $\mathbf V_{T,0}$ thus imply
constants $0<\underline v_*\leq\overline v_*<\infty$ such that
\begin{align*}
 \underline v_*\leq v_{T,0,\mathrm{SR}}\leq\overline v_*
\end{align*}
uniformly for all sufficiently large $T$.
On the common neighborhood event from Step~1, we may expand
\begin{align*}
 |\widehat v_{T,\mathrm{SR}}-v_{T,0,\mathrm{SR}}|
 &\leq C\norm{\widehat{\mathbf V}_T-\mathbf V_{T,0}}_{\mathrm{op}}
 +C\norm{\nabla g(\widehat{\boldsymbol\theta}_{T})
 -\nabla g(\boldsymbol\theta_{T,0})}_2.
\end{align*}
Both terms on the right are $o_P(1)$ uniformly. Since the
probability of the complementary event tends to zero uniformly,
\begin{align*}
 \widehat v_{T,\mathrm{SR}}-v_{T,0,\mathrm{SR}}=o_P(1)
\end{align*}
uniformly over $P\in\mathcal P$ and implies relative
variance consistency. Moreover, the probability of a nonpositive estimated scalar variance vanishes uniformly.
Hence,
\begin{align}\label{eq:sharpe-decomp}
     \sqrt{T_\mathrm{eval}}\{\widehat{\mathrm{SR}}^\mathrm{DB}_{T}-\mathrm{SR}_{T,0}\}
     &=\nabla g(\boldsymbol\theta_{T,0})'
     \sqrt{T_\mathrm{eval}}(\widehat{\boldsymbol\theta}^\mathrm{DB}_{T}
     -\boldsymbol\theta_{T,0})+o_P(1)
\end{align}
 uniformly over $P\in\mathcal P$. 

\textit{Step 3 (Projection and studentization).}
Consider the deterministic unit vector
\begin{align*}
 \mathbf c_T(P)
 :=\frac{\mathbf V_{T,0}^{1/2}\nabla g(\boldsymbol\theta_{T,0})}
 {\sqrt{v_{T,0,\mathrm{SR}}}}.
\end{align*}
The linear term in (\ref{eq:sharpe-decomp}), divided
by $\sqrt{v_{T,0,\mathrm{SR}}}$, equals
\begin{align*}
 \mathbf c_T(P)'\mathbf V_{T,0}^{-1/2}\sqrt{T_\mathrm{eval}}
 \{\widehat{\boldsymbol\theta}_{T}-\boldsymbol\theta_{T,0}\}.
\end{align*}
For any scalar bounded-Lipschitz test function, composition with this
unit-vector projection is a vector bounded-Lipschitz test function
with the same bounds. In light of the Portmanteau lemma (e.g. Lemma 2.2 in \citet{van2000asymptotic}), the uniform vector CLT therefore implies a
uniform scalar standard normal limit even though $\mathbf c_T(P)$
may vary with $T$ and $P$. The Taylor remainder is negligible after
standardization because the scalar variance is bounded away from zero.
Relative variance consistency and uniform Slutsky now yield the final result.

\subsection{Theorem~\ref{thm:cross-fitted-time-series-clt}}
We proceed as in the proof of Theorem~\ref{thm:cross-fitted-smooth-score-clt}.
Recall from that proof the simplified notation
$\boldsymbol\xi_{T,\tau,t}$,
$\mathbf D_{T,k}$, and $\overline{\mathbf D}_{T,k}$, with exactly the
same definitions
All constants and stochastic bounds are uniform over $P\in\mathcal P$
and $T$. Sums over $k$ range over $K_\mathrm{train}+1,\ldots,K$.

\textit{Step 1 (Oracle mixing CLT).}
Step 1 in Theorem~\ref{thm:cross-fitted-smooth-score-clt} carries over if we replace the IID triangular-array CLT by a counterpart for triangular arrays with absolutely regular rows. We use the Wooldridge--White CLT (see Theorem 5.20 in \citet{White2001}). Since $\beta$-mixing implies $\alpha$-mixing with the same rate bound, Assumption~\ref{ass:beta-mixing-rows}, alongside the other imposed assumptions, ensures the conditions of the CLT are met. Uniformity follows from the same argument as before.

\textit{Step 2 (Estimator decomposition and centered nuisance remainder).}
Recall the estimator decomposition in Step 2 of the proof of Theorem~\ref{thm:cross-fitted-smooth-score-clt}. We obtain the same conclusion under Assumption~\ref{ass:beta-mixing-rows} instead of Assumption~\ref{ass:iid-rows} by applying Lemma~\ref{lem:empirical-process-time-series}.
Indeed, applying this lemma componentwise to
\begin{align*}
f_{T,k}(\mathbf w)
={}&\boldsymbol\psi_T^\mathrm{DB}
\bigl(\mathbf w,\boldsymbol\theta_{T,0},\widehat{\boldsymbol\eta}_T^{(k)}\bigr)
-\boldsymbol\psi_T^\mathrm{DB}
\bigl(\mathbf w,\boldsymbol\theta_{T,0},\boldsymbol\eta_{T,0}\bigr)\\
&-\mathbb E_P\left[
\boldsymbol\psi_T^\mathrm{DB}
\bigl(\mathbf W_T,\boldsymbol\theta_{T,0},\widehat{\boldsymbol\eta}_T^{(k)}\bigr)
-\boldsymbol\psi_T^\mathrm{DB}
\bigl(\mathbf W_T,\boldsymbol\theta_{T,0},\boldsymbol\eta_{T,0}\bigr)
\right],
\end{align*}
where the population expectation holds the fitted nuisance fixed, gives
\begin{align*}
\sup_{P\in\mathcal P}
P\left(\norm{\mathbf S_{T,k}}_2>\varepsilon\right)
=o(1)
\end{align*}
for every $\varepsilon>0$ and each
$k=K_\mathrm{train}+1,\ldots,K$.
As $K$ is fixed, a union bound and the triangle inequality give the same conclusion for the sum over the evaluation folds.

\textit{Step 3 (Bias).}
Step 3 in the proof of Theorem~\ref{thm:cross-fitted-smooth-score-clt} carries over unchanged.

\textit{Step 4 (Uniform HAC consistency).}
We apply Lemma~\ref{lem:hac-inference} to
\begin{align*}
f_{T,\mathbf a}(\boldsymbol\eta,\mathbf w)
:=\mathbf a'\{\boldsymbol\psi_T^\mathrm{DB}
\bigl(\mathbf w,\boldsymbol\theta_{T,0},\boldsymbol\eta\bigr)
+\boldsymbol\theta_{T,0}\},
\end{align*}
which is independent of $\boldsymbol\theta_{T,0}$ due to the particular affine specification of $\boldsymbol\psi_T^\mathrm{DB}$.
Here $\mathbf a$ ranges over the coordinate vectors and their sums. The corresponding full-sample estimator is
$\widehat\theta_{T,\mathbf a}
:=\mathbf a'\widehat{\boldsymbol\theta}_T^\mathrm{DB}
$. Under the imposed conditions, the mean of
$f_{T,\mathbf a}(\boldsymbol\eta,\mathbf W_{T,t})$ is zero at
$\boldsymbol\eta=\boldsymbol\eta_{T,0}$, its marginal $4+\nu$ norm is bounded, and the fitted second moment on the evaluation block is bounded.
Finally, with probability at least $1-K_T=1-\Delta_T$,
the squared marginal nuisance error is at most
$\kappa_T=\norm{\mathbf a}_2^2\delta_T$.
Hence, in light of the rate restriction on $m_T$ and the $\sqrt{T}$ rate
$\widehat{\boldsymbol\theta}_T^\mathrm{DB}
-\boldsymbol\theta_{T,0}$,
Lemma~\ref{lem:hac-inference} gives, for every $\varepsilon>0$ and each $\mathbf a$ under consideration,
\[
\sup_{P\in\mathcal P}
P\left(
\left|\mathbf a'(\widehat{\mathbf V}_{T,k}^{\mathrm{HAC}}-\mathbf V_T^\mathrm{LR})\mathbf a\right|
>\varepsilon
\right)\longrightarrow0,
\]
for each fold $k = K_\mathrm{train}+1,\ldots,K$. Since $K$ is fixed we obtain from a union bound and the triangle inequality   
\[
\sup_{P\in\mathcal P}
P\left(
\left|\mathbf a'(\widehat{\mathbf V}_{T}^{\mathrm{HAC}}-\mathbf V_T^\mathrm{LR})\mathbf a\right|
>\varepsilon
\right)\longrightarrow0,
\]
Scalar consistency and polarization
yield the corresponding uniform convergence of
$\norm{\widehat{\mathbf V}_{T}^{\mathrm{HAC}}-\mathbf V_T^\mathrm{LR}}_{\mathrm{op}}$.

\textit{Step 5 (Studentization).}
The uniform consistency of $\widehat{\mathbf V}_{T}^{\mathrm{HAC}}$ and Assumption~\ref{ass:ts-oracle-covariance} imply
\begin{align*}
 \sup_{P\in\mathcal P}P\left(
 \lambda_{\min}(\widehat{\mathbf V}_{T}^{\mathrm{HAC}}-\mathbf V_T^\mathrm{LR})<\underline\ell/2\right)\to 1.
\end{align*}
Continuity of the inverse square-root map on uniformly positive-definite matrices gives the final result in light of the uniform Slutsky theorem.

\subsection{Lemma~\ref{lem:pooled-nn-residual-covariance-rates}}
The bound on $\widehat{\boldsymbol\mu}_T$ follows directly from the network output
restriction. Application of Theorem \ref{thm:farrell-pool} gives, with probability at least
$
1-\exp\!\left\{
-T_{\mathrm{block}}^{p/(\beta+p)}
\log^7T_{\mathrm{block}}\,\log(NT_{\mathrm{block}})
\right\}
$
that 
\begin{align}
\frac1N
\left\|\widehat{\boldsymbol\mu}_T-\boldsymbol\mu_{T,0}
\right\|_{P,2}^2
\leq q_T,
\label{eq:nn-population-event}\\
\frac1{NT_{\mathrm{block}}}
\sum_{t=1}^{T_{\mathrm{block}}}
\left\|
\widehat{\boldsymbol\mu}_T(\mathbf Z_{T,t})
-\boldsymbol\mu_{T,0}(\mathbf Z_{T,t})
\right\|_2^2
\leq q_T.
\label{eq:nn-empirical-event}
\end{align}

The errors $\boldsymbol\varepsilon_{T,t}$ are centered and uniformly bounded under the imposed conditions. In light of its iid coordinates, we obtain the vector's sub-Gaussianity as required in Theorem 4.7.1 and Exercise 4.7.3 in \citet{vershynin2018high}. Let
\begin{align*}
\mathbf U_T
&:=\frac1{T_{\mathrm{block}}}
\sum_{t=1}^{T_{\mathrm{block}}}
\boldsymbol\varepsilon_{T,t}\boldsymbol\varepsilon_{T,t}',
\qquad
\overline{\boldsymbol\varepsilon}_T^{\,0}
:=\frac1{T_{\mathrm{block}}}
\sum_{t=1}^{T_{\mathrm{block}}}\boldsymbol\varepsilon_{T,t},\\
\widetilde{\boldsymbol\Sigma}_T
&:=\frac1{T_{\mathrm{block}}}
\sum_{t=1}^{T_{\mathrm{block}}}
(\boldsymbol\varepsilon_{T,t}
-\overline{\boldsymbol\varepsilon}_T^{\,0})
(\boldsymbol\varepsilon_{T,t}
-\overline{\boldsymbol\varepsilon}_T^{\,0})'
=\mathbf U_T
-\overline{\boldsymbol\varepsilon}_T^{\,0}
 \overline{\boldsymbol\varepsilon}_T^{\,0\prime}.
\end{align*}
The conditional return model gives
$\mathbb E_P[\boldsymbol\varepsilon_{T,t}
\boldsymbol\varepsilon_{T,t}']=\boldsymbol\Sigma_{T,0}$.
The  high-probability covariance bound for sub-Gaussian vectors in Exercise 4.7.3 in \citet{vershynin2018high}, with tail parameter $N$, gives
\[
\|\mathbf U_T-\boldsymbol\Sigma_{T,0}\|_{\mathrm{op}}
\leq C\left\{
\sqrt{\frac{N}{T_{\mathrm{block}}}}
+\frac{N}{T_{\mathrm{block}}}
\right\}
\]
with probability at least $1-2\exp(-N)$, where the required sub-Gaussianity of $\boldsymbol\varepsilon_{T,t}$ follows by its uniformly bounded coordinates and their independence under Assumption~\ref{ass:FLM-main-text}, and the eigenvalue conditions in Assumption~\ref{ass:conditional-return-model}. The corresponding
sample-mean bound gives
\[
\|\overline{\boldsymbol\varepsilon}_T^{\,0}\|_2^2
\leq C\frac{N}{T_{\mathrm{block}}}
\]
with probability at least $1-2\exp(-N)$. Consequently, on an event wih probability at least $1-4\exp(-N)$,
\[
\|\widetilde{\boldsymbol\Sigma}_T
-\boldsymbol\Sigma_{T,0}\|_{\mathrm{op}}
\leq\|\mathbf U_T-\boldsymbol\Sigma_{T,0}\|_{\mathrm{op}}
+\|\overline{\boldsymbol\varepsilon}_T^{\,0}\|_2^2
\leq b_{\Sigma,T},
\]
after enlarging $C$.

To control fitted residuals, define
\[
\mathbf e_{T,t}
:=\boldsymbol\mu_{T,0}(\mathbf Z_{T,t})
-\widehat{\boldsymbol\mu}_T(\mathbf Z_{T,t}),
\qquad
\overline{\mathbf e}_T
:=\frac1{T_{\mathrm{block}}}
\sum_{t=1}^{T_{\mathrm{block}}}\mathbf e_{T,t}.
\]
Since $\widehat{\boldsymbol\varepsilon}_{T,t}
=\boldsymbol\varepsilon_{T,t}+\mathbf e_{T,t}$, the exact centered
covariance decomposition is
\begin{align*}
\widehat{\boldsymbol\Sigma}_T-\boldsymbol\Sigma_{T,0}
&=\widetilde{\boldsymbol\Sigma}_T-\boldsymbol\Sigma_{T,0}
+\mathbf H_T+\mathbf C_T+\mathbf C_T',\\
\mathbf H_T
&:=\frac1{T_{\mathrm{block}}}
\sum_{t=1}^{T_{\mathrm{block}}}
(\mathbf e_{T,t}-\overline{\mathbf e}_T)
(\mathbf e_{T,t}-\overline{\mathbf e}_T)',\\
\mathbf C_T
&:=\frac1{T_{\mathrm{block}}}
\sum_{t=1}^{T_{\mathrm{block}}}
(\boldsymbol\varepsilon_{T,t}
-\overline{\boldsymbol\varepsilon}_T^{\,0})
(\mathbf e_{T,t}-\overline{\mathbf e}_T)'.
\end{align*}
Positive semidefiniteness and
\eqref{eq:nn-empirical-event} give
\[
\|\mathbf H_T\|_{\mathrm{op}}
\leq\operatorname{tr}(\mathbf H_T)
\leq\frac1{T_{\mathrm{block}}}
\sum_{t=1}^{T_{\mathrm{block}}}\|\mathbf e_{T,t}\|_2^2
\leq Nq_T,
\]
with probability at least
$1-\exp\!\left\{
-T_{\mathrm{block}}^{p/(\beta+p)}
\log^7T_{\mathrm{block}}\,\log(NT_{\mathrm{block}})
\right\}$.

The operator-norm product inequality applied to the two centered data
matrices gives
\[
\|\mathbf C_T\|_{\mathrm{op}}
\leq\sqrt{
\|\widetilde{\boldsymbol\Sigma}_T\|_{\mathrm{op}}
\|\mathbf H_T\|_{\mathrm{op}}}.
\]
Thus, on the joint event,
\begin{align*}
\|\widehat{\boldsymbol\Sigma}_T
-\boldsymbol\Sigma_{T,0}\|_{\mathrm{op}}
&\leq b_{\Sigma,T}
+2\sqrt{(C+b_{\Sigma,T})Nq_T}
+Nq_T
=r_{\Sigma,T},
\end{align*}
where the covariance upper bound has been absorbed into $C$. A union bound gives
the stated probability $1-\Delta_T$.

\section{Intermediate lemmas} \label{sec:app_lemmas}
This section contains intermediate lemmas for the debiased score. We show for the market-timing and long-short portfolios that the required estimation rates and moment bounds are satisfied.

\begin{mylem}[Expected debiased-score difference]
\label{lem:expected-score-difference}
Let Assumptions \ref{ass:conditional-return-model}-\ref{ass:signal-and-nuisance-rates} hold. Then, for the market-timing portfolio satisfying
$\max_i a_{T,i,t}\leq C/N_T$ and for the smoothed long-short
portfolio with fixed $\tau\in(0,\infty)$,it holds with $\mathbb{P}_P$-probability at least $1-\Delta_T$, that 
\[
\left\|
\E_P\!\left[
\boldsymbol\psi_T^{\mathrm{DB}}\!\left(
\mathbf W_T,\boldsymbol\theta,\widehat{\boldsymbol\eta}_T^{(k)}
\right)
-
\boldsymbol\psi_T^{\mathrm{DB}}\!\left(
\mathbf W_T,\boldsymbol\theta,
\boldsymbol\eta_{T,0}
\right)
\right]
\right\|_2
\leq
C\left(
\left(\frac{r_{\mu,T}}{\sqrt{N_T}}\right)
^2+
\left(\frac{r_{\mu,T}}{\sqrt{N_T}}\right)
\left(\frac{\delta_{\Sigma,T}}{N_T}
\right)
\right),
\]
where the expectation on the left-hand side is with respect to the marginal distribution of the fresh observation $\mW_T$.
\end{mylem}

\begin{proof}
For each deterministic $\boldsymbol\mu\in\mathcal M_T$ satisfying $P(\|\vmu(\mW_T)\|_\infty \leq C) = 1$  and
$\boldsymbol\Sigma\in\mathcal S_T$, define 
\[
\delta_{\mu,T}
:=
\frac{
\left\|
\boldsymbol\mu(\mathbf Z_T)
-\boldsymbol\mu_{T,0}(\mathbf Z_T)
\right\|_{P,2}
}{\sqrt{N_T}},
\qquad
\delta_{\Sigma,T}
:=
\left\|
\boldsymbol\Sigma-\boldsymbol\Sigma_{T,0}
\right\|_{\mathrm{op}}
\]
and write
\[
\mathbf e_T(\mathbf Z_T)
=
\boldsymbol\mu(\mathbf Z_T)
-\boldsymbol\mu_{T,0}(\mathbf Z_T),
\qquad
\mathbf m_{T,\kappa}(\mathbf Z_T)
=
\boldsymbol\mu_{T,0}(\mathbf Z_T)
+\kappa\mathbf e_T(\mathbf Z_T),
\quad 0\leq\kappa\leq1.
\]
The conditional mean and covariance restrictions give
\begin{align*}
&\E_P\!\left[
\psi_{T,1}^{\mathrm{DB}}\!\left(
\mathbf W_T,\boldsymbol\theta,
(\boldsymbol\mu,\boldsymbol\Sigma)
\right)
-
\psi_{T,1}^{\mathrm{DB}}\!\left(
\mathbf W_T,\boldsymbol\theta,
(\boldsymbol\mu_{T,0},\boldsymbol\Sigma_{T,0})
\right)
\right]
\\
&\quad=
\E_P\!\left[
\left\{
\boldsymbol\omega_T(\mathbf Z_T,\boldsymbol\mu(\mathbf Z_T))
-
\boldsymbol\omega_T(
  \mathbf Z_T,\boldsymbol\mu_{T,0}(\mathbf Z_T))
\right\}'
\boldsymbol\mu_{T,0}(\mathbf Z_T)
\right.
\\
&\hspace{8em}\left.
-
\boldsymbol\mu(\mathbf Z_T)'
\mathbf J_T(\mathbf Z_T,\boldsymbol\mu(\mathbf Z_T))
\mathbf e_T(\mathbf Z_T)
\right]
\\
&\quad=
\E_P\!\left[
\int_0^1
\mathbf e_T(\mathbf Z_T)'
\left\{
\mathbf J_T(\mathbf Z_T,\mathbf m_{T,\kappa}(\mathbf Z_T))'
-
\mathbf J_T(\mathbf Z_T,\boldsymbol\mu(\mathbf Z_T))'
\right\}
\boldsymbol\mu_{T,0}(\mathbf Z_T)\,d\kappa
\right.
\\
&\hspace{8em}\left.
-
\mathbf e_T(\mathbf Z_T)'
\mathbf J_T(\mathbf Z_T,\boldsymbol\mu(\mathbf Z_T))'
\mathbf e_T(\mathbf Z_T)
\right].
\end{align*}
For the second component, the corresponding decomposition is
\begin{align*}
&\E_P\!\left[
\psi_{T,2}^{\mathrm{DB}}\!\left(
\mathbf W_T,\boldsymbol\theta,
(\boldsymbol\mu,\boldsymbol\Sigma)
\right)
-
\psi_{T,2}^{\mathrm{DB}}\!\left(
\mathbf W_T,\boldsymbol\theta,
(\boldsymbol\mu_{T,0},\boldsymbol\Sigma_{T,0})
\right)
\right]
\\
&\quad=
2\,\E_P\!\Biggl[
\int_0^1
\mathbf e_T(\mathbf Z_T)'
\Biggl\{
\begin{aligned}
&\mathbf J_T\!\left(
  \mathbf Z_T,\mathbf m_{T,\kappa}(\mathbf Z_T)
\right)'
\left[
  \boldsymbol\mu_{T,0}(\mathbf Z_T)
  \boldsymbol\mu_{T,0}(\mathbf Z_T)'
  +\boldsymbol\Sigma_{T,0}
\right]
\boldsymbol\omega_T\!\left(
  \mathbf Z_T,\mathbf m_{T,\kappa}(\mathbf Z_T)
\right)
\\
&-
\mathbf J_T\!\left(
  \mathbf Z_T,\boldsymbol\mu(\mathbf Z_T)
\right)'
\left[
  \boldsymbol\mu_{T,0}(\mathbf Z_T)
  \boldsymbol\mu_{T,0}(\mathbf Z_T)'
  +\boldsymbol\Sigma_{T,0}
\right]
\boldsymbol\omega_T\!\left(
  \mathbf Z_T,\boldsymbol\mu(\mathbf Z_T)
\right)
\end{aligned}
\Biggr\}\,d\kappa
\\
&\qquad
-
\left\{
\boldsymbol\omega_T\!\left(
  \mathbf Z_T,\boldsymbol\mu(\mathbf Z_T)
\right)'
\mathbf e_T(\mathbf Z_T)
\right\}
\left\{
\boldsymbol\mu(\mathbf Z_T)'
\mathbf J_T\!\left(
  \mathbf Z_T,\boldsymbol\mu(\mathbf Z_T)
\right)
\mathbf e_T(\mathbf Z_T)
\right\}
\\
&\qquad
-
\left\{
\boldsymbol\omega_T\!\left(
  \mathbf Z_T,\boldsymbol\mu(\mathbf Z_T)
\right)'
\boldsymbol\mu_{T,0}(\mathbf Z_T)
\right\}
\left\{
\mathbf e_T(\mathbf Z_T)'
\mathbf J_T\!\left(
  \mathbf Z_T,\boldsymbol\mu(\mathbf Z_T)
\right)
\mathbf e_T(\mathbf Z_T)
\right\}
\\
&\qquad
-
\boldsymbol\omega_T\!\left(
  \mathbf Z_T,\boldsymbol\mu(\mathbf Z_T)
\right)'
\left(\boldsymbol\Sigma-\boldsymbol\Sigma_{T,0}\right)
\mathbf J_T\!\left(
  \mathbf Z_T,\boldsymbol\mu(\mathbf Z_T)
\right)
\mathbf e_T(\mathbf Z_T)
\Biggr].
\end{align*}
These expressions separate the path remainders, the terms quadratic in
the signal error, and the interaction between the signal and covariance errors.

For market timing,
\[
\mathbf J_T(\mathbf z,\mathbf m)
=\mathbf a_{T,t}(\mathbf z)\mathbf a_{T,t}(\mathbf z)',
\qquad
\|\mathbf J_T(\mathbf z,\mathbf m)\|_{\mathrm{op}}
\leq \frac{C}{N_T},
\]
and the Jacobian does not depend on $\mathbf m$. Also,
$\|\boldsymbol\omega_T(\mathbf z,\mathbf m)\|_2
\leq C/\sqrt{N_T}$ by the diversification and coordinatewise signal
bounds. The first component simplifies directly to
\[
-\E_P\!\left[
\left\{
\mathbf a_{T,t}(\mathbf Z_T)'
\mathbf e_T(\mathbf Z_T)
\right\}^{2}
\right],
\]
which is bounded in absolute value by
$C\|\mathbf e_T(\mathbf Z_T)\|_{P,2}^2/N_T$.
Applying the same operator-norm bounds to the second-component
decomposition gives, for both components, an upper bound of
\begin{align*}
C\left[
\frac{
\|\mathbf e_T(\mathbf Z_T)\|_{P,2}^{2}
}{N_T}
+
\frac{
\delta_{\Sigma,T}
\|\mathbf e_T(\mathbf Z_T)\|_{P,2}
}{N_T^{3/2}}
\right].
\end{align*}

For the fixed-$\tau$ smoothed long-short portfolio,
\[
\|\boldsymbol\omega_T(\mathbf z,\mathbf m)\|_1\leq C,
\qquad
\|\boldsymbol\omega_T(\mathbf z,\mathbf m)\|_2
\leq \frac{10}{\sqrt{N_T}},
\qquad
\|\mathbf J_T(\mathbf z,\mathbf m)\|_{\mathrm{op}}
\leq \frac{5}{N_T\tau}.
\]

The calibrated-logistic Jacobian also satisfies a contracted derivative
bound
\begin{align*}
\left|
\mathbf u'
D\mathbf J_T(\mathbf z,\mathbf m)[\mathbf u]\mathbf v
\right|
\leq
\frac{C}{N_T\tau^2}
\|\mathbf v\|_\infty\|\mathbf u\|_2^2,
\end{align*}
where $\mathbf u$ is the perturbation direction and also the vector
contracting on the left and $\mathbf v$ is the vector in the right-hand
slot. To see this bound, write $\mathbf D=\operatorname{diag}(d_1,\ldots,d_{N_T})$,
$\mathbf d=(d_1,\ldots,d_{N_T})'$, and
$\bar u_d=(\mathbf d'\mathbf u)/(\mathbf 1'\mathbf d)$. Each
calibrated-logistic Jacobian has the form
\[
\mathbf H
=
\mathbf D-\frac{\mathbf d\mathbf d'}{\mathbf 1'\mathbf d},
\qquad
d_i=\frac{p_i(1-p_i)}{\tau},
\]
where the rank-one term accounts for the calibration cutoffs
$c_+(\vmu)$ and $c_-(\vmu)$ changing with $\vmu$. Moreover,
$\mathbf u'\mathbf H\mathbf u
=\sum_i d_i(u_i-\bar u_d)^2 \leq \frac{1}{4\tau}\|\mathbf u\|_2^2.$
Differentiating $\mathbf H$ and applying this bound gives the stated
contraction, recalling the definitions of $p_i^+$ and $p_i^+$ and their dependence on $\tau$ through the logistic function. For the long-short portfolio, apply the same argument separately to the two branches, replacing $p_i$, $\mathbf d$, $\mathbf D$, and $\mathbf H$ by $p_i^\pm$, $\mathbf d^\pm$, $\mathbf D^\pm$, and $\mathbf H^\pm$, respectively, where $d_i^\pm=p_i^\pm(1-p_i^\pm)/\tau$, and sum the resulting bounds using $\widetilde{\mathbf J}=(10/N_T)(\mathbf H^++\mathbf H^-)$. The coordinatewise signal bound implies
\begin{align*}
\left\|
\left[
\boldsymbol\mu_{T,0}(\mathbf Z_T)
\boldsymbol\mu_{T,0}(\mathbf Z_T)'
+\boldsymbol\Sigma_{T,0}
\right]
\boldsymbol\omega_T(\mathbf Z_T,\mathbf m)
\right\|_\infty
\leq C.
\end{align*}
Hence, the path-integral terms in the decomposition are bounded
by $C_\tau\|\mathbf e_T(\mathbf Z_T)\|_{P,2}^2/N_T$.

The two signal-product terms have the same bound, using
$\|\boldsymbol\mu(\mathbf Z_T)\|_2+
\|\boldsymbol\mu_{T,0}(\mathbf Z_T)\|_2=O(\sqrt{N_T})$.
The covariance-error term is bounded by
\[
\frac{C_\tau\delta_{\Sigma,T}}{N_T^{3/2}}
\|\mathbf e_T(\mathbf Z_T)\|_{P,2}.
\]
Thus, for the smoothed long-short portfolio as well, the norm of the
expected score difference is bounded by
\begin{align*}
C_\tau\left[
\frac{
\|\mathbf e_T(\mathbf Z_T)\|_{P,2}^{2}
}{N_T}
+
\frac{
\delta_{\Sigma,T}
\|\mathbf e_T(\mathbf Z_T)\|_{P,2}
}{N_T^{3/2}}
\right].
\end{align*}
Hence, there is a constant $C < \infty$ independent of $P$ 
and $T$, such that 
\[
\left\|
\E_P\!\left[
\boldsymbol\psi_T^{\mathrm{DB}}\!\left(
\mathbf W_T,\boldsymbol\theta,
(\boldsymbol\mu,\boldsymbol\Sigma)
\right)
-
\boldsymbol\psi_T^{\mathrm{DB}}\!\left(
\mathbf W_T,\boldsymbol\theta,
(\boldsymbol\mu_{T,0},\boldsymbol\Sigma_{T,0})
\right)
\right]
\right\|_2
\leq
C\left(
\delta_{\mu,T}^2+
\dfrac{\delta_{\Sigma,T}\delta_{\mu,T}}{N_T}
\right).
\]
The final result now follows under the rate conditions in Assumption~\ref{ass:signal-and-nuisance-rates}, recalling that the expectation in the previous display is with respect to the marginal distribution of the fresh observation $\mW_T$.
\end{proof}

\begin{mylem}[Score-difference and oracle-score moment bounds]
\label{lem:score-moment-bounds}
Suppose Assumptions~\ref{ass:conditional-return-model}-\ref{ass:signal-and-nuisance-rates} hold. For either
the market-timing portfolio satisfying
$\max_i a_{T,i,t}\leq C/N_T$ or the fixed-$\tau$ smoothed long-short
portfolio, it holds, for all $P \in \mathcal{P}$ and all folds $k=1,\ldots,K$, that 
\begin{align*}
&
\left\|
\boldsymbol\psi_T^{\mathrm{DB}}\!\left(
\mathbf W_T,\boldsymbol\theta,
(\widehat{\boldsymbol\mu}_{T,k},
 \widehat{\boldsymbol\Sigma}_{T,k})
\right)
-
\boldsymbol\psi_T^{\mathrm{DB}}\!\left(
\mathbf W_T,\boldsymbol\theta,
(\boldsymbol\mu_{T,0},\boldsymbol\Sigma_{T,0})
\right)
\right\|_{P,2}
\leq
C\left(
\frac{r_{\mu,T}}{\sqrt{N_T}}
+\frac{r_{\Sigma,T}}{N_T}
\right),
\\[1ex]
&
\left\|
\boldsymbol\psi_T^{\mathrm{DB}}\!\left(
\mathbf W_T,\boldsymbol\theta,
(\widehat{\boldsymbol\mu}_{T,k},
 \widehat{\boldsymbol\Sigma}_{T,k})
\right)
-
\boldsymbol\psi_T^{\mathrm{DB}}\!\left(
\mathbf W_T,\boldsymbol\theta,
(\boldsymbol\mu_{T,0},\boldsymbol\Sigma_{T,0})
\right)
\right\|_{P,3}
\leq C,
\\[1ex]
&
\left\|
\boldsymbol\psi_T^{\mathrm{DB}}\!\left(
\mathbf W_T,\boldsymbol\theta,
(\boldsymbol\mu_{T,0},\boldsymbol\Sigma_{T,0})
\right)
\right\|_{P,5}
\leq C,
\end{align*}
where the first and second inequalities have $\mathbb{P}_P$-probability at least $1-\Delta_T$.
\end{mylem}

\begin{proof}
Subtract the realized score formulas componentwise and telescope their
products. The portfolio weight and Jacobian bounds, together with the
coordinatewise signal and data bounds, as derived in the proof of Lemma~\ref{lem:expected-score-difference} or provided in Assumption~\ref{ass:signal-and-nuisance-rates}, give
\begin{align*}
&
\left\|
\boldsymbol\psi_T^{\mathrm{DB}}\!\left(
\mathbf W_T,\boldsymbol\theta,
(\widehat{\boldsymbol\mu}_{T,k},
 \widehat{\boldsymbol\Sigma}_{T,k})
\right)
-
\boldsymbol\psi_T^{\mathrm{DB}}\!\left(
\mathbf W_T,\boldsymbol\theta,
(\boldsymbol\mu_{T,0},\boldsymbol\Sigma_{T,0})
\right)
\right\|_2
\\
&\quad\leq
C\left[
\frac{
\|\widehat{\boldsymbol\mu}_{T,k}(\mathbf W_T)
-\boldsymbol\mu_{T,0}(\mathbf W_T)\|_2
}{\sqrt{N_T}}
+
\frac{
\|\widehat{\boldsymbol\Sigma}_{T,k}
-\boldsymbol\Sigma_{T,0}\|_{\mathrm{op}}
}{N_T}
\right].
\end{align*}
Taking the $L_2(P)$ norm gives the first
inequality.

The same uniform bounds give 
$\|\boldsymbol\omega_T(\mathbf W_T,\boldsymbol\mu)\|_1\leq C$,
$\|\boldsymbol\omega_T(\mathbf W_T,\boldsymbol\mu)\|_2\leq C/\sqrt{N_T}$,
and $\|\mathbf J_T(\mathbf W_T,\boldsymbol\mu)\|_{\mathrm{op}}\leq C/N_T$.
Hence, the score components are uniformly bounded. For example,
\begin{align*}
&\left|
\boldsymbol\mu(\mathbf W_T)'
\mathbf J_T(\mathbf W_T,\boldsymbol\mu(\mathbf W_T))
\bigl(\boldsymbol\mu(\mathbf W_T)-\mathbf R_T\bigr)
\right|
\leq C,
\\
&\left|
\boldsymbol\omega_T(\mathbf W_T,\boldsymbol\mu(\mathbf W_T))'
\left\{
\boldsymbol\mu(\mathbf W_T)\boldsymbol\mu(\mathbf W_T)'
+\boldsymbol\Sigma
\right\}
\mathbf J_T(\mathbf W_T,\boldsymbol\mu(\mathbf W_T))
\bigl(\boldsymbol\mu(\mathbf W_T)-\mathbf R_T\bigr)
\right|
\leq C.
\end{align*}
We obtain that the score difference is bounded in $L_3(P)$, and the oracle score
is bounded in $L_5(P)$. The nuisance bounds hold for all folds
with $\mathbb P_P$-probability at least $1-\Delta_T$, giving the stated
probability for the first two inequalities.
\end{proof}

\section{Additional Results}
\subsection{Additional time-series results}
This section presents additional time-series results on convergence in probability and HAC consistency of measurable functions of triangular arrays with $\beta$-mixing rows. These results may be of independent interest.

\subsection{Training-dependent empirical-process negligibility}
\begin{mylem}[Uniform negligibility of a training-dependent empirical process]
\label{lem:empirical-process-time-series}
Suppose that Assumption~\ref{ass:beta-mixing-rows} holds. Let $T_1+T_2=T$, where
$T_1/T\to\zeta\in(0,1)$. Let $\widehat{\boldsymbol\eta}_T$ be a nuisance
estimator measurable with respect to
$\sigma(\mathbf W_{T,t}:1\leq t\leq T_1)$, and let
$f_T(\boldsymbol\eta,\mathbf w)$ be jointly measurable and real-valued.
Suppose that, for every candidate nuisance parameter and every
$P\in\mathcal P$,
\begin{align*}
 \int f_T(\boldsymbol\eta,\mathbf w)\,dP(\mathbf W_T)(\mathbf w)=0.
\end{align*}
Suppose that there exist deterministic sequences $\kappa_T\in(0,1)$
and $K_T\geq0$, with $\kappa_T\to0$ and $K_T\to0$, such that
\begin{align*}
 \sup_{P\in\mathcal P}P\left(
 \norm{f_T(\widehat{\boldsymbol\eta}_T,\mathbf W_T)}
 _{P,2}^{\,2}>\kappa_T\right)\leq K_T.
\end{align*}
Suppose further that there exist a finite constant $C$, uniform over
$P\in\mathcal P$ and $T$, and a deterministic sequence
$\Xi_T\geq0$, with $\Xi_T\to0$, such that
\begin{align*}
 \sup_{P\in\mathcal P}P\left(
 \norm{f_T(\widehat{\boldsymbol\eta}_T,\mathbf W_T)}
 _{P,3}>C\right)&\leq\Xi_T,\\
 \sup_{P\in\mathcal P}\sup_{1\leq t\leq T}\E_P\left[
 |f_T(\widehat{\boldsymbol\eta}_T,\mathbf W_{T,t})|\right]&\leq C.
\end{align*}
The marginal norms integrate only over $\mathbf W_T$, keeping
$\widehat{\boldsymbol\eta}_T$ fixed, and are training-measurable.
Then, for every $\varepsilon>0$,
\begin{align*}
 \sup_{P\in\mathcal P}P\left(
 \left|\frac1{\sqrt{T_2}}\sum_{t=T_1+1}^{T}
 f_T(\widehat{\boldsymbol\eta}_T,\mathbf W_{T,t})\right|
 >\varepsilon\right)\longrightarrow0.
\end{align*}
\end{mylem}

\begin{proof}
Fix $P\in\mathcal P$. All constants below are uniform over $P$ and $T$.
Write
\begin{align*}
 f_{T,t}&:=f_T(\widehat{\boldsymbol\eta}_T,\mathbf W_{T,t}),
 \qquad\mathcal F_{T,\mathrm{train}}
 :=\sigma(\mathbf W_{T,t}:1\leq t\leq T_1),\\
 A_T&:=\{\norm{f_T(\widehat{\boldsymbol\eta}_T,\mathbf W_T)}
 _{P,2}^{\,2}\leq\kappa_T\},\\
 B_T&:=\{\norm{f_T(\widehat{\boldsymbol\eta}_T,\mathbf W_T)}
 _{P,3}\leq C\}.
\end{align*}
Then $A_T,B_T\in\mathcal F_{T,\mathrm{train}}$ and
$P((A_T\cap B_T)^c)\leq K_T+\Xi_T$.

\textit{Step 1 (Separation and coupling).}
Choose $l_T=\lceil\log T_2\rceil<T_2$ for all sufficiently large $T$.
Decompose
\begin{align*}
 \frac1{\sqrt{T_2}}\sum_{t=T_1+1}^{T}f_{T,t}&=D_T+S_T,\\
 D_T&:=\frac1{\sqrt{T_2}}\sum_{t=T_1+1}^{T_1+l_T}f_{T,t},
 \qquad S_T:=\frac1{\sqrt{T_2}}\sum_{t=T_1+l_T+1}^{T}f_{T,t}.
\end{align*}
Note that measurable transformations cannot increase absolute-regularity
coefficients. By Berbee's coupling lemma, on an extension of the
probability space there is a copy of the future sequence
$\{\mathbf W_{T,t}^*:t\geq T_1+l_T+1\}$, with its original joint law,
independent of $\mathcal F_{T,\mathrm{train}}$, such that
\begin{align*}
 P\left((\mathbf W_{T,t})_{t\geq T_1+l_T+1}
 \neq(\mathbf W_{T,t}^*)_{t\geq T_1+l_T+1}\right)
 \leq\beta_{P,T}(l_T+1)\leq C_\beta\rho_\beta^{l_T+1}.
\end{align*}
Training and its nuisance estimator are unchanged. Define
\begin{align*}
 f_{T,t}^*&:=f_T(\widehat{\boldsymbol\eta}_T,\mathbf W_{T,t}^*),
 \qquad S_T^*:=\frac1{\sqrt{T_2}}
 \sum_{t=T_1+l_T+1}^{T}f_{T,t}^*.
\end{align*}
For every $u>0$,
$P(|S_T|>u)\leq P(|S_T^*|>u)+C_\beta\rho_\beta^{l_T+1}$.

\textit{Step 2 (Conditional covariance bound).}
Conditional on training, the estimator is fixed and the coupled future
retains its original joint law. On $B_T$, marginal centering gives
$\E_P[f_{T,t}^*\mid\mathcal F_{T,\mathrm{train}}]=0$.
Davydov's inequality and the bound of strong-mixing coefficients by
absolute-regularity coefficients give, on $B_T$,
\begin{align*}
 |\operatorname{Cov}_P(f_{T,t}^*,f_{T,t+h}^*
 \mid\mathcal F_{T,\mathrm{train}})|\leq C\rho_\beta^{h/3},
 \qquad h\geq1.
\end{align*}
On $A_T\cap B_T$, the Cauchy--Schwarz inequality also bounds these covariances by
$\kappa_T$ for $h\geq0$. Choose
$p_T=\lceil3\log(1/\kappa_T)/|\log\rho_\beta|\rceil$.
Then, on $A_T\cap B_T$,
\begin{align*}
 &\sum_{h=0}^{\infty}|\operatorname{Cov}_P
 (f_{T,t}^*,f_{T,t+h}^*\mid\mathcal F_{T,\mathrm{train}})|\\
 &\quad\leq(p_T+1)\kappa_T+C\sum_{h=l_T+1}^{\infty}\rho_\beta^{h/3}
 \leq C\kappa_T\{1+\log(1/\kappa_T)\}.
\end{align*}
Conditional stationarity and the retained block length, at most $T_2$,
therefore give
\begin{align*}
 \Var_P(S_T^*\mid\mathcal F_{T,\mathrm{train}})
 \leq C\kappa_T\{1+\log(1/\kappa_T)\}
\end{align*}
on that event.

\textit{Step 3 (Probability bound for the retained sum).}
Conditional Chebyshev gives
\begin{align*}
 P(|S_T^*|>u)
 &\leq K_T+\Xi_T+
 \E_P[\mathbf1_{A_T\cap B_T}
 P(|S_T^*|>u\mid\mathcal F_{T,\mathrm{train}})]\\
 &\leq K_T+\Xi_T+
 \frac{C\kappa_T\{1+\log(1/\kappa_T)\}}{u^2}.
\end{align*}
The final display vanishes uniformly and  the coupling error does as well because of  $l_T\to\infty$.

\textit{Step 4 (Removal of the gap).}
The unconditional first-moment bound yields
$\E_P|D_T|\leq Cl_T/\sqrt{T_2}$.
Combining Markov's inequality and the preceding steps,
\begin{align*}
 &P\left(\left|\frac1{\sqrt{T_2}}\sum_{t=T_1+1}^{T}f_{T,t}\right|
 >\varepsilon\right)\\
 &\quad\leq\frac{Cl_T}{\varepsilon\sqrt{T_2}}+K_T+\Xi_T
 +\frac{C\kappa_T\{1+\log(1/\kappa_T)\}}{\varepsilon^2}
 +C_\beta\rho_\beta^{l_T+1}\longrightarrow0
\end{align*}
uniformly over $P\in\mathcal P$.
\end{proof}

\begin{definition}[Feasible HAC estimator]
\label{def:feasible-hac-estimator}
Let $f_{T,t}:=f_T(\widehat{\boldsymbol\eta}_T,\mathbf W_{T,t})$, 
for $t=T_1+1,\ldots,T$, for $f_T$ some jointly measurable scalar function, where $T_1+T_2=T$ and
$\widehat{\boldsymbol\eta}_T$ uses only training observations
$t\leq T_1$. Let $\widehat\theta_T$ be a centering estimator measurable
with respect to the full sample. For $h=0,\ldots,T_2-1$, define
\begin{align*}
 \widehat\gamma_T(h):=\frac1{T_2}\sum_{t=T_1+h+1}^{T}
 (f_{T,t}-\widehat\theta_T)(f_{T,t-h}-\widehat\theta_T).
\end{align*}
Let $m_T\in\{0,\ldots,T_2-1\}$ be a deterministic lag truncation parameter
and let $w_{T_2,h}$ be deterministic weights, with $w_{T_2,0}=1$.
The feasible heteroskedasticity and autocorrelation consistent (HAC)
estimator is
\begin{align}
 \widehat \Gamma_T:=\widehat\gamma_T(0)
 +2\sum_{h=1}^{m_T}w_{T_2,h}\widehat\gamma_T(h).
 \label{eq:feasible-hac-estimator}
\end{align}
Finally, let
\begin{align*}
\Gamma_{T,0}&:=\V_P\left(\frac1{\sqrt{T_2}}\sum_{t=T_1+1}^{T}
 \{f_T(\boldsymbol\eta_{T,0},\mathbf W_{T,t})-\theta_{T,0}\}\right),\\
 \theta_{T,0}&:=\E_P[f_T(\boldsymbol\eta_{T,0},\mathbf W_{T,t})].
\end{align*}
\end{definition}

\begin{mylem}[Uniform consistency of the feasible HAC estimator]
\label{lem:hac-inference}
Suppose that Assumption~\ref{ass:beta-mixing-rows} holds. Let $T_1+T_2=T$, where
$T_1/T\to\zeta\in(0,1)$. Let $\widehat{\boldsymbol\eta}_T$ be measurable
with respect to $\sigma(\mathbf W_{T,t}:1\leq t\leq T_1)$, let
$\boldsymbol\eta_{T,0}$ be nonrandom, and let
$f_T(\boldsymbol\eta,\mathbf w)$ be jointly measurable and real-valued function (which may be different from the choice in Lemma~\ref{lem:empirical-process-time-series}).

Suppose that there exist deterministic sequences $\kappa_T\in(0,1)$
and $K_T\geq0$, with $\kappa_T\to0$ and $K_T\to0$, such that
\begin{align*}
 \sup_{P\in\mathcal P}P\left(
 \norm{f_T(\widehat{\boldsymbol\eta}_T,\mathbf W_T)
 -f_T(\boldsymbol\eta_{T,0},\mathbf W_T)}_{P,2}^{\,2}
 >\kappa_T\right)\leq K_T.
\end{align*}
Suppose further that, for some fixed $\nu>0$, a finite uniform constant
$C$ satisfies
\begin{align*}
 \sup_{P\in\mathcal P}\sup_T
 \norm{f_T(\boldsymbol\eta_{T,0},\mathbf W_T)}_{P,4+\nu}
 &\leq C,\\
 \sup_{P\in\mathcal P}\sup_T\sup_{1\leq t\leq T}\E_P[f_T(\widehat{\boldsymbol\eta}_T,\mathbf W_{T,t})^{\,2}]&\leq C.
\end{align*}

Let the deterministic bandwidth and weights satisfy
\begin{align*}
 m_T&\to\infty,\qquad m_T=o(T_2^{1/3}),
 \qquad m_T\sqrt{\kappa_T}\to0,\\
 w_{T_2,0}&=1,\qquad
 \sup_T\max_{0\leq h\leq m_T}|w_{T_2,h}|\leq C,
 \qquad w_{T_2,h}\to1\quad\text{for every fixed }h\geq0.
\end{align*}
Finally, let $\widehat\theta_T$ be full-sample measurable and suppose that, for
every $\varepsilon>0$,
\begin{align*}
 \sup_{P\in\mathcal P}P\left(
 m_T|\widehat\theta_T-\theta_{T,0}(P)|>\varepsilon\right)\to0.
\end{align*}
Then, for every $\varepsilon>0$,
\begin{align*}
 \sup_{P\in\mathcal P}P(|\widehat \Gamma_T-\Gamma_T(P)|>\varepsilon)\to0.
\end{align*}
\end{mylem}

\begin{proof}
Fix $P\in\mathcal P$. All constants and stochastic bounds below are uniform over $P$ and $T$. 
Define
\begin{align*} f_{T,t}&:=f_T(\widehat{\boldsymbol\eta}_T,\mathbf W_{T,t}),
 \qquad f_{T,0,t}:=f_T(\boldsymbol\eta_{T,0},\mathbf W_{T,t}).
\end{align*} and write
\begin{align*}
 Y_{T,t}:=f_{T,0,t}-\theta_{T,0},
 \qquad e_{T,t}:=f_{T,t}-f_{T,0,t},
 \qquad d_T:=\widehat\theta_T-\theta_{T,0}.
\end{align*}

\textit{Step 1 (Convergence of the infeasible estimator).}
Define
\begin{align*}
 \bar \Gamma_T:=\frac1{T_2}\sum_{t=T_1+1}^{T}Y_{T,t}^2
 +\frac2{T_2}\sum_{h=1}^{m_T}w_{T_2,h}
 \sum_{t=T_1+h+1}^{T}Y_{T,t}Y_{T,t-h}.
\end{align*}
Let $\gamma_{T}(h):=\operatorname{Cov}_P(Y_{T,t},Y_{T,t+h})$ and
$a_\nu:=(2+\nu)/(4+\nu)>0$. Stationarity, the oracle moment bound, and
Davydov's inequality give
\begin{align*}
 |\gamma_{T}(0)|\leq C,
 \qquad |\gamma_{T}(h)|\leq C\rho_\beta^{a_\nu h},\quad h\geq1.
\end{align*}
Moreover,
\begin{align*}
 \Gamma_T&=\gamma_{T}(0)+2\sum_{h=1}^{T_2-1}
 \left(1-\frac h{T_2}\right)\gamma_{T}(h),\\
 \E_P[\bar \Gamma_T]&=\gamma_{T}(0)+2\sum_{h=1}^{m_T}w_{T_2,h}
 \left(1-\frac h{T_2}\right)\gamma_{T}(h).
\end{align*}
Hence
\begin{align*}
 |\E_P[\bar \Gamma_T]-\Gamma_T|
\leq2\sum_{h=1}^{m_T}|w_{T_2,h}-1|\,|\gamma_{T}(h)|
 +2\sum_{h=m_T+1}^{T_2-1}|\gamma_{T}(h)|.
\end{align*}
The last term is at most $C\rho_\beta^{a_\nu(m_T+1)}$.
For all sufficiently large $T$, define
\begin{align*}
 H_T:=\max\left\{j\in\{1,\ldots,m_T\}:
 \max_{1\leq h\leq j}|w_{T_2,h}-1|\leq\frac1j\right\}.
\end{align*}
For every fixed $j$, eventually $m_T\geq j$ and the displayed weight
condition holds. Hence the set is eventually nonempty, $H_T\to\infty$,
and the maximum discrepancy through $H_T$ is at most $H_T^{-1}$.
Splitting at $H_T$ yields
\begin{align*}
 \sum_{h=1}^{m_T}|w_{T_2,h}-1|\,|\gamma_{T}(h)|
 &\leq\frac C{H_T}\sum_{h=1}^{H_T}\rho_\beta^{a_\nu h}
 +C\sum_{h=H_T+1}^{\infty}\rho_\beta^{a_\nu h}\\
 &\leq\frac C{H_T}+C\rho_\beta^{a_\nu(H_T+1)}\to0.
\end{align*}
Consequently,
\begin{align*}
 \sup_{P\in\mathcal P}|\E_P[\bar \Gamma_T]-\Gamma_T|
 \leq\frac C{H_T}+C\rho_\beta^{a_\nu(H_T+1)}
 +C\rho_\beta^{a_\nu(m_T+1)}\to0.
\end{align*}
For the fluctuation set $U_{T,h,t}:=Y_{T,t}Y_{T,t-h}$.
H\"older's inequality gives $\norm{U_{T,h,t}}_{P,2+\nu/2}\leq C$, while the Cauchy--Schwarz inequality bounds product covariances by $C$ for lag distances
$0\leq k\leq h$. For $k>h$, Davydov's inequality gives, with
$b_\nu:=\nu/(4+\nu)>0$,
\begin{align*}
 |\operatorname{Cov}_P(U_{T,h,t},U_{T,h,t+k})|
 \leq C\rho_\beta^{b_\nu(k-h)}.
\end{align*}
The sum of absolute covariances is bounded by $C(h+1)$, so
\begin{align*}
 \norm{\frac1{T_2}\sum_{t=T_1+h+1}^{T}
 \{U_{T,h,t}-\E_P[U_{T,h,t}]\}}_{P,2}
 \leq C\sqrt{\frac{h+1}{T_2}}.
\end{align*}
Minkowski's inequality and bounded weights imply
\begin{align*}
 \norm{\bar \Gamma_T-\E_P[\bar \Gamma_T]}_{P,2}
 \leq\frac C{\sqrt{T_2}}\sum_{h=0}^{m_T}\sqrt{h+1}
 \leq\frac{C(m_T+1)^{3/2}}{\sqrt{T_2}}\to0.
\end{align*}
Thus $\bar \Gamma_T-\Gamma_T=o_P(1)$ uniformly.

\textit{Step 2 (Feasible-to-infeasible decomposition).}
Expanding products gives
\begin{align*}
 \widehat \Gamma_T-\bar \Gamma_T
 &=\frac1{T_2}\sum_{t=T_1+1}^{T}\{2Y_{T,t}e_{T,t}+e_{T,t}^2\}\\
 &\quad+\frac2{T_2}\sum_{h=1}^{m_T}w_{T_2,h}
 \sum_{t=T_1+h+1}^{T}
 \{Y_{T,t}e_{T,t-h}+Y_{T,t-h}e_{T,t}+e_{T,t}e_{T,t-h}\}\\
 &\quad-2d_T\left\{\frac1{T_2}\sum_{t=T_1+1}^{T}(Y_{T,t}+e_{T,t})\right.\\
 &\hspace{23mm}\left.+\frac1{T_2}\sum_{h=1}^{m_T}w_{T_2,h}
 \sum_{t=T_1+h+1}^{T}
 (Y_{T,t}+Y_{T,t-h}+e_{T,t}+e_{T,t-h})\right\}\\
 &\quad+d_T^2\left\{1+2\sum_{h=1}^{m_T}w_{T_2,h}
 \left(1-\frac h{T_2}\right)\right\}.
\end{align*}

\textit{Step 3 (Estimated centering).}
Let $F_T:=T_2^{-1}\sum_{t=T_1+1}^{T}
\{f_{T,t}-\theta_{T,0}(P)\}^2$.
The fitted second-moment and oracle moment bounds imply $\E_P[F_T]\leq C$,
hence $F_T=O_P(1)$ uniformly. In light if the Cauchy--Schwarz inequality, we may bound each normalized
subset sum of $Y_{T,t}+e_{T,t}$ by $\sqrt{F_T}$.
The linear and quadratic centering contributions are therefore bounded by
\begin{align*}
 C(m_T+1)|d_T|\sqrt{F_T}=o_P(1),
 \qquad C(m_T+1)d_T^2=o_P(1).
\end{align*}
utilizing $m_T|d_T|=o_P(1)$.

\textit{Step 4 (Products of nuisance errors).}
Define $Q_T:=T_2^{-1}\sum_{t=T_1+1}^{T}e_{T,t}^2$.
Choose $l_T=\lceil\log T_2\rceil$ and split
\begin{align*}
 Q_T&=Q_T^{\mathrm{gap}}+Q_T^{\mathrm{rem}},\\
 Q_T^{\mathrm{gap}}&:=\frac1{T_2}\sum_{t=T_1+1}^{T_1+l_T}e_{T,t}^2,
 \qquad Q_T^{\mathrm{rem}}:=\frac1{T_2}
 \sum_{t=T_1+l_T+1}^{T}e_{T,t}^2.
\end{align*}
The unconditional bounds give $\E_P[Q_T^{\mathrm{gap}}]\leq Cl_T/T_2$.
Let $\mathcal F_{T,\mathrm{train}}:=\sigma(\mathbf W_{T,t}:1\leq t\leq T_1)$
and
\begin{align*}
 A_T:=\{\norm{f_T(\widehat{\boldsymbol\eta}_T,\mathbf W_T)
 -f_T(\boldsymbol\eta_{T,0},\mathbf W_T)}_{P(\mathbf W_T),2}^2
 \leq\kappa_T\}.
\end{align*}
Then $P(A_T^c)\leq K_T$. Berbee's lemma supplies a copy of the remainder
block, preserving its joint law and independent of training, with
mismatch probability at most $C_\beta\rho_\beta^{l_T+1}$.
Keeping training unchanged, define
\begin{align*}
 e_{T,t}^*&:=f_T(\widehat{\boldsymbol\eta}_T,\mathbf W_{T,t}^*)
 -f_T(\boldsymbol\eta_{T,0},\mathbf W_{T,t}^*),\\
 Q_T^*&:=\frac1{T_2}\sum_{t=T_1+l_T+1}^{T}(e_{T,t}^*)^2.
\end{align*}
On $A_T$, $\E_P[Q_T^*\mid\mathcal F_{T,\mathrm{train}}]\leq\kappa_T$.
Conditional Markov, coupling, and the gap bound give, for $u>0$,
\begin{align*}
 P(m_T^2Q_T>u)
 \leq\frac{Cm_T^2l_T}{uT_2}+K_T+C_\beta\rho_\beta^{l_T+1}
 +\frac{2m_T^2\kappa_T}{u}\to0.
\end{align*}
The convergence follows as $m_T\sqrt{\kappa_T}\to0$ and 
\begin{align*}
 \frac{m_T^2l_T}{T_2}
 =\left(\frac{m_T}{T_2^{1/3}}\right)^2
 \frac{l_T}{T_2^{1/3}}\to0.
\end{align*}
Thus $m_T\sqrt{Q_T}=o_P(1)$ uniformly.
For every $h\leq m_T$, the Cauchy--Schwarz inequality gives
\begin{align*}
 \frac1{T_2}\left|\sum_{t=T_1+h+1}^{T}e_{T,t}e_{T,t-h}\right|\leq Q_T.
\end{align*}
The nuisance-product contribution is at most $C(m_T+1)Q_T=o_P(1)$.

\textit{Step 5 (Mixed terms and conclusion).}
Set $R_T:=T_2^{-1}\sum_{t=T_1+1}^{T}Y_{T,t}^2=O_P(1)$ uniformly.
Each normalized lagged mixed sum, in either orientation and including
lag zero, is at most $\sqrt{R_TQ_T}$ in absolute value.
Their total contribution is bounded by
$C(m_T+1)\sqrt{R_TQ_T}=o_P(1)$.
Steps~3--5 give $\widehat \Gamma_T-\bar \Gamma_T=o_P(1)$ uniformly.
Combining this with Step~1 proves the lemma.
\end{proof}

\subsection{Pooled neural-network rates}\label{sec:supp-FLM-pooled}
In this section, we extend the result in \cite{FarrellLiangMisra2021}, henceforth FLM, to pooled least-squares regression. Throughout this section, we use FLM's notation, which is independent of the notation used elsewhere in this paper. The key consequence of unrestricted cross-sectional dependence is that the independent sampling units are the $T$ dates rather than the $NT$ date-asset observations. Using FLM's proof strategy, adapted to independent date blocks by averaging observations within each date, the guaranteed squared-error rate is $T^{-\beta/(\beta+d)}\log^7T\log(NT)$. Relative to FLM's univariate bound, the leading term includes the additional factor $\log(NT)/\log T$. Here $\beta$ denotes the smoothness of the population target and $d$ the dimensionality of the covariates.

Unrestricted cross-sectional dependence is a realistic assumption in asset-pricing studies, since most feature sets contain state variables, like market volatility, that are unchanged between assets. In contrast, when observations are mutually independent in the cross-section over time, the corresponding pooled-regression argument can use $NT$ instead of $T$ as the sample size, in light of FLM's univariate result.

Let $\mathcal B=(\boldsymbol Z_1,\ldots,\boldsymbol Z_N)$ denote a generic cross-sectional block, where $\boldsymbol Z_i=(Y_i,\boldsymbol X_i')'$. We define the population target as the minimizer of block-averaged expected squared-error loss:
\begin{align*}
\ell(f,\boldsymbol z)&=\frac{1}{2}\{y-f(\boldsymbol x)\}^2,
& f_* \in\underset{\boldsymbol{f:\,\norm{f}_{\mathrm{pool}}<\infty}}{\arg\min}
\ \mathbb E\!\left[\frac{1}{N}\sum_{i=1}^N\ell(f,\boldsymbol Z_i)\right],
\end{align*}
where minimization is over square-integrable measurable functions under the pooled population norm defined below, not over the neural-network class. We suppress dependence of the generic block and the population target on $T$ is suppressed.

\begin{assumption}[1P: Panel sampling and smoothness]
\label{ass:panel}
For each $T$, let $N=N_T\geq1$ and observe $\boldsymbol Z_{i,t}=(Y_{i,t},\boldsymbol X_{i,t}')'$, $1\leq i\leq N$, $1\leq t\leq T$. Assume:
\begin{enumerate}
\item The full blocks $\mathcal B_t=(\boldsymbol Z_{1,t},\ldots,\boldsymbol Z_{N,t})$ are independent and identically distributed copies of $\mathcal B$ over $t$. Dependence within each block is unrestricted.
\item For every $i$ and $t$, $\boldsymbol Z_{i,t}\in\mathcal Y\times[-1,1]^d$ almost surely, where $\boldsymbol X_{i,t}$ is continuously distributed. Marginal distributions may differ across $i$.
\item For a fixed constant $M>0$, $\norm{f_*}_\infty\leq M$ and $\mathcal Y\subseteq[-M,M]$.
\item FLM's Assumption 2 holds 
with smoothness parameter $\beta\in\mathbb N_+$.
\end{enumerate}
The dimension $d$, the smoothness parameter $\beta$, and the constant $M$ are independent of $N$ and $T$.
\end{assumption}

\begin{theorem}[Panel version of FLM Theorem 1]
\label{thm:farrell-pool}
Under Assumption~\ref{ass:panel}, let
\begin{align*}
\widehat f_{\mathrm{MLP,pool}}
\in\underset{f\in\mathcal F_{\mathrm{MLP}},\,\norm{f}_\infty\leq2M}{\arg\min}
\sum_{t=1}^T\sum_{i=1}^N\{Y_{i,t}-f(\boldsymbol X_{i,t})\}^2,
\end{align*}
using a ReLU multilayer perceptron with width $H\asymp T^{d/[2(\beta+d)]}\log^2T$ and depth $L\asymp\log T$.

For all sufficiently large $T$, with probability at least $1-\exp\{-T^{d/(\beta+d)}\log^8T\}$, both
\begin{align*}
\textnormal{(a)}\quad
\norm{\widehat f_{\mathrm{MLP,pool}}-f_*}_{\mathrm{pool}}^2
&\leq C\left\{T^{-\beta/(\beta+d)}\log^7T\log(NT)+\frac{\log\log T}{T}\right\},\\
\textnormal{(b)}\quad
\E_{NT}\bigl[(\widehat f_{\mathrm{MLP,pool}}-f_*)^2\bigr]
&\leq C\left\{T^{-\beta/(\beta+d)}\log^7T\log(NT)+\frac{\log\log T}{T}\right\}.
\end{align*}
where the pooled population norm is given by
\[
\norm{h}_{\mathrm{pool}}^2=\E\!\left[\frac1N\sum_{i=1}^Nh(\boldsymbol X_i)^2\right],
\]
and $\E_{NT}[h^2]=(NT)^{-1}\sum_{t=1}^T\sum_{i=1}^Nh(\boldsymbol X_{i,t})^2$.
The constant $C$ does not depend on $N$ or $T$.
\end{theorem}

\begin{proof}
The proof follows along the steps of the univariate-regression proof of FLM, with several notable changes. We outline these changes below.
Following FLM's notation, write $\widehat f=\widehat f_{\mathrm{MLP,pool}}$ and let $\mathcal F_{\mathrm{DNN}}$ denote its deterministic network class restricted by $\norm{f}_\infty\leq2M$. Choose a deterministic approximant $f_T\in\mathcal F_{\mathrm{DNN}}$ with $\norm{f_T-f_*}_\infty\leq\epsilon_T$, and define the block empirical operator and block loss difference as
\begin{align*}
\E_T^{\mathrm{block}}[a]&=\frac1T\sum_{t=1}^Ta(\mathcal B_t),
& g_f^{\mathrm{block}}(\mathcal B_t)&=\frac1N\sum_{i=1}^N\{\ell(f,\boldsymbol Z_{i,t})-\ell(f_*,\boldsymbol Z_{i,t})\}.
\end{align*}
Since $f_*$ is the pooled population squared-loss minimizer, for every fixed $f$,
\[
\E[g_f^{\mathrm{block}}]=\frac12\norm{f-f_*}_{\mathrm{pool}}^2.
\]
Thus, empirical optimality gives the block counterpart of FLM's decomposition (A.1):
\begin{align*}
\frac12\norm{\widehat f-f_*}_{\mathrm{pool}}^2
\leq (\E-\E_T^{\mathrm{block}})g_{\widehat f}^{\mathrm{block}}
+\E_T^{\mathrm{block}}g_{f_T}^{\mathrm{block}},
\end{align*}
where population expectations integrate over a generic block with the function held fixed.

For FLM's Step 1, Bernstein's inequality is applied to the independent cross-sectional averages $g_{f_T}^{\mathrm{block}}(\mathcal B_t)$, rather than to individual observations as in FLM. Jensen's inequality gives $\Var(g_{f_T}^{\mathrm{block}})\leq C\epsilon_T^2$ without imposing within-date independence; the population mean is at most $\epsilon_T^2/2$ and the envelope is bounded by a fixed constant. Thus the analogue of FLM's (A.2) is, with probability at least $1-e^{-\tilde\gamma}$,
\begin{align*}
\E_T^{\mathrm{block}}g_{f_T}^{\mathrm{block}}
\leq C\left(\epsilon_T^2+\frac{\tilde\gamma}{T}\right).
\end{align*}

For FLM's Step 2, FLM's localized loss class becomes
\[
\mathcal G_r^{\mathrm{block}}=\{g_f^{\mathrm{block}}:f\in\mathcal F_{\mathrm{DNN}},\ \norm{f-f_*}_{\mathrm{pool}}\leq r\},
\]
and its quadratic process is replaced by the block averages
\[
q_f^{\mathrm{block}}(\mathcal B_t)=N^{-1}\sum_i\{f(\boldsymbol X_{i,t})-f_*(\boldsymbol X_{i,t})\}^2,
\]
such that we have
\begin{align*}
(\E_T^{\mathrm{block}}-\E)q_f^{\mathrm{block}}
&=\E_{NT}[(f-f_*)^2]-\norm{f-f_*}_{\mathrm{pool}}^2,
\end{align*}
and we use the following block-version of FLM's `$R_n\mathcal{F}$' function:
\begin{align*}
R_T^{\mathrm{block}}\mathcal A
&=\sup_{a\in\mathcal A}\frac1T\sum_{t=1}^T\eta_ta(\mathcal B_t).
\end{align*}
Here, the independent Rademacher signs $\eta_t$ are attached to dates, not to individual observations. Jensen's inequality preserves the radius-dependent variance bounds for both block classes and supplies their empirical $L_2$ metric comparisons with the pooled metric.
Specifically, write $\norm{a}_{T,\mathrm{block}}^2=\E_T^{\mathrm{block}}[a^2]$ and $\norm{h}_{NT}^2=\E_{NT}[h^2]$. For $a=g,q$, boundedness and Jensen's inequality give
\begin{align*}
\E[(a_f^{\mathrm{block}})^2]
&\leq C\norm{f-f_*}_{\mathrm{pool}}^2,
& \norm{a_f^{\mathrm{block}}-a_{\widetilde f}^{\mathrm{block}}}_{T,\mathrm{block}}
&\leq C\norm{f-\widetilde f}_{NT},
\end{align*}
with $C$ independent of $N$ and $T$.
Because a single sign multiplies an entire block, these metric comparisons replace scalar contraction before applying FLM's Dudley and symmetrization lemmas. FLM's original Lemma 4 can be applied to all $NT$ covariate evaluations and a deterministic covering step requires no within-date independence, such that
\begin{align*}
\mathcal N\!\left(\delta,\mathcal F_{\mathrm{DNN}}|_{\{\boldsymbol X_{i,t}\}_{i,t}},\infty\right)
\leq\left\{\frac{4eM\,NT}{\delta\,\mathrm{Pdim}(\mathcal F_{\mathrm{DNN}})}\right\}^{\mathrm{Pdim}(\mathcal F_{\mathrm{DNN}})},
\qquad NT\geq\mathrm{Pdim}(\mathcal F_{\mathrm{DNN}}).
\end{align*}
The block metric comparisons transfer these covers to the localized loss and quadratic classes, up to fixed rescaling of covering radii. Symmetrizing the block quadratic process and solving the resulting empirical-radius inequality supplies the radius control needed for chaining.
For each fixed $r\geq T^{-1}$, this gives, with probability at least $1-2e^{-\tilde\gamma}$,
\begin{align*}
\sup_{\substack{f\in\mathcal F_{\mathrm{DNN}}\\
                \norm{f-f_*}_{\mathrm{pool}}\leq r}}
\E_{NT}[(f-f_*)^2]
\leq C\left\{r^2+
\frac{\mathrm{Pdim}(\mathcal F_{\mathrm{DNN}})\log(NT)+\tilde\gamma}{T}\right\}.
\end{align*}
FLM's norm comparison, one-step improvement, and dyadic localization in Appendix A.2 then give both population and pooled empirical bounds, where we note that concentration uses $T$ independent blocks, whereas the entropy logarithm is now $\log(NT)$ rather than $\log T$. The approximation and MLP specialization follow FLM's Appendix A.3 unchanged, with `$n$' replaced by $T$ in the architecture and stochastic denominators and the modified entropy logarithm retained. Since $\log(NT)\geq\log T$, FLM's original confidence choice is absorbed by the modified bound, proving the theorem.
\end{proof}

\section{Additional Simulation Results} \label{sec:app_simulation}
\subsection{Neural network implementation}\label{sec:app_sim_training}
\subsubsection{OLS initialization}\label{sec:app_sim_ols_init}
Let $f_{\mathrm{OLS}}$ be the pooled OLS fit with an intercept and all 25 predictors, estimated on the fitting dates. To initialize the network, we use 16 signed pairs of affine indices
\[
h_k(\mathbf x)=f_{\mathrm{OLS}}(\mathbf x)+\mathbf d_k'(\mathbf x-\bar{\mathbf x}_f)+\kappa_k,
\qquad k=1,\ldots,16,
\]
where $\bar{\mathbf x}_f$ is the fitting-sample predictor mean. We draw the slope and intercept offsets independently from $\mathrm{Unif}[-1,1]$, center each coordinate across pairs, and scale all offsets jointly so that the root mean square of the perturbations over pairs and fitting observations is $5\%$ of the fitting-sample standard deviation of $f_{\mathrm{OLS}}$ (divisor $360N$). The pair for $h_k$ consists of $\operatorname{ReLU}(h_k)$ and $\operatorname{ReLU}(-h_k)$ with output weights $1/16$ and $-1/16$ and zero output bias for the network. Their sum over pairs reproduces $f_{\mathrm{OLS}}$. For a unit with slope vector $\mathbf w$, bias $b$ and output weight $v$, we balance the initialization by rescaling $(\mathbf w,b,v)$ to $(s\mathbf w,sb,v/s)$ for $s=(|v|/\|\mathbf w\|_2)^{1/2}$.

\subsubsection{Optimization and validation}\label{sec:sim_optimization}
From the same initialization, we use full-batch L-BFGS to minimize fitting mean squared error plus $\alpha$ times the sum of squared weights, for $\alpha\in\{0,10^{-5},10^{-4},10^{-3}\}$. The memory is 10, the line search uses weak Wolfe conditions and at most 20 iterations, and the cap is 250 updates with gradient and step tolerances of $10^{-5}$. We perform early stopping with a patience parameter of 25, or stop upon reaching the epoch cap, a tolerance, or a failed update. The penalty with the lowest validation error is chosen.

\subsection{Population Sharpe ratios}\label{sec:app_sim_nulls}
We set $a=0.5\times360^{-3/8}$. For market timing, let $A_0=N^{-1}\sum_{i=1}^N\mu_0(\mathbf x_{it})$ and $M_j=\mathbb E[A_0^j]$ for $j=2,4$. The oracle return is $A_0(A_0+\bar\varepsilon_t)$, with $\bar\varepsilon_t=N^{-1}\sum_i\varepsilon_{it}$. Integrating the DGP polynomials gives the finite-$N$ target
\[
\operatorname{SR}_{0,\mathrm{MT}}
=\frac{M_2}{\sqrt{M_4-M_2^2+M_2\sigma_\varepsilon^2/N}},
\qquad \sigma_\varepsilon^2=\operatorname{Var}(\varepsilon_{it}).
\]
The exact clipped-law variance differs from the stored Gaussian reference variance $\sigma^2$ by about $3.02\times10^{-23}$ in relative terms. The clipping therefore leaves the reported numerical references unchanged. The analytical market-timing targets agree with the stored values to numerical precision.

For long--short, use the oracle weights $\mathbf w$ for the chosen bandwidth and set $d=\mathbf w'\boldsymbol\mu_0$. We apply the Sharpe map to the averages of $d$ and $d^2+\sigma_\varepsilon^2\|\mathbf w\|_2^2$ over 8,192 independent predictor panels per $N$. We reuse the same panels across loadings and bandwidths. Recomputing the Sharpe map and its Monte Carlo standard error from the stored conditional moments and covariances reproduces the reported references. The largest smoothed long--short reference standard error is $0.001053$. The passive EW population Sharpe ratio is zero, so the strategy and differential nulls coincide numerically.

\subsection{Results for other values of $\tau$}\label{sec:app_sim_size}
\begin{table}[!htbp]
\centering
\footnotesize
\begin{threeparttable}
\caption{LS rejection frequencies ($\tau=0.0025$)}
\label{tab:nonlinear:tau0p0025}
\setlength{\tabcolsep}{8pt}
\begin{tabular}{rrrrrr}
\toprule
& & \multicolumn{2}{c}{LS} & \multicolumn{2}{c}{LS--EW} \\
\cmidrule(lr){3-4}\cmidrule(lr){5-6}
$\lambda$ & $N$ & C & D & C & D \\
\midrule
0.25 & 50 & 9.2 & 13.3 & 8.2 & 11.5 \\
0.25 & 150 & 6.1 & 13.2 & 6.6 & 12.3 \\
0.25 & 250 & 4.8 & 9.2 & 4.1 & 10.0 \\
0.25 & 500 & 5.7 & 8.9 & 6.5 & 8.2 \\
\addlinespace
0.40 & 50 & 9.0 & 15.0 & 7.1 & 13.5 \\
0.40 & 150 & 5.9 & 9.4 & 5.7 & 9.3 \\
0.40 & 250 & 4.5 & 8.9 & 4.4 & 7.9 \\
0.40 & 500 & 4.7 & 8.6 & 5.5 & 8.3 \\
\addlinespace
0.55 & 50 & 9.0 & 16.4 & 7.3 & 13.2 \\
0.55 & 150 & 6.9 & 12.0 & 7.3 & 11.1 \\
0.55 & 250 & 6.2 & 9.7 & 5.5 & 8.8 \\
0.55 & 500 & 5.6 & 9.5 & 6.1 & 9.5 \\
\bottomrule
\end{tabular}
\begin{tablenotes}[flushleft]\footnotesize
\item Entries are rejection percentages from 1,000 Monte Carlo replications, with a nominal size of 5\%. C and D denote conventional and debiased tests.
\end{tablenotes}
\end{threeparttable}
\end{table}

\begin{table}[!htbp]
\centering
\footnotesize
\begin{threeparttable}
\caption{LS rejection frequencies ($\tau=0.005$)}
\label{tab:nonlinear:tau0p005}
\setlength{\tabcolsep}{8pt}
\begin{tabular}{rrrrrr}
\toprule
& & \multicolumn{2}{c}{LS} & \multicolumn{2}{c}{LS--EW} \\
\cmidrule(lr){3-4}\cmidrule(lr){5-6}
$\lambda$ & $N$ & C & D & C & D \\
\midrule
0.25 & 50 & 9.5 & 12.3 & 8.6 & 10.0 \\
0.25 & 150 & 6.8 & 10.6 & 6.6 & 9.7 \\
0.25 & 250 & 5.5 & 7.8 & 4.0 & 6.4 \\
0.25 & 500 & 6.1 & 7.3 & 6.5 & 6.0 \\
\addlinespace
0.40 & 50 & 9.5 & 13.4 & 7.5 & 10.8 \\
0.40 & 150 & 6.2 & 7.9 & 5.6 & 7.5 \\
0.40 & 250 & 4.8 & 8.2 & 4.4 & 6.6 \\
0.40 & 500 & 4.9 & 7.0 & 6.0 & 6.2 \\
\addlinespace
0.55 & 50 & 9.6 & 13.2 & 7.7 & 11.2 \\
0.55 & 150 & 7.2 & 9.5 & 7.3 & 8.4 \\
0.55 & 250 & 6.1 & 7.8 & 5.5 & 7.3 \\
0.55 & 500 & 5.6 & 7.9 & 6.2 & 7.8 \\
\bottomrule
\end{tabular}
\begin{tablenotes}[flushleft]\footnotesize
\item Entries are rejection percentages from 1,000 Monte Carlo replications, with a nominal size of 5\%. C and D denote conventional and debiased tests.
\end{tablenotes}
\end{threeparttable}
\end{table}

\clearpage
\begin{table}[!htbp]
\centering
\footnotesize
\begin{threeparttable}
\caption{LS rejection frequencies ($\tau=0.01$)}
\label{tab:nonlinear:tau0p01}
\setlength{\tabcolsep}{8pt}
\begin{tabular}{rrrrrr}
\toprule
& & \multicolumn{2}{c}{LS} & \multicolumn{2}{c}{LS--EW} \\
\cmidrule(lr){3-4}\cmidrule(lr){5-6}
$\lambda$ & $N$ & C & D & C & D \\
\midrule
0.25 & 50 & 10.0 & 10.8 & 9.3 & 9.4 \\
0.25 & 150 & 7.3 & 8.4 & 6.9 & 7.2 \\
0.25 & 250 & 5.9 & 7.1 & 4.5 & 6.9 \\
0.25 & 500 & 6.2 & 5.8 & 6.5 & 5.5 \\
\addlinespace
0.40 & 50 & 10.2 & 11.1 & 8.1 & 7.8 \\
0.40 & 150 & 6.4 & 6.9 & 6.2 & 6.3 \\
0.40 & 250 & 5.8 & 6.6 & 4.9 & 6.0 \\
0.40 & 500 & 6.5 & 6.3 & 5.6 & 5.9 \\
\addlinespace
0.55 & 50 & 10.8 & 12.0 & 8.8 & 9.3 \\
0.55 & 150 & 8.2 & 7.3 & 7.6 & 5.8 \\
0.55 & 250 & 6.8 & 6.8 & 5.8 & 6.5 \\
0.55 & 500 & 6.4 & 7.2 & 6.8 & 7.5 \\
\bottomrule
\end{tabular}
\begin{tablenotes}[flushleft]\footnotesize
\item Entries are rejection percentages from 1,000 Monte Carlo replications, with a nominal size of 5\%. C and D denote conventional and debiased tests.
\end{tablenotes}
\end{threeparttable}
\end{table}

\clearpage
\begin{table}[!htbp]
\centering
\footnotesize
\begin{threeparttable}
\caption{LS rejection frequencies ($\tau=0.02$)}
\label{tab:nonlinear:tau0p02}
\setlength{\tabcolsep}{8pt}
\begin{tabular}{rrrrrr}
\toprule
& & \multicolumn{2}{c}{LS} & \multicolumn{2}{c}{LS--EW} \\
\cmidrule(lr){3-4}\cmidrule(lr){5-6}
$\lambda$ & $N$ & C & D & C & D \\
\midrule
0.25 & 50 & 11.6 & 9.6 & 10.6 & 8.6 \\
0.25 & 150 & 8.8 & 8.2 & 7.4 & 7.3 \\
0.25 & 250 & 7.0 & 6.8 & 5.7 & 6.7 \\
0.25 & 500 & 7.6 & 5.7 & 7.0 & 6.2 \\
\addlinespace
0.40 & 50 & 11.4 & 8.4 & 9.3 & 6.9 \\
0.40 & 150 & 8.1 & 7.0 & 7.5 & 5.8 \\
0.40 & 250 & 7.0 & 6.0 & 6.1 & 5.3 \\
0.40 & 500 & 7.6 & 5.4 & 6.7 & 4.9 \\
\addlinespace
0.55 & 50 & 12.8 & 10.1 & 9.7 & 7.3 \\
0.55 & 150 & 10.4 & 5.3 & 9.8 & 4.8 \\
0.55 & 250 & 8.8 & 6.7 & 7.5 & 5.8 \\
0.55 & 500 & 9.0 & 6.1 & 7.6 & 6.1 \\
\bottomrule
\end{tabular}
\begin{tablenotes}[flushleft]\footnotesize
\item Entries are rejection percentages from 1,000 Monte Carlo replications, with a nominal size of 5\%. C and D denote conventional and debiased tests.
\end{tablenotes}
\end{threeparttable}
\end{table}

\clearpage
\begin{table}[!htbp]
\centering
\footnotesize
\begin{threeparttable}
\caption{LS rejection frequencies ($\tau=0.05$)}
\label{tab:nonlinear:tau0p05}
\setlength{\tabcolsep}{8pt}
\begin{tabular}{rrrrrr}
\toprule
& & \multicolumn{2}{c}{LS} & \multicolumn{2}{c}{LS--EW} \\
\cmidrule(lr){3-4}\cmidrule(lr){5-6}
$\lambda$ & $N$ & C & D & C & D \\
\midrule
0.25 & 50 & 18.6 & 9.9 & 14.3 & 8.2 \\
0.25 & 150 & 15.8 & 6.4 & 12.5 & 6.7 \\
0.25 & 250 & 12.0 & 6.9 & 10.6 & 6.1 \\
0.25 & 500 & 13.2 & 6.6 & 11.4 & 5.5 \\
\addlinespace
0.40 & 50 & 19.5 & 10.6 & 12.8 & 8.0 \\
0.40 & 150 & 14.6 & 7.4 & 13.4 & 7.1 \\
0.40 & 250 & 13.4 & 6.0 & 10.9 & 5.7 \\
0.40 & 500 & 14.4 & 4.8 & 12.6 & 5.0 \\
\addlinespace
0.55 & 50 & 19.8 & 10.9 & 15.2 & 7.9 \\
0.55 & 150 & 18.4 & 6.2 & 13.2 & 5.5 \\
0.55 & 250 & 16.5 & 7.5 & 12.2 & 5.9 \\
0.55 & 500 & 15.2 & 4.9 & 13.5 & 5.3 \\
\bottomrule
\end{tabular}
\begin{tablenotes}[flushleft]\footnotesize
\item Entries are rejection percentages from 1,000 Monte Carlo replications, with a nominal size of 5\%. C and D denote conventional and debiased tests.
\end{tablenotes}
\end{threeparttable}
\end{table}

\clearpage
\begin{table}[!htbp]
\centering
\footnotesize
\begin{threeparttable}
\caption{LS rejection frequencies ($\tau=0.1$)}
\label{tab:nonlinear:tau0p1}
\setlength{\tabcolsep}{8pt}
\begin{tabular}{rrrrrr}
\toprule
& & \multicolumn{2}{c}{LS} & \multicolumn{2}{c}{LS--EW} \\
\cmidrule(lr){3-4}\cmidrule(lr){5-6}
$\lambda$ & $N$ & C & D & C & D \\
\midrule
0.25 & 50 & 26.3 & 10.4 & 20.1 & 7.2 \\
0.25 & 150 & 24.2 & 5.1 & 17.6 & 6.1 \\
0.25 & 250 & 19.2 & 5.1 & 16.4 & 5.2 \\
0.25 & 500 & 19.2 & 5.8 & 17.4 & 4.6 \\
\addlinespace
0.40 & 50 & 28.7 & 9.8 & 19.0 & 8.5 \\
0.40 & 150 & 23.7 & 7.1 & 18.6 & 6.1 \\
0.40 & 250 & 21.2 & 5.0 & 17.5 & 4.6 \\
0.40 & 500 & 20.4 & 4.8 & 18.4 & 5.3 \\
\addlinespace
0.55 & 50 & 29.4 & 11.1 & 18.8 & 7.5 \\
0.55 & 150 & 24.8 & 5.6 & 19.1 & 4.7 \\
0.55 & 250 & 23.5 & 5.6 & 18.0 & 4.5 \\
0.55 & 500 & 20.2 & 4.2 & 18.6 & 4.6 \\
\bottomrule
\end{tabular}
\begin{tablenotes}[flushleft]\footnotesize
\item Entries are rejection percentages from 1,000 Monte Carlo replications, with a nominal size of 5\%. C and D denote conventional and debiased tests.
\end{tablenotes}
\end{threeparttable}
\end{table}

\clearpage
\begin{table}[!htbp]
\centering
\footnotesize
\begin{threeparttable}
\caption{LS rejection frequencies ($\tau=0.2$)}
\label{tab:nonlinear:tau0p2}
\setlength{\tabcolsep}{8pt}
\begin{tabular}{rrrrrr}
\toprule
& & \multicolumn{2}{c}{LS} & \multicolumn{2}{c}{LS--EW} \\
\cmidrule(lr){3-4}\cmidrule(lr){5-6}
$\lambda$ & $N$ & C & D & C & D \\
\midrule
0.25 & 50 & 29.1 & 6.8 & 22.8 & 6.5 \\
0.25 & 150 & 24.0 & 3.9 & 18.3 & 4.3 \\
0.25 & 250 & 18.1 & 4.5 & 15.9 & 4.3 \\
0.25 & 500 & 16.7 & 4.8 & 15.8 & 4.6 \\
\addlinespace
0.40 & 50 & 28.8 & 5.6 & 20.5 & 5.3 \\
0.40 & 150 & 22.0 & 5.3 & 18.5 & 5.3 \\
0.40 & 250 & 19.8 & 3.7 & 16.2 & 3.2 \\
0.40 & 500 & 16.0 & 5.7 & 14.8 & 4.6 \\
\addlinespace
0.55 & 50 & 29.4 & 7.1 & 18.9 & 5.2 \\
0.55 & 150 & 21.3 & 4.1 & 15.5 & 3.7 \\
0.55 & 250 & 18.5 & 4.5 & 14.5 & 4.2 \\
0.55 & 500 & 14.8 & 4.0 & 14.8 & 4.1 \\
\bottomrule
\end{tabular}
\begin{tablenotes}[flushleft]\footnotesize
\item Entries are rejection percentages from 1,000 Monte Carlo replications, with a nominal size of 5\%. C and D denote conventional and debiased tests.
\end{tablenotes}
\end{threeparttable}
\end{table}

\clearpage
\begin{table}[!htbp]
\centering
\footnotesize
\begin{threeparttable}
\caption{LS rejection frequencies ($\tau=0.3$)}
\label{tab:nonlinear:tau0p3}
\setlength{\tabcolsep}{8pt}
\begin{tabular}{rrrrrr}
\toprule
& & \multicolumn{2}{c}{LS} & \multicolumn{2}{c}{LS--EW} \\
\cmidrule(lr){3-4}\cmidrule(lr){5-6}
$\lambda$ & $N$ & C & D & C & D \\
\midrule
0.25 & 50 & 23.6 & 4.9 & 18.0 & 5.9 \\
0.25 & 150 & 16.5 & 3.8 & 13.7 & 4.1 \\
0.25 & 250 & 13.2 & 4.6 & 10.6 & 4.8 \\
0.25 & 500 & 11.2 & 6.1 & 11.2 & 5.2 \\
\addlinespace
0.40 & 50 & 22.1 & 4.5 & 14.7 & 4.5 \\
0.40 & 150 & 14.7 & 4.7 & 13.2 & 5.3 \\
0.40 & 250 & 13.4 & 3.6 & 10.9 & 4.2 \\
0.40 & 500 & 11.5 & 6.0 & 9.7 & 5.4 \\
\addlinespace
0.55 & 50 & 21.9 & 5.6 & 14.1 & 4.8 \\
0.55 & 150 & 13.4 & 5.1 & 11.3 & 5.0 \\
0.55 & 250 & 12.3 & 4.5 & 9.2 & 5.0 \\
0.55 & 500 & 9.4 & 4.8 & 8.8 & 4.8 \\
\bottomrule
\end{tabular}
\begin{tablenotes}[flushleft]\footnotesize
\item Entries are rejection percentages from 1,000 Monte Carlo replications, with a nominal size of 5\%. C and D denote conventional and debiased tests.
\end{tablenotes}
\end{threeparttable}
\end{table}

\clearpage
\begin{table}[!htbp]
\centering
\footnotesize
\begin{threeparttable}
\caption{LS rejection frequencies ($\tau=0.4$)}
\label{tab:nonlinear:tau0p4}
\setlength{\tabcolsep}{8pt}
\begin{tabular}{rrrrrr}
\toprule
& & \multicolumn{2}{c}{LS} & \multicolumn{2}{c}{LS--EW} \\
\cmidrule(lr){3-4}\cmidrule(lr){5-6}
$\lambda$ & $N$ & C & D & C & D \\
\midrule
0.25 & 50 & 18.4 & 4.7 & 15.4 & 5.3 \\
0.25 & 150 & 12.5 & 4.0 & 10.2 & 4.6 \\
0.25 & 250 & 8.3 & 5.3 & 7.5 & 5.7 \\
0.25 & 500 & 8.3 & 5.9 & 8.7 & 5.4 \\
\addlinespace
0.40 & 50 & 17.4 & 4.8 & 11.5 & 4.7 \\
0.40 & 150 & 11.0 & 5.3 & 9.6 & 5.3 \\
0.40 & 250 & 9.6 & 4.4 & 7.7 & 4.7 \\
0.40 & 500 & 8.5 & 6.1 & 7.7 & 5.4 \\
\addlinespace
0.55 & 50 & 16.6 & 4.6 & 11.7 & 5.1 \\
0.55 & 150 & 10.7 & 5.4 & 8.6 & 5.9 \\
0.55 & 250 & 9.2 & 4.7 & 6.8 & 5.1 \\
0.55 & 500 & 7.0 & 5.7 & 7.1 & 5.9 \\
\bottomrule
\end{tabular}
\begin{tablenotes}[flushleft]\footnotesize
\item Entries are rejection percentages from 1,000 Monte Carlo replications, with a nominal size of 5\%. C and D denote conventional and debiased tests.
\end{tablenotes}
\end{threeparttable}
\end{table}

\clearpage
\begin{table}[!htbp]
\centering
\footnotesize
\begin{threeparttable}
\caption{LS rejection frequencies ($\tau=0.8$)}
\label{tab:nonlinear:tau0p8}
\setlength{\tabcolsep}{8pt}
\begin{tabular}{rrrrrr}
\toprule
& & \multicolumn{2}{c}{LS} & \multicolumn{2}{c}{LS--EW} \\
\cmidrule(lr){3-4}\cmidrule(lr){5-6}
$\lambda$ & $N$ & C & D & C & D \\
\midrule
0.25 & 50 & 10.9 & 5.4 & 9.9 & 4.5 \\
0.25 & 150 & 6.8 & 5.1 & 7.2 & 5.3 \\
0.25 & 250 & 4.5 & 5.5 & 6.1 & 5.6 \\
0.25 & 500 & 6.8 & 5.0 & 7.0 & 5.6 \\
\addlinespace
0.40 & 50 & 10.7 & 5.9 & 8.0 & 4.7 \\
0.40 & 150 & 7.0 & 4.6 & 6.4 & 5.1 \\
0.40 & 250 & 5.9 & 4.8 & 5.9 & 4.7 \\
0.40 & 500 & 7.2 & 6.0 & 6.6 & 5.4 \\
\addlinespace
0.55 & 50 & 12.3 & 4.9 & 8.8 & 4.7 \\
0.55 & 150 & 6.7 & 4.9 & 6.1 & 6.1 \\
0.55 & 250 & 6.5 & 5.7 & 4.8 & 4.6 \\
0.55 & 500 & 5.9 & 5.6 & 6.5 & 5.7 \\
\bottomrule
\end{tabular}
\begin{tablenotes}[flushleft]\footnotesize
\item Entries are rejection percentages from 1,000 Monte Carlo replications, with a nominal size of 5\%. C and D denote conventional and debiased tests.
\end{tablenotes}
\end{threeparttable}
\end{table}

\FloatBarrier

\begin{landscape}
\subsection{Full-grid test-statistic histograms}\label{sec:app_sim_histograms}
The figures cover all 12 settings. Each histogram uses the matching population null and overlays the standard normal density.
\begin{figure}[H]
\centering
\includegraphics[width=\linewidth,height=0.80\textheight,keepaspectratio]{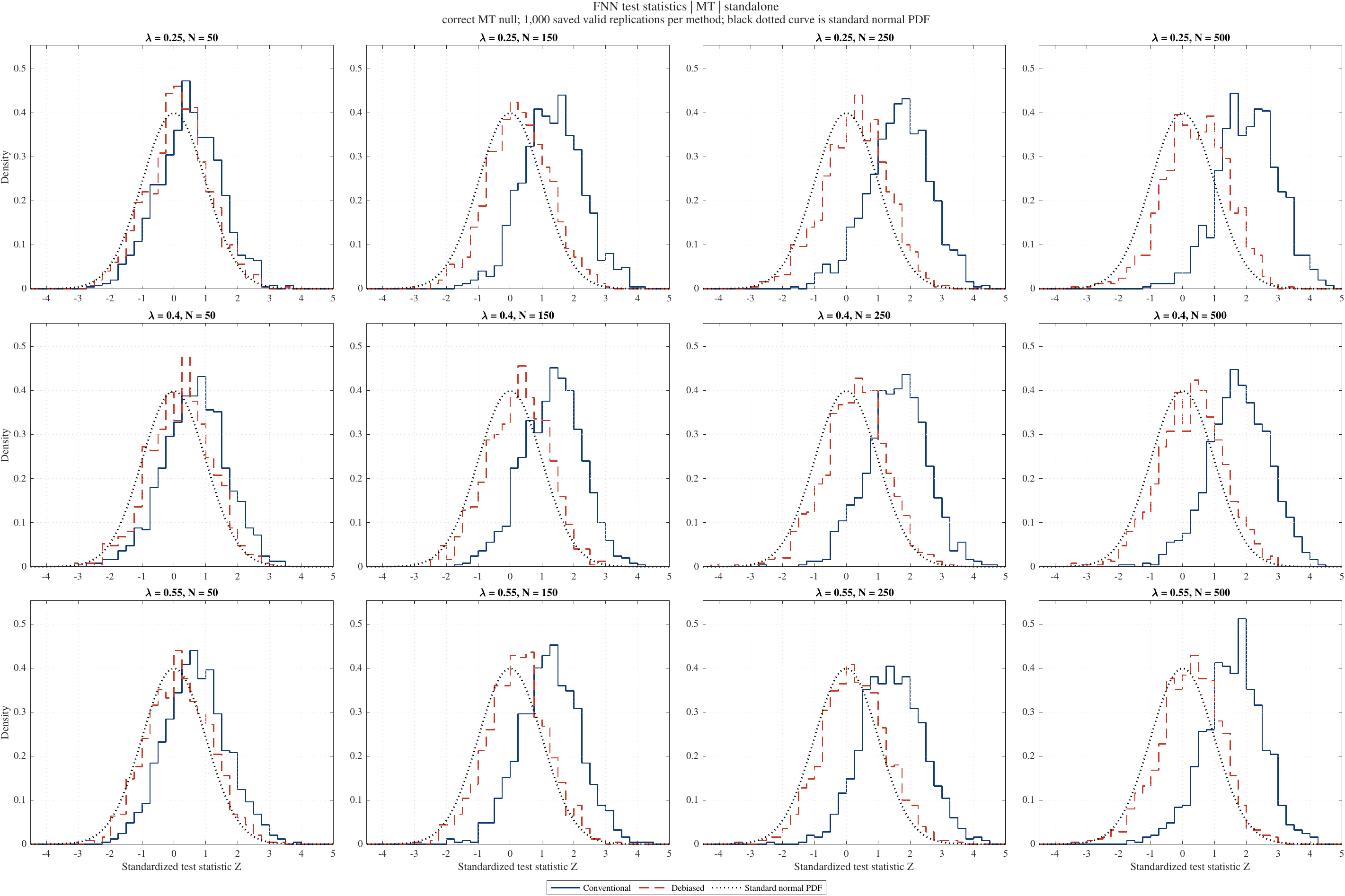}
\caption{MT test-statistic histograms, all settings. The dotted curve is the standard normal density.}\label{fig:nonlinear:hist_mt_full}
\end{figure}
\newpage
\begin{figure}[H]
\centering
\includegraphics[width=\linewidth,height=0.89\textheight,keepaspectratio]{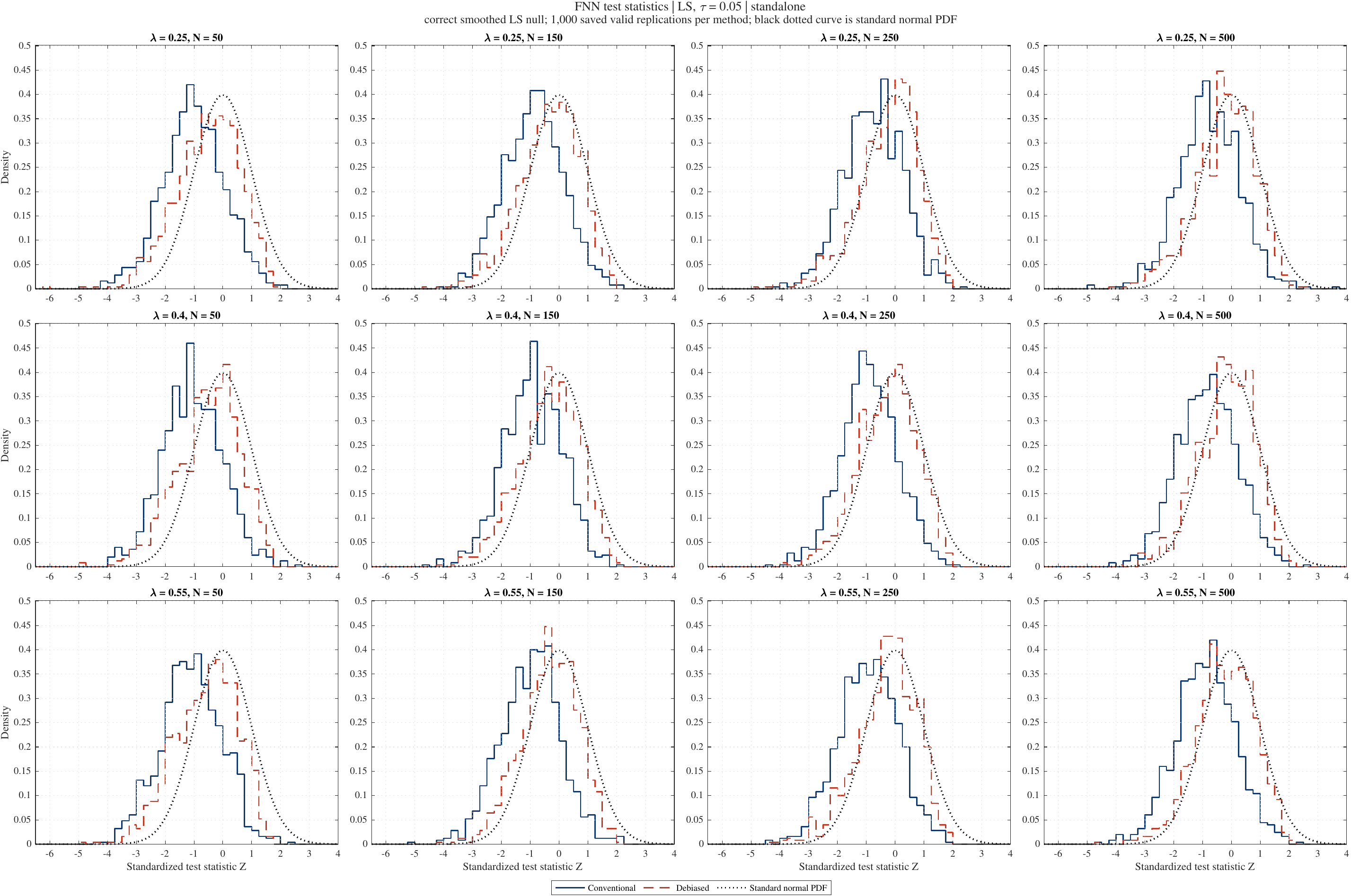}
\caption{Smoothed-LS test-statistic histograms at $\tau=0.05$, all settings. The dotted curve is the standard normal density.}\label{fig:nonlinear:hist_ls_full}
\end{figure}
\end{landscape}
\FloatBarrier

\FloatBarrier
\section{Additional Empirical Results} \label{sec:app_empirical}
\subsection{Forecast tuning}\label{supp:empirical-tuning}
For elastic net we use a mixing parameter of 0.5 and 20 logarithmically spaced penalties from $10^{-4}$ to $10^{-1}$. PCR uses $1$, $2$, $3$, $5$, $10$, $20$, $30$, $40$, $50$, $75$, or $100$ components. PLS uses $1$ through $6$, $8$, or $10$ components. The neural networks use Adam, a batch size of 10,000, a 100-epoch cap, and patience of five validation checks. The learning rate and $\ell_2$ penalty each range over $\{0.01,0.1\}$. We select the pair by validation mean squared error of the three-network ensemble.

\subsection{Passive benchmarks and comparative Sharpe inference}
Table~\ref{tab:review_passive_benchmarks} reports the annualized Sharpe ratios of the three passive benchmarks. Tables~\ref{tab:review_sp_diff_estimates} and~\ref{tab:review_ls_diff_estimates} report the estimated annualized strategy-minus-benchmark Sharpe gaps for S\&P 500 timing and long--short at $\tau=0.001$. Tables~\ref{tab:review_sp_differences} and~\ref{tab:review_ls_diff_main} in the main text report the corresponding $t$-statistics and their changes after debiasing.
\begin{table}[!htbp]
\centering
\caption{Annualized Sharpe ratios of passive benchmarks}
\label{tab:review_passive_benchmarks}
\small
\begin{tabular}{lrrr}
\toprule
Benchmark & Annual $\widehat{SR}$ & $t$ against zero & Months \\
\midrule
CRSP EW & 0.517 & 2.57 & 420 \\
CRSP VW & 0.564 & 2.90 & 420 \\
Observed-panel S\&P & 0.608 & 3.16 & 420 \\
\bottomrule
\end{tabular}
\begin{minipage}{0.98\linewidth}
\vspace{6pt}\footnotesize
\textit{Notes:} The evaluation sample is January 1987--December 2021 ($T=420$). Sharpe ratios are annualized by $\sqrt{12}$. All $t$-statistics use Newey--West HAC standard errors with five lags. Each \(t\)-statistic tests whether the corresponding benchmark’s Sharpe ratio is zero. Observed-panel S\&P is the passive portfolio of matched S\&P constituents weighted by lagged market equity from the forecasting data. CRSP EW and VW use the supplied dividend-inclusive index series. The supplied index series are converted to excess returns using the 30-day Treasury-bill rate.
\end{minipage}
\end{table}

\begin{table}[!htbp]
\centering
\caption{Annualized Sharpe differences between S\&P 500 timing and passive benchmarks}
\label{tab:review_sp_diff_estimates}
\small
\setlength{\tabcolsep}{5pt}
\begin{adjustbox}{max width=\linewidth}
\begin{tabular}{lrrrrrrr}
\toprule
 & \multicolumn{2}{c}{CRSP EW} & \multicolumn{2}{c}{CRSP VW} & \multicolumn{2}{c}{Observed-panel S\&P} &  \\
\cmidrule(lr){2-3} \cmidrule(lr){4-5} \cmidrule(lr){6-7}
Method & Conv. & Deb. & Conv. & Deb. & Conv. & Deb. & $\Delta$ estimate \\
\midrule
OLS-3 & $-0.062$ & $-0.075$ & $-0.109$ & $-0.122$ & $-0.153$ & $-0.166$ & $-0.014$ \\
ENet & $0.008$ & $0.032$ & $-0.039$ & $-0.015$ & $-0.083$ & $-0.059$ & $+0.024$ \\
PCR & $-0.008$ & $0.029$ & $-0.055$ & $-0.017$ & $-0.099$ & $-0.061$ & $+0.037$ \\
PLS & $-0.047$ & $-0.042$ & $-0.094$ & $-0.088$ & $-0.138$ & $-0.133$ & $+0.006$ \\
\midrule
NN1 & $0.096$ & $0.148$ & $0.050$ & $0.101$ & $0.005$ & $0.057$ & $+0.052$ \\
NN2 & $-0.026$ & $-0.030$ & $-0.072$ & $-0.077$ & $-0.116$ & $-0.121$ & $-0.005$ \\
NN3 & $0.072$ & $0.071$ & $0.025$ & $0.024$ & $-0.019$ & $-0.020$ & $-0.001$ \\
NN4 & $-0.006$ & $-0.021$ & $-0.052$ & $-0.067$ & $-0.096$ & $-0.111$ & $-0.015$ \\
NN5 & $0.004$ & $-0.013$ & $-0.043$ & $-0.059$ & $-0.087$ & $-0.103$ & $-0.017$ \\
\bottomrule
\end{tabular}
\end{adjustbox}
\begin{minipage}{0.98\linewidth}
\vspace{6pt}\footnotesize
\textit{Notes:} Conv. and Deb. report the conventional and debiased annualized strategy-minus-benchmark Sharpe estimates. Positive values favor the strategy. The change is debiased minus conventional, computed before rounding. It is common to all three comparisons because both procedures subtract the same passive Sharpe ratio for each benchmark (Table~\ref{tab:review_passive_benchmarks}).
\end{minipage}
\end{table}

\begin{table}[!htbp]
\centering
\caption{Annualized long--short Sharpe differences from passive benchmarks ($\tau=0.001$)}
\label{tab:review_ls_diff_estimates}
\small
\setlength{\tabcolsep}{5pt}
\begin{adjustbox}{max width=\linewidth}
\begin{tabular}{lrrrrrrr}
\toprule
 & \multicolumn{2}{c}{CRSP EW} & \multicolumn{2}{c}{CRSP VW} & \multicolumn{2}{c}{Observed-panel S\&P} &  \\
\cmidrule(lr){2-3} \cmidrule(lr){4-5} \cmidrule(lr){6-7}
Method & Conv. & Deb. & Conv. & Deb. & Conv. & Deb. & $\Delta$ estimate \\
\midrule
OLS-3 & $0.273$ & $0.247$ & $0.227$ & $0.201$ & $0.183$ & $0.156$ & $-0.026$ \\
ENet & $1.175$ & $1.179$ & $1.128$ & $1.133$ & $1.084$ & $1.089$ & $+0.004$ \\
PCR & $1.107$ & $1.084$ & $1.060$ & $1.037$ & $1.016$ & $0.993$ & $-0.023$ \\
PLS & $1.361$ & $1.298$ & $1.315$ & $1.251$ & $1.271$ & $1.207$ & $-0.063$ \\
\midrule
NN1 & $1.685$ & $1.655$ & $1.638$ & $1.608$ & $1.594$ & $1.564$ & $-0.030$ \\
NN2 & $0.950$ & $1.421$ & $0.903$ & $1.374$ & $0.859$ & $1.330$ & $+0.471$ \\
NN3 & $0.435$ & $0.419$ & $0.389$ & $0.373$ & $0.345$ & $0.329$ & $-0.016$ \\
NN4 & $-0.407$ & $-0.174$ & $-0.453$ & $-0.221$ & $-0.497$ & $-0.265$ & $+0.232$ \\
NN5 & $-0.524$ & $-0.602$ & $-0.571$ & $-0.648$ & $-0.615$ & $-0.692$ & $-0.078$ \\
\bottomrule
\end{tabular}
\end{adjustbox}
\begin{minipage}{0.98\linewidth}
\vspace{6pt}\footnotesize
\textit{Notes:} Conv. and Deb. report the conventional and debiased annualized strategy-minus-benchmark Sharpe estimates. Positive values favor the strategy. The change is debiased minus conventional, computed before rounding. It is common to all three comparisons because both procedures subtract the same passive Sharpe ratio for each benchmark (Table~\ref{tab:review_passive_benchmarks}).
\end{minipage}
\end{table}

\FloatBarrier

\subsection{Long--short bandwidth sensitivity}\label{supp:empirical-bandwidth}
Tables~\ref{tab:review_ls_tau_sr} and~\ref{tab:review_ls_tau_t} report Sharpe estimates and $t$-statistics for tests against zero at all five bandwidths. The remaining tables report $t$-statistics for comparisons with the three passive benchmarks at the four alternative bandwidths. The table notes explain the undefined debiased statistics at the three smallest bandwidths for NN4.
\begin{table}[!htbp]
\centering
\caption{Annualized long--short Sharpe ratios by bandwidth}
\label{tab:review_ls_tau_sr}
\small
\begin{tabular}{lrrrrr}
\toprule
Method & $0.0001$ & $0.00025$ & $0.0005$ & $0.001$ & $0.0025$ \\
\midrule
\multicolumn{6}{l}{\textit{Panel A: Conventional}} \\
\addlinespace[3pt]
OLS-3 & 0.792 & 0.795 & 0.796 & 0.791 & 0.767 \\
ENet & 1.718 & 1.715 & 1.707 & 1.692 & 1.638 \\
PCR & 1.467 & 1.541 & 1.596 & 1.624 & 1.613 \\
PLS & 1.891 & 1.893 & 1.891 & 1.879 & 1.828 \\
\midrule
NN1 & 2.204 & 2.201 & 2.205 & 2.202 & 2.112 \\
NN2 & 1.685 & 1.652 & 1.595 & 1.467 & 1.197 \\
NN3 & 1.033 & 1.014 & 1.005 & 0.952 & 0.810 \\
NN4 & 0.200 & 0.152 & 0.132 & 0.111 & 0.102 \\
NN5 & 0.121 & 0.042 & 0.001 & -0.007 & -0.001 \\
\midrule
\multicolumn{6}{l}{\textit{Panel B: Debiased}} \\
\addlinespace[3pt]
OLS-3 & 0.858 & 0.814 & 0.795 & 0.764 & 0.610 \\
ENet & 1.761 & 1.723 & 1.709 & 1.697 & 1.599 \\
PCR & 1.530 & 1.564 & 1.598 & 1.601 & 1.478 \\
PLS & 1.794 & 1.798 & 1.819 & 1.815 & 1.676 \\
\midrule
NN1 & 2.044 & 2.122 & 2.149 & 2.172 & 2.102 \\
NN2 & 1.507 & 1.801 & 1.918 & 1.938 & 1.668 \\
NN3 & 1.364 & 0.837 & 0.824 & 0.937 & 0.944 \\
NN4 & --- & --- & --- & 0.343 & 0.123 \\
NN5 & 0.043 & 0.041 & 0.017 & -0.084 & -0.105 \\
\bottomrule
\end{tabular}
\begin{minipage}{0.98\linewidth}
\vspace{6pt}\footnotesize
\textit{Notes:} The evaluation sample is January 1987--December 2021 ($T=420$). Sharpe ratios are annualized by $\sqrt{12}$. Conventional inference uses realized returns and their squares; debiased inference uses the orthogonal first- and second-moment scores. Column headings are smoothing bandwidths $\tau$. Dashes at the three smallest bandwidths for NN4 denote negative corrected moment variance, meaning the SR estimate is undefined. Entries are annualized Sharpe estimates. 
\end{minipage}
\end{table}

\begin{table}[!htbp]
\centering
\caption{Zero-Sharpe tests of long--short portfolios by bandwidth}
\label{tab:review_ls_tau_t}
\small
\begin{tabular}{lrrrrr}
\toprule
Method & $0.0001$ & $0.00025$ & $0.0005$ & $0.001$ & $0.0025$ \\
\midrule
\multicolumn{6}{l}{\textit{Panel A: Conventional}} \\
\addlinespace[3pt]
OLS-3 & 4.02 & 4.02 & 4.03 & 4.01 & 3.90 \\
ENet & 7.90 & 7.90 & 7.87 & 7.80 & 7.59 \\
PCR & 6.48 & 6.76 & 6.92 & 7.02 & 7.06 \\
PLS & 7.47 & 7.48 & 7.47 & 7.43 & 7.38 \\
\midrule
NN1 & 9.90 & 9.93 & 9.95 & 9.90 & 9.47 \\
NN2 & 6.14 & 6.12 & 6.01 & 5.60 & 4.69 \\
NN3 & 5.17 & 5.12 & 5.02 & 4.51 & 3.77 \\
NN4 & 1.02 & 0.77 & 0.66 & 0.55 & 0.51 \\
NN5 & 0.77 & 0.23 & 0.00 & -0.04 & -0.01 \\
\midrule
\multicolumn{6}{l}{\textit{Panel B: Debiased}} \\
\addlinespace[3pt]
OLS-3 & 4.52 & 4.54 & 4.43 & 4.04 & 2.61 \\
ENet & 8.13 & 8.73 & 8.91 & 8.82 & 7.98 \\
PCR & 6.85 & 7.56 & 7.69 & 7.64 & 6.87 \\
PLS & 8.32 & 8.62 & 8.59 & 8.39 & 7.46 \\
\midrule
NN1 & 9.88 & 10.29 & 10.61 & 10.95 & 10.00 \\
NN2 & 5.87 & 6.75 & 6.99 & 6.63 & 5.31 \\
NN3 & 0.81 & 3.26 & 4.70 & 5.20 & 4.13 \\
NN4 & --- & --- & --- & 0.37 & 0.31 \\
NN5 & 1.70 & 0.83 & 0.17 & -0.22 & -0.21 \\
\bottomrule
\end{tabular}
\begin{minipage}{0.98\linewidth}
\vspace{6pt}\footnotesize
\textit{Notes:} The evaluation sample is January 1987--December 2021 ($T=420$). Sharpe ratios are annualized by $\sqrt{12}$. All $t$-statistics use Newey--West HAC standard errors with five lags. Conventional inference uses realized returns and their squares; debiased inference uses the orthogonal first- and second-moment scores. Column headings are smoothing bandwidths $\tau$. Dashes at the three smallest bandwidths for NN4 denote negative corrected moment variance, meaning the SR estimate is undefined. Entries are \(t\)-statistics for testing whether the strategy’s Sharpe ratio is zero.
\end{minipage}
\end{table}

\begin{table}[!htbp]
\centering
\caption{$t$-statistics for long--short Sharpe differences from passive benchmarks ($\tau=0.0001$)}
\label{tab:review_ls_diff_7}
\small
\begin{tabular}{lrrr}
\toprule
Method & \shortstack{CRSP\\EW} & \shortstack{CRSP\\VW} & \shortstack{Observed-panel\\S\&P} \\
\midrule
\multicolumn{4}{l}{\textit{Panel A: Conventional}} \\
\addlinespace[3pt]
OLS-3 & 0.87 & 0.75 & 0.61 \\
ENet & 3.65 & 3.75 & 3.67 \\
PCR & 2.91 & 2.78 & 2.66 \\
PLS & 3.91 & 3.97 & 3.91 \\
\midrule
NN1 & 5.38 & 5.53 & 5.46 \\
NN2 & 3.16 & 3.14 & 3.09 \\
NN3 & 1.75 & 1.63 & 1.50 \\
NN4 & -1.27 & -1.53 & -1.70 \\
NN5 & -1.64 & -1.81 & -1.98 \\
\midrule
\multicolumn{4}{l}{\textit{Panel B: Debiased}} \\
\addlinespace[3pt]
OLS-3 & 1.10 & 1.00 & 0.86 \\
ENet & 3.84 & 3.94 & 3.86 \\
PCR & 3.02 & 2.96 & 2.86 \\
PLS & 3.96 & 4.04 & 3.98 \\
\midrule
NN1 & 5.04 & 5.16 & 5.07 \\
NN2 & 2.72 & 2.68 & 2.59 \\
NN3 & 0.50 & 0.48 & 0.45 \\
NN4 & --- & --- & --- \\
NN5 & -2.37 & -2.66 & -2.91 \\
\bottomrule
\end{tabular}
\begin{minipage}{0.98\linewidth}
\vspace{6pt}\footnotesize
\textit{Notes:} The evaluation sample is January 1987--December 2021 ($T=420$). Sharpe ratios are annualized by $\sqrt{12}$. All $t$-statistics use Newey--West HAC standard errors with five lags. Conventional inference uses realized returns and their squares; debiased inference uses the orthogonal first- and second-moment scores. Positive differences favor the strategy. Each entry tests $H_0:SR_{\mathrm{strategy}}-SR_{\mathrm{benchmark}}=0$. These are two-sided tests; the usual asymptotic 5\% critical values are $\pm1.96$. Dashes for NN4 denote nonpositive debiased moment variance; the SR and its test are undefined. Observed-panel S\&P is the passive portfolio of matched S\&P constituents weighted by lagged market equity from the forecasting data. CRSP EW and VW use the supplied dividend-inclusive index series. The supplied index series are converted to excess returns using the 30-day Treasury-bill rate.
\end{minipage}
\end{table}

\begin{table}[!htbp]
\centering
\caption{$t$-statistics for long--short Sharpe differences from passive benchmarks ($\tau=0.00025$)}
\label{tab:review_ls_diff_8}
\small
\begin{tabular}{lrrr}
\toprule
Method & \shortstack{CRSP\\EW} & \shortstack{CRSP\\VW} & \shortstack{Observed-panel\\S\&P} \\
\midrule
\multicolumn{4}{l}{\textit{Panel A: Conventional}} \\
\addlinespace[3pt]
OLS-3 & 0.88 & 0.76 & 0.62 \\
ENet & 3.65 & 3.74 & 3.67 \\
PCR & 3.12 & 3.02 & 2.91 \\
PLS & 3.92 & 3.98 & 3.92 \\
\midrule
NN1 & 5.39 & 5.53 & 5.46 \\
NN2 & 3.07 & 3.06 & 3.01 \\
NN3 & 1.69 & 1.55 & 1.43 \\
NN4 & -1.50 & -1.75 & -1.91 \\
NN5 & -1.91 & -2.04 & -2.20 \\
\midrule
\multicolumn{4}{l}{\textit{Panel B: Debiased}} \\
\addlinespace[3pt]
OLS-3 & 0.98 & 0.86 & 0.72 \\
ENet & 3.86 & 3.96 & 3.88 \\
PCR & 3.28 & 3.24 & 3.14 \\
PLS & 4.07 & 4.15 & 4.09 \\
\midrule
NN1 & 5.31 & 5.43 & 5.34 \\
NN2 & 3.48 & 3.48 & 3.39 \\
NN3 & 1.04 & 0.95 & 0.81 \\
NN4 & --- & --- & --- \\
NN5 & -2.36 & -2.63 & -2.86 \\
\bottomrule
\end{tabular}
\begin{minipage}{0.98\linewidth}
\vspace{6pt}\footnotesize
\textit{Notes:} The evaluation sample is January 1987--December 2021 ($T=420$). Sharpe ratios are annualized by $\sqrt{12}$. All $t$-statistics use Newey--West HAC standard errors with five lags. Conventional inference uses realized returns and their squares; debiased inference uses the orthogonal first- and second-moment scores. Positive differences favor the strategy. Each entry tests $H_0:SR_{\mathrm{strategy}}-SR_{\mathrm{benchmark}}=0$. These are two-sided tests; the usual asymptotic 5\% critical values are $\pm1.96$. Dashes for NN4 denote nonpositive debiased moment variance; the SR and its test are undefined. Observed-panel S\&P is the passive portfolio of matched S\&P constituents weighted by lagged market equity from the forecasting data. CRSP EW and VW use the supplied dividend-inclusive index series. The supplied index series are converted to excess returns using the 30-day Treasury-bill rate.
\end{minipage}
\end{table}

\begin{table}[!htbp]
\centering
\caption{$t$-statistics for long--short Sharpe differences from passive benchmarks ($\tau=0.0005$)}
\label{tab:review_ls_diff_9}
\small
\begin{tabular}{lrrr}
\toprule
Method & \shortstack{CRSP\\EW} & \shortstack{CRSP\\VW} & \shortstack{Observed-panel\\S\&P} \\
\midrule
\multicolumn{4}{l}{\textit{Panel A: Conventional}} \\
\addlinespace[3pt]
OLS-3 & 0.89 & 0.77 & 0.63 \\
ENet & 3.63 & 3.72 & 3.64 \\
PCR & 3.25 & 3.18 & 3.09 \\
PLS & 3.92 & 3.97 & 3.91 \\
\midrule
NN1 & 5.41 & 5.57 & 5.50 \\
NN2 & 2.91 & 2.91 & 2.86 \\
NN3 & 1.63 & 1.50 & 1.38 \\
NN4 & -1.58 & -1.81 & -1.96 \\
NN5 & -2.04 & -2.16 & -2.31 \\
\midrule
\multicolumn{4}{l}{\textit{Panel B: Debiased}} \\
\addlinespace[3pt]
OLS-3 & 0.91 & 0.80 & 0.65 \\
ENet & 3.87 & 3.98 & 3.90 \\
PCR & 3.45 & 3.41 & 3.30 \\
PLS & 4.11 & 4.20 & 4.14 \\
\midrule
NN1 & 5.45 & 5.58 & 5.49 \\
NN2 & 3.76 & 3.79 & 3.72 \\
NN3 & 1.20 & 1.10 & 0.93 \\
NN4 & --- & --- & --- \\
NN5 & -2.34 & -2.54 & -2.74 \\
\bottomrule
\end{tabular}
\begin{minipage}{0.98\linewidth}
\vspace{6pt}\footnotesize
\textit{Notes:} The evaluation sample is January 1987--December 2021 ($T=420$). Sharpe ratios are annualized by $\sqrt{12}$. All $t$-statistics use Newey--West HAC standard errors with five lags. Conventional inference uses realized returns and their squares; debiased inference uses the orthogonal first- and second-moment scores. Positive differences favor the strategy. Each entry tests $H_0:SR_{\mathrm{strategy}}-SR_{\mathrm{benchmark}}=0$. These are two-sided tests; the usual asymptotic 5\% critical values are $\pm1.96$. Dashes for NN4 denote nonpositive debiased moment variance; the SR and its test are undefined. Observed-panel S\&P is the passive portfolio of matched S\&P constituents weighted by lagged market equity from the forecasting data. CRSP EW and VW use the supplied dividend-inclusive index series. The supplied index series are converted to excess returns using the 30-day Treasury-bill rate.
\end{minipage}
\end{table}

\begin{table}[!htbp]
\centering
\caption{$t$-statistics for long--short Sharpe differences from passive benchmarks ($\tau=0.0025$)}
\label{tab:review_ls_diff_10}
\small
\begin{tabular}{lrrr}
\toprule
Method & \shortstack{CRSP\\EW} & \shortstack{CRSP\\VW} & \shortstack{Observed-panel\\S\&P} \\
\midrule
\multicolumn{4}{l}{\textit{Panel A: Conventional}} \\
\addlinespace[3pt]
OLS-3 & 0.81 & 0.68 & 0.54 \\
ENet & 3.41 & 3.49 & 3.41 \\
PCR & 3.26 & 3.23 & 3.15 \\
PLS & 3.76 & 3.81 & 3.75 \\
\midrule
NN1 & 5.14 & 5.31 & 5.24 \\
NN2 & 1.85 & 1.81 & 1.72 \\
NN3 & 0.94 & 0.81 & 0.69 \\
NN4 & -1.70 & -1.90 & -2.03 \\
NN5 & -2.06 & -2.18 & -2.33 \\
\midrule
\multicolumn{4}{l}{\textit{Panel B: Debiased}} \\
\addlinespace[3pt]
OLS-3 & 0.27 & 0.14 & 0.01 \\
ENet & 3.42 & 3.48 & 3.40 \\
PCR & 2.99 & 2.93 & 2.83 \\
PLS & 3.53 & 3.56 & 3.48 \\
\midrule
NN1 & 5.34 & 5.48 & 5.41 \\
NN2 & 2.81 & 2.85 & 2.79 \\
NN3 & 1.33 & 1.24 & 1.13 \\
NN4 & -0.98 & -1.10 & -1.19 \\
NN5 & -1.22 & -1.27 & -1.35 \\
\bottomrule
\end{tabular}
\begin{minipage}{0.98\linewidth}
\vspace{6pt}\footnotesize
\textit{Notes:} The evaluation sample is January 1987--December 2021 ($T=420$). Sharpe ratios are annualized by $\sqrt{12}$. All $t$-statistics use Newey--West HAC standard errors with five lags. Conventional inference uses realized returns and their squares; debiased inference uses the orthogonal first- and second-moment scores. Positive differences favor the strategy. Each entry tests $H_0:SR_{\mathrm{strategy}}-SR_{\mathrm{benchmark}}=0$. These are two-sided tests; the usual asymptotic 5\% critical values are $\pm1.96$. Observed-panel S\&P is the passive portfolio of matched S\&P constituents weighted by lagged market equity from the forecasting data. CRSP EW and VW use the supplied dividend-inclusive index series. The supplied index series are converted to excess returns using the 30-day Treasury-bill rate.
\end{minipage}
\end{table}

\FloatBarrier

\end{document}